\documentclass[letterpaper,twocolumn,10pt]{article}
\usepackage{zhanggroup}

\usepackage{hyperref}
\hypersetup{
  colorlinks=true,
  linkcolor={blue!70!green},
  citecolor={green!70!blue},
  urlcolor={orange!70!red}
}
\usepackage[normalem]{ulem}
\usepackage{multirow}
\usepackage{subcaption}
\usepackage{caption}
\usepackage{mathtools}
\usepackage{bbm}
\usepackage{booktabs}
\usepackage{tikz}
\usepackage{xspace}
\usepackage{amsmath,amsfonts}
\usepackage{graphicx}
\usepackage{textcomp}
\usepackage{xcolor}
\usepackage{cleveref}

\newcommand{\MembershipDoctor}{\textsc{Membership-Doctor}}
\newcommand{\mypara}[1]{\smallskip\noindent\textbf{#1.}}

\DeclareMathOperator*{\argmax}{argmax}
\DeclareMathOperator*{\argmin}{argmin}

\newcommand{\MIA}{\mathcal{I}}
\newcommand{\MLModel}{\mathcal{M}}

\newcommand{\DataPoint}{x}
\newcommand{\Dataset}{\mathcal{D}}
\newcommand{\Label}{y}
\newcommand{\Posteriors}{\mathcal{P}}
\newcommand{\AttackModel}{\mathcal{A}}
\newcommand{\TargetTrain}{\mathcal{D}_{\textit{target}}^{\textit{train}}}
\newcommand{\TargetReference}{\mathcal{D}_{\textit{target}}^{\textit{reference}}}
\newcommand{\TargetTest}{\mathcal{D}_{\textit{target}}^{\textit{test}}}
\newcommand{\ShadowTrain}{\mathcal{D}_{\textit{shadow}}^{\textit{train}}}
\newcommand{\ShadowReference}{\mathcal{D}_{\textit{shadow}}^{\textit{reference}}}
\newcommand{\ShadowTest}{\mathcal{D}_{\textit{shadow}}^{\textit{test}}}

\begin{document}

\date{}

\title{\Large \bf \MembershipDoctor: Comprehensive Assessment of Membership Inference Against Machine Learning Models}

\author{
Xinlei He\textsuperscript{1}\thanks{The first two authors made equal contributions.}\ \ \
Zheng Li\textsuperscript{1}\textsuperscript{\textcolor{blue!60!green}{$\ast$}}\ \ \
Weilin Xu\textsuperscript{2}\ \ \
Cory Cornelius\textsuperscript{2}\ \ \
Yang Zhang\textsuperscript{1}
\\
\\
\textsuperscript{1}\textit{CISPA Helmholtz Center for Information Security}\ \ \ 
\textsuperscript{2}\textit{Intel Labs}
}

\maketitle

\begin{abstract}

Machine learning models are prone to memorizing sensitive data, making them vulnerable to membership inference attacks in which an adversary aims to infer whether an input sample was used to train the model.
Over the past few years, researchers have produced many membership inference attacks and defenses, making it an emerging and rapidly growing research topic.
These attacks and defenses employ a variety of strategies and are conducted in different models and datasets.
The lack of comprehensive benchmark, however, means we do not understand the strengths and weaknesses of existing attacks and defenses.
This impedes new research in this area and makes it difficult for practitioners to select the right attack or defense.

We fill this gap by presenting a large-scale measurement of different membership inference attacks and defenses.
We systematize membership inference through the study of nine attacks and six defenses, and couple them with the most comprehensive evaluation (thus far) using six benchmark datasets and four popular model architectures.
We first measure the performance of different attacks and defenses in the holistic evaluation, and quantify the impact of the threat model on the results of these attacks.
We find that some assumptions of the threat model, such as same-architecture and same-distribution between shadow and target models, are unnecessary.
We are also the first to execute attacks on the real-world data collected from the Internet, instead of laboratory datasets.
We further investigate what determines the performance of membership inference attacks and reveal that the commonly believed overfitting level is not sufficient for the success of the attacks.
Instead, the Jensen-Shannon distance of entropy/cross-entropy between member and non-member samples correlates with attack performance much better.
This gives us a new way to accurately predict membership inference risks without running the attack.
Finally, we find that data augmentation degrades the performance of existing attacks to a larger extent, and we propose an adaptive attack using augmentation to train shadow and attack models that improve attack performance.
Our analysis relies on a modular framework named \MembershipDoctor{}, which integrates all existing attacks and defenses.
We envision that \MembershipDoctor{} will help researchers and practitioners pick the suitable attack or defense, and facilitate new research in the field.

\end{abstract}

\section{Introduction}

Machine Learning (ML), and in particular Deep Learning, has progressed tremendously over the last decade.
Despite being popular, ML models are shown to be vulnerable to various privacy attacks~\cite{FJR15,GL162,SSSS17,SS20}, represented by membership inference attacks~\cite{SSSS17,NSH18,SZHBFB19,NSH19}.
In particular, the goal of membership inference attacks is to determine whether a candidate data sample was used to train a certain ML model.
This may raise privacy concerns as the membership can reveal sensitive information about people.
For instance, identifying a person's participation in a hospital's health analytic training set reveals that this individual was once a patient in that hospital.
Membership inference attacks and corresponding countermeasures have seen remarkable advancements in the past several years, awarding researchers and model owners with a variety of methods to assess the vulnerability of models to membership leakage.

However, existing membership inference attacks and corresponding defenses have been conducted in different models (mostly trained or fine-tuned by the authors themselves) and datasets.
To gain a deeper insight on membership inference, a comprehensive benchmark is still missing.
This gives rise to the need for a holistic understanding of the risks caused by these attacks, such as the best attack and defense in the same experimental setting, the necessity of the assumption on the adversary, the intrinsic factors that affect the attacks' performance, and the attack effectiveness under ML models trained in practice, i.e., with data augmentations.
For these reasons, we conduct a holistic privacy risk assessment of ML models through the lens of membership inference attacks.
Note that in this paper, we focus on the images classification tasks which have been most widely used in previous works~\cite{YGFJ18,HMDC19,SZHBFB19,NSH19,SSM19,LF20,HWWBSZ21,LZ21}.

\subsection{Our Contributions}

\mypara{Attacks and Defenses}
In this work, we systematize membership inference research through the study of different attacks and defenses.
Concretely, regarding the attacks, we consider nine state-of-the-art attacks that leverage different information (prediction posteriors, predicted and ground truth label, or both of them) and different attack methodologies (neural network-based, metric-based, and query-based attacks) as shown in \Cref{table:mia_summary}.
Regarding the defenses, we consider six different defense mechanisms, as shown in \Cref{table:defense_summary}, namely training time defenses (label smoothing~\cite{SVISW16}, AdvReg~\cite{NSH18}, MixupMMD~\cite{LLR21}, DP-SGD~\cite{ACGMMTZ16}, and data augmentation~\cite{ZK16,CZSL20}) and inference time defenses (MemGuard~\cite{JSBZG19}).
Regarding the defense, we first evaluate them in the default setting, and then we further adjust the hyperparameters in those defenses to better study their privacy-utility trade-off (see also \Cref{figure:defense_tradeoff}).

\mypara{Threat Model Taxonomy}
We categorize the adversary's background knowledge into two dimensions, including knowledge of the target model and an auxiliary dataset (called shadow dataset).
Regarding the knowledge of the target model, we assume the adversary may or may not know its architecture/hyperparameters.
Regarding the shadow dataset, we consider it may come from the same or different distributions of the target model's training dataset.
We also investigate the case whereby the adversary leverages a shadow dataset collected from the Internet to launch the attack, which makes the attacks more realistic.

\mypara{Evaluation}
We perform the most comprehensive measurement of membership inference (thus far), jointly over four model architectures (VGG11~\cite{SZ15}, MobileNetV2~\cite{SHZZC18}, ResNet18~\cite{HZRS16}, and ResNet34~\cite{HZRS16}), six benchmark image datasets (CIFAR10~\cite{CIFAR}, CIFAR100~\cite{CIFAR}, Place20~\cite{ZLKOT18}, Place40~\cite{ZLKOT18}, Place60~\cite{ZLKOT18} and Place80~\cite{ZLKOT18}), nine state-of-the-art attacks, and six defenses, building a total of 1,296 attack scenarios.
Besides, we also investigate the necessity of different assumptions in the threat model, the confounding factors that are related to the attack performance, and the effect of data augmentation on the attack performance.

To sum up, our measurement aims to answer four research questions:

\begin{description}
    \item[\textbf{RQ1}] What is the best attack and the best defense in a fixed scenario?
    \item[\textbf{RQ2}] What are the implications of different assumptions in the threat model?
    \item[\textbf{RQ3}] What determines the performance of membership inference attacks?
    \item[\textbf{RQ4}] What is the effect of training machine learning models with data augmentations?
\end{description}

\mypara{Main Findings}
Our evaluation shows that more knowledge leads to stronger attacks in general.
Specially, considering the attacks we show in \Cref{table:mia_summary}, we show for the first time that the \texttt{NN-normal+label} attacks achieve the best attack performance and \texttt{Metric-corr} attacks perform the worst.
The reason is that the former considers both posteriors and the ground truth label while the latter only considers prediction correctness.
By jointly considering defense effectiveness and the utility of the target model, we also demonstrate for the first time that MixupMMD performs as the best defense since it reduces membership inference attack's performance to a larger extent while preserving reasonable utility.

Our measurement about the impact of the threat model on the performance of membership inference attacks shows that some assumptions of the threat model, such as the shadow dataset/model having the same distribution/architecture as the target dataset/model, are unnecessary.
Particularly, we perform this measurement on a real-world dataset collected from the Internet (called WildCIFAR100).
We discover that an adversary can still launch effective attacks even with a shadow dataset collected from the Internet: against a ResNet18 trained on CIFAR100, the attack accuracy is similar regardless of whether one uses a shadow dataset of the same distribution (0.982 on CIFAR100) or one collected from the internet (0.983 on WildCIFAR100).

\begin{figure}[!t]
\centering
\includegraphics[width=0.85\columnwidth]{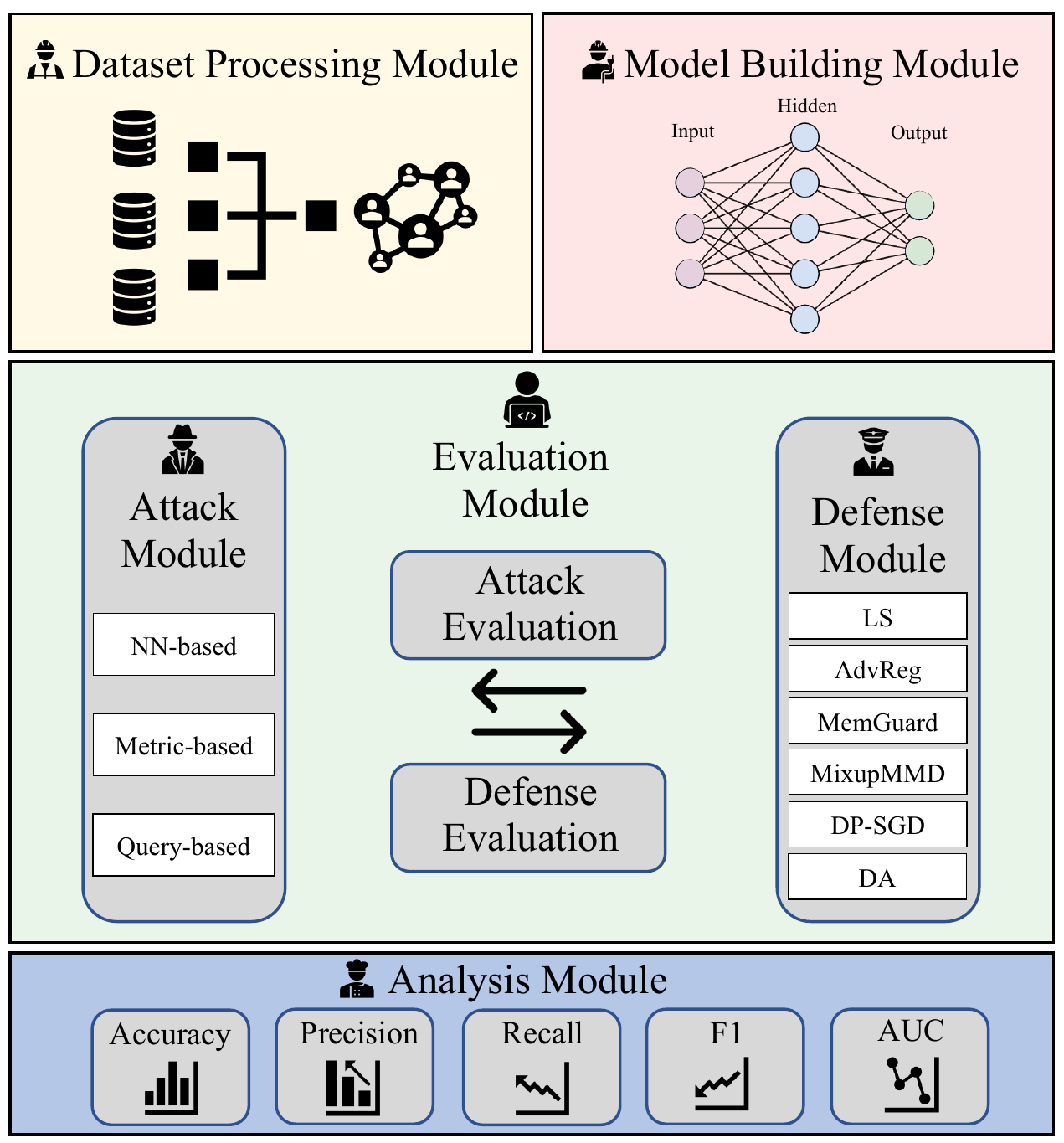}
\caption{Overview of \MembershipDoctor’s modules}
\label{figure:system}
\end{figure}

A more interesting observation is that although the overfitting level is positively correlated with the attack performance, we discover that the Jensen-Shannon distance (JS distance) of entropy/cross-entropy between members and non-members has a much higher correlation to the attack performance, and such correlation is model and dataset agnostic as shown in \Cref{figure:RQ3_scatter_correlation_sorted}.
This renders us a new way to accurately estimate membership inference risks even without running the attack.
For instance, the estimated attack performance when the target model is VGG11 train for 100 epochs on CIFAR10 is 0.768, which is very close to the ground truth 0.760
with less than 1\% bias.

Last but not the least, we find that compared to other defenses, data augmentation can be considered as a better defense as it not only mitigates existing MIAs to the largest extent but also significantly increases the target model's utility.
Moreover, data augmentation can be entangled with existing defenses to further reduce membership inference risks.
We propose an augmented attack based on data augmentation and show this improves attack performance against models that employ data augmentation.
However, even under the augmented attack, the membership privacy of models trained with data augmentation is still better preserved compared to those trained without data augmentation (see also \Cref{subsection:evaluation_RQ4} for more details).

Note that our preliminary goal is to establish a benchmark for membership inference, thus we perform similar experiments as previous works~\cite{SZHBFB19, LZ21, KD21} (e.g., \textbf{RQ2} and \textbf{RQ4}) to validate and confirm their results.
In addition, we investigate membership leakages from various new angles, such as collecting a real-world dataset (from the Internet) as the shadow dataset to conduct membership inference, and first demonstrate that JS distance correlates with attack performance above the overfitting level, which can be used to directly estimate the membership risks.
Also, we discover that data augmentations can serve as a more effective mechanism to improve the model's utility while reducing membership leakage.

\mypara{\MembershipDoctor}
We implement our code as a modular framework named \MembershipDoctor{} to evaluate state-of-the-art membership inference attacks and defenses.
\MembershipDoctor{} (shown in \Cref{figure:system}) can easily integrate additional attacks and defenses with new datasets and models by adding the functional code to the attack/defense module.
Our code will be publicly available which will facilitate researchers to benchmark new attacks and defenses in the field of membership inferences.

\section{Preliminary}
\label{section:preliminary}

\subsection{Machine Learning}

In this paper, we focus on training a classification ML model with supervised learning, as many previous works do~\cite{SSSS17,SZHBFB19,SM21}.
We consider $\Dataset$ as a set of data samples where each pair $(\DataPoint,\Label) \in \Dataset$ corresponds to a data sample $\DataPoint$ and its label $\Label$.
The goal of an ML classifier $\MLModel$ is to map a data sample $\DataPoint$ to its label $\Label$.
The model $\MLModel$'s output, given a data sample $\DataPoint$, is denoted as $\Posteriors=\MLModel(\DataPoint)$, which is a vector that indicates the probability of $\DataPoint$ belonging to a specific class.
The sum of all values in $\Posteriors$ is 1 by definition.
We refer $\Posteriors$ as posteriors in this paper.

Given a training dataset $\Dataset^{train}$, the parameters of model $\MLModel$'s are optimized by minimizing the loss function $\mathcal{L}(\Label, \MLModel(\DataPoint))$ over all training samples.
Formally, we have:

\begin{equation}
\argmin_{\MLModel} \frac{1}{|\Dataset^{train}|} \sum_{(\DataPoint,\Label) \in \Dataset^{train}} \mathcal{L}(\Label, \MLModel(\DataPoint))
\end{equation}

In this paper, we leverage cross-entropy as the loss function as it is one of the most common loss functions used for classification tasks.

\subsection{Membership Inference Attack}

As one of the most popular privacy attacks against ML models, membership inference attacks have raised the attention of various researchers~\cite{SSSS17,NSH18,NSH19,SZHBFB19,LZ21,SM21}.
The goal of membership inference attacks is to determine whether a specific sample is used to train the target model or not.
More formally, given a data sample $x$, a trained ML model $\MLModel$, and external knowledge of an adversary, denoted by $\MIA$, a membership inference attack $\AttackModel$ can be defined as the following function.
\[
\AttackModel: x, \MLModel, \MIA \rightarrow \{0, 1\}.
\]
Here, 0 means the data sample $x$ is not a member of $\MLModel$'s training dataset and 1 otherwise.
The attack model $\AttackModel$ is essentially a binary classifier.

\mypara{Threat Model}
Given a target model, we assume that the adversary only has black-box access to the target model, which means that the adversary can only query the target model with data samples and obtain the prediction posteriors.
We also assume that the adversary has a shadow dataset to train the shadow model, where the shadow dataset comes from the same distribution as the training dataset (called target dataset) of the target model.
The goal of the shadow model is to mimic the behavior of the target model to perform the attack.
In addition, we assume the shadow model has the same architecture and hyperparameters as the target model following previous works~\cite{SSSS17,SM21,LZ21}.
We further show in \Cref{subsection:relax_threat_model} that the assumption about the same distribution shadow dataset and same architecture shadow model can be relaxed.
Note that in this paper, we only focus on the image dataset, as some of our evaluated attacks~\cite{LZ21} and defenses~\cite{ZK16,CZSL20} are mainly designed for the image data.
We leave it as future work to investigate other types of datasets.

\section{Attacks and Defenses}
\label{section:MIAs}

In this section, we first discuss nine representative membership inference attacks.
Then, we summarize six defenses to mitigate membership inference attacks.

\begin{table}[!t]
\caption{Summary of membership inference attacks. For features, P denotes the use of posteriors and L denotes the use of label.}
\label{table:mia_summary}
\centering
\setlength{\tabcolsep}{1.2pt}
\begin{tabular}{l l c c}
\toprule
Attacks  & Features & Methodology & Cost \\
\midrule
\texttt{NN-top3}~\cite{SZHBFB19} & P & NN-based & Medium \\
\texttt{NN-sorted}~\cite{SZHBFB19}  & P & NN-based & Medium \\
\texttt{NN-normal}~\cite{SSSS17}  & P & NN-based & Medium \\
\texttt{NN-normal+label}~\cite{NSH18}  & P, L & NN-based & Medium \\
\texttt{Metric-corr}~\cite{LF20,SM21}  & L & Metric-based & Low \\
\texttt{Metric-conf}~\cite{SSM19,SM21}  & P, L & Metric-based & Low \\
\texttt{Metric-ent}~\cite{SSSS17,SM21}  & P, L & Metric-based & Low \\
\texttt{Metric-ment}~\cite{SM21}  & P, L & Metric-based & Low \\
\texttt{Label-only}~\cite{LZ21,CTCP21} & L & Query-based & High \\
\bottomrule
\end{tabular}
\end{table}

\begin{table*}[t]
\caption{Summary of defense methods. Note that the Low/Medium/High denotes the magnitude of the effect.}
\label{table:defense_summary}
\centering
\resizebox{1.0\linewidth}{!}{
\begin{tabular}{l c c c c c}
\toprule
Method & Extra Training Steps & Extra Inference Steps & Extra Data & Defense Effectiveness & Utility \\
\midrule
Original                                      & -         & -         & -         & -                     & -                     \\
Label Smoothing~\cite{SVISW16}              & -         & -         & -         & Decrease (Low)       & Decrease (Low)       \\
AdvReg~\cite{NSH18}                         &\checkmark & -         &\checkmark & Increase (Low)         & Same                     \\
MemGuard~\cite{JSBZG19}                   & -         &\checkmark & -         & Increase (Medium)     & Same                     \\
MixupMMD~\cite{LLR21}                       &\checkmark & -         &\checkmark & Increase (High)        & Decrease (Medium)     \\
DP-SGD~\cite{ACGMMTZ16}                         & -         & -         & -         & Increase (High)        & Decrease (High)   \\
Data Augmentation~\cite{ZK16,CZSL20}  & -         & -         & -         & Increase (High)        & Increase (High)       \\
\bottomrule
\end{tabular}
}
\end{table*}

\subsection{Existing Attacks}

Membership inference attacks can be categorized into three types: neural network-based attacks~\cite{SSSS17,NSH18,SZHBFB19}, metric-based attacks~\cite{SSSS17,YGFJ18,SSM19,LF20,SM21}, and query-based attacks~\cite{LZ21,CTCP21}.

\mypara{Neural Network-based Attacks}
In neural network-based (NN-based) attacks~\cite{SSSS17,NSH18,SZHBFB19}, given the posteriors and/or their ground truth labels, the adversary's goal is to train an attack model to distinguish the members and non-members.
Concretely, the adversary first trains a shadow model to mimic the behavior of the target model using the shadow training dataset that comes from the same distribution of the target dataset.
Then, the adversary derives an attack training dataset by querying the shadow model with the shadow training dataset (labeled as members) and shadow testing dataset (labeled as non-members).
With the attack training dataset, the adversary can train the attack model, which is a multi-layer perceptron (MLP).
Once the attack model is trained, the adversary can perform the attack over the target dataset to differentiate members and non-members.
In this paper, we investigate 4 different NN-based attacks: \texttt{NN-top3}~\cite{SZHBFB19}, \texttt{NN-sorted}~\cite{SZHBFB19}, \texttt{NN-normal}~\cite{SSSS17}, and \texttt{NN-normal+label}~\cite{NSH18}.
\texttt{NN-top3} and \texttt{NN-sorted} leverage the top 3 and whole sorted posteriors (ranked from large to small) as the input to the attack model.
\texttt{NN-normal} takes the original prediction posteriors as the input to the attack model.
For \texttt{NN-normal+label}, besides the prediction posteriors, we also add an extra bit to denote whether the prediction is correct or not and combine these two parts as the input to the attack model.

\mypara{Metric-based Attacks}
Metric-based attacks~\cite{SSSS17,YGFJ18,SSM19,LF20,SM21} need to train the shadow model as well.
However, different from NN-based attacks, metric-based attacks do not require training the attack model, instead, they may leverage a certain metric and a predefined threshold over the metric (calculated over the shadow dataset by querying the shadow model) to differentiate members and non-members.

\texttt{Metric-corr} attacks~\cite{LF20,SM21} use the prediction correctness (i.e., whether the sample is correctly classified or not) to predict membership status, which can be comparable to the NN-based attacks especially for the cases where the overfitting level, i.e., the difference between training and testing accuracy, is large.
The \texttt{Metric-corr} attack is defined as follows:
\begin{equation}
\MIA_{corr}(\MLModel(\DataPoint), \Label) = \mathbbm{1}\{\argmax_i \MLModel(x)_i=\Label\}
\end{equation}
Where $\mathbbm{1}(\cdot)$ is the indicator function.

\texttt{Metric-conf} attacks~\cite{YGFJ18,SSM19,SM21} use the prediction confidence of the correct class under the assumption that the confidence should be high for the member samples as the target model is optimized with this objective.
Song and Mittal~\cite{SM21} further improve the performance by setting class-dependant threshold $\tau_i$ for each target class.
The \texttt{Metric-conf} attack is defined as follows:
\begin{equation}
\MIA_{conf}(\MLModel(\DataPoint), \Label) = \mathbbm{1}\{\MLModel(x)_i \ge \tau_i\}
\end{equation}

\texttt{Metric-ent} attacks~\cite{SSSS17, SM21} use the entropy of prediction posteriors as a metric to decide whether a sample is a member or not, under the assumption that member samples should have a lower entropy value since the target model ought to be more confident with member samples.
Song and Mittal~\cite{SM21} further improve the performance by setting class-dependant threshold $\tau_\Label$ for each target class.
The \texttt{Metric-ent} attack is defined as follows:
\begin{equation}
\MIA_{ent}(\MLModel(\DataPoint), \Label) = \mathbbm{1}\{\ -\sum_{i}\MLModel(x)_i\log(\MLModel(x)_i) \le \tau_\Label\}
\end{equation}

\texttt{Metric-ment} attacks~\cite{SM21} use a modified entropy metric that takes advantage of both the information of prediction posteriors' entropy and the ground truth label to predict membership status.
The \texttt{Metric-ment} attack is defined as follows:
\begin{equation}
\begin{aligned}
\textsc{Ment}(\MLModel(x),y) &= -(1-\MLModel(x)_\Label)\log(\MLModel(x)_\Label)\\
& -\sum_{i \neq \Label}\MLModel(x)_i\log(\MLModel(x)_i) \\
\MIA_{ment}(\MLModel(\DataPoint), \Label) &= \mathbbm{1}\{\ \textsc{Ment}(\MLModel(x),y) \le \tau_\Label\}    
\end{aligned}
\end{equation}
Where \textsc{Ment} denotes the modified entropy.

\mypara{Query-based Attacks}
Rather than using posteriors, query-based attacks restricted the attack to using only predicted labels from the target model.
For example, \texttt{Label-only} attacks~\cite{LZ21,CTCP21} determine membership status by sending multiple queries to the target model.
Concretely, \texttt{Label-only} attacks add adversarial perturbations to the input sample until the predicted label has been changed.
The attack measures the magnitude of the perturbation and considers the data sample as a member if its magnitude is larger than a predefined threshold, which can be derived by training a shadow model.
For a given target model, the attack assumes the robustness to the adversarial perturbations is higher for a member sample as it is used to train the model.

Note that in this paper we do not consider white-box attacks~\cite{NSH19,SDSOJ19} as white-box attacks require full access to the target model's parameters, which is less realistic in the real-world scenario.
Also, as pointed out by Nasr et al.~\cite{NSH19} and Sablayroll et al.~\cite{SDSOJ19}, white-box attacks provide only limited benefits compared to the black-box attacks.

\subsection{Defense Taxonomy}

We investigate six defense mechanisms: label smoothing~\cite{SVISW16}, adversarial regularization (AdvReg)~\cite{NSH18}, MemGuard~\cite{JSBZG19}, MixupMMD~\cite{LLR21}, DP~\cite{ACGMMTZ16}, and data augmentation~\cite{ZK16,CZSL20}.
For defenses, we first keep the hyperparameters to specific values as the default settings, and then we vary those hyperparameters to investigate the trade-off between privacy and utility (see \Cref{figure:defense_tradeoff}).

\mypara{Label Smoothing}
Label smoothing (LS)~\cite{SVISW16} is a form of regularization during training the target model.
Instead of using the one-hot ground-truth label, where the probability is 1 for the true class, LS sets the probability of true class to $1-\epsilon$ and the remaining classes to $\epsilon/(k-1)$ where $k$ is the total number of classes.
We consider LS~\cite{SVISW16} as a defense against membership inference attacks as it can reduce the overfitting by making the target model less confident on training samples.
In our evaluation, we set $\epsilon=0.8$ as the default setting and vary it to $(0.2, 0.4, 0.6, 0.8)$ later in the privacy-utility trade-off analysis.

\mypara{AdvReg}
AdvReg~\cite{NSH18} uses adversarial training to reduce the efficacy of membership inference attacks.
During the target model's training, in addition to training the target model with the training dataset, the model owner also trains an adversarial classifier using the training dataset (serves as members) and a reference dataset (serves as non-members) that aims to successfully infer the membership status of the data using the posteriors generated from the target model.
The target model is updated by minimizing the original classification loss with the training dataset and maximizing the adversarial classifier's loss with both training and reference datasets.
An parameter $\lambda$ is used to control the importance of the two losses.
We set $\lambda=1$ following Nasr et al.~\cite{NSH18} as the default setting and vary it to $(1, 2, 5, 10, 20)$ later in the privacy-utility trade-off analysis.

\mypara{MemGuard}
Different from previous defenses, MemGuard~\cite{JSBZG19} does not temper the target model's training procedure.
Instead, it crafts adversarial perturbation to the prediction posteriors returned by the target model.
MemGuard is a two-phase defense.
In the first phase, the defender generates a bounded noise vector for a given posterior without changing its prediction label.
In the second phase, the defender adds the noise vector to a given posterior with a certain probability.
Since MemGuard does not change the prediction label provided by the target model, it preserves the target model utility.

\mypara{MixupMMD}
MixupMMD~\cite{LLR21} reduces the generalization gap (training accuracy minus testing accuracy) since this gap is believed to be related to the membership inference attack performance.
In addition to the normal training procedure that optimizes the target model with the training dataset, MixupMMD matches the training and validation accuracy by regularizing the Maximum Mean Discrepancy of the posterior distributions between the training and reference datasets.
We balance the two terms by setting the balance parameter $\lambda=3$ following Li et al.~\cite{LLR21} as the default setting and vary it to $(1, 2, 5, 10, 20)$ later in the privacy-utility trade-off analysis.

\mypara{DP-SGD}
Differential Privacy (DP)~\cite{DR14,LLSY16} is a general method to protect privacy by limiting the impact of data samples to the final output.
During target model training, DP-SGD~\cite{ACGMMTZ16} adds Gaussian noise to the gradient.
Regarding DP-SGD, we implement it with Opacus~\cite{OPACUS}.
The noise scale is set to $0.001$ and the max gradient norm is set to 1 as the default setting and we very the noise scale to $(0.0001, 0.001, 0.01, 0.1, 1)$ later in the privacy-utility trade-off analysis.

\mypara{Data Augmentation}
Data augmentation (DA) is a widely applied method to generate extra training samples and improve the model's utility~\cite{KSH12,ZK16,ZCDL18,CZSL20}.
Data augmentation creates additional views of the original data samples during the target model training procedure, which may reduce the target model's memorization of the original training samples.
We show later in~\Cref{subsection:evaluation_RQ4} that data augmentation is a better defense and can be entangled with other defenses to further decrease the performance of membership inference attacks.

\subsection{Adaptive Attacks}

To evaluate the defense, we consider an adaptive adversary~\cite{JSBZG19,SM21,HZ21}.
More specifically, we assume that the adversary knows the training details of target models including defense mechanisms, model architectures, and hyperparameters.
This means that the shadow model can be trained in the same way to better mimic the behavior of the target model.

\section{Evaluation}
\label{section:evaluation}

\subsection{Experimental Settings}
\label{subsection:experimental_setting}

\mypara{Datasets}
We consider six benchmark image datasets of different complexity--CIFAR10~\cite{CIFAR}, CIFAR100~\cite{CIFAR}, Place20~\cite{ZLKOT18}, Place40~\cite{ZLKOT18}, Place60~\cite{ZLKOT18}, and Place80~\cite{ZLKOT18}--to conduct our experiments.
The CIFAR10 and CIFAR100~\cite{CIFAR} datasets contain 60,000 images in 10 and 100 classes, respectively, with 6,000 training and 600 testing images per class.
The Place20, Place40, Place60, and Place80 datasets were derived from the Places365 dataset~\cite{ZLKOT18}, which originally contains more than 1.8 million images with 365 scene categories.
We randomly selected 20, 40, 60, and 80 scene categories and 3,000, 1,500, 1000, and 750 images of each category to form the Place20, Place40, Place60, and Place80 datasets (each dataset has 60,000 images).
We also consider a more realistic scenario, as described in \Cref{subsection:relax_threat_model}, where the adversary collects images from the Internet as the shadow dataset to train the shadow model.
To do this, we created a WildCIFAR100 dataset (with the same classes as CIFAR100) by collecting images from the Internet.\footnote{We collected the images from \url{https://www.naver.com/}.
Each class has roughly 550 images.}
We reshape the image size to 32$\times$32 for all datasets.

\mypara{Datasets Configuration}
For a given dataset, we randomly split it into six equal parts: $\TargetTrain$, $\TargetReference$, $\TargetTest$, $\ShadowTrain$, $\ShadowReference$, and $\ShadowTest$.
We use the $\TargetTrain$ split to train the target model and treat it as the members of the target model.
We treat the $\TargetTest$ split as non-members of the target model.
Some defense mechanisms (e.g., AdvReg and MixupMMD) use the $\TargetReference$ split to mimic the behaviors of non-members.
We use the $\ShadowTrain$ split to train the shadow model and treat it as the members of the shadow model.
We treat the $\ShadowTest$ split as the non-members of the shadow model.
We use both $\ShadowTrain$ and $\ShadowTest$ to create an attack training dataset to train the attack models.
Finally, adaptive attacks use the $\ShadowReference$ split to mimic the behaviors of the specific defense mechanism applied to the target model.
Note that the target model's performance is lower than the SOTA performance since we use fewer data in the training dataset.
For instance, we only use 10,000 images from CIFAR10 as the training dataset while the original training dataset has 50,000 images.
We also train our target model with full training dataset and the performance and the testing accuracy is close to the SOTA performance.

\mypara{Target Models}
We adopt four popular neural network architectures as our target model: VGG11~\cite{SZ15}, MobileNetV2~\cite{SHZZC18}, ResNet18~\cite{HZRS16}, and ResNet34~\cite{HZRS16}.
We trained these target models from scratch for 100 epochs with batches of 128 and use cross-entropy as the loss function.
We used SGD with an initial learning rate of 0.1 and a cosine annealing learning rate schedule.

\mypara{Attack Models}
For NN-based attacks, the attack model is a 3-layer MLP with 64, 32, and 2 hidden neurons for each layer.
We use the same hyperparameters setting as the target model but set the starting learning rate to 0.01.
For metric-based attacks, we follow the implementation of Song and Mittal~\cite{SM21}.
For \texttt{Label-only} attacks, we follow the implementation from ART~\cite{ART}.

\mypara{Metrics}
Following previous works~\cite{SSSS17,JSBZG19,NSH19,SZHBFB19,LZ21,SM21}, we use accuracy (attack success rate) as the main evaluation metric for both original classification tasks and membership inference attacks.

\subsection{Evaluate Membership Inference Attacks and Defenses}
\label{subsection:best_attack_and_defense}

Following previous works~\cite{SSSS17,SZHBFB19,SM21}, we first assume that the adversary has the same distribution shadow dataset as the target dataset and can train a shadow model which has the same architecture, hyperparameters, and defense mechanism as the target model.
We later show, in \Cref{subsection:relax_threat_model}, that these assumptions about the shadow dataset, shadow model's architecture, and hyperparameters can be relaxed.

Note that here we only consider the target model trained without data augmentation and we further study the effect of data augmentation in \Cref{subsection:evaluation_RQ4}.
For completeness, we also evaluate the attack performance with different metrics including Precision, Recall, F1, and AUC (see \Cref{appendix:diff_evaluate_metric} for more details).

\begin{figure}[!t]
\centering
\includegraphics[width=0.85\columnwidth]{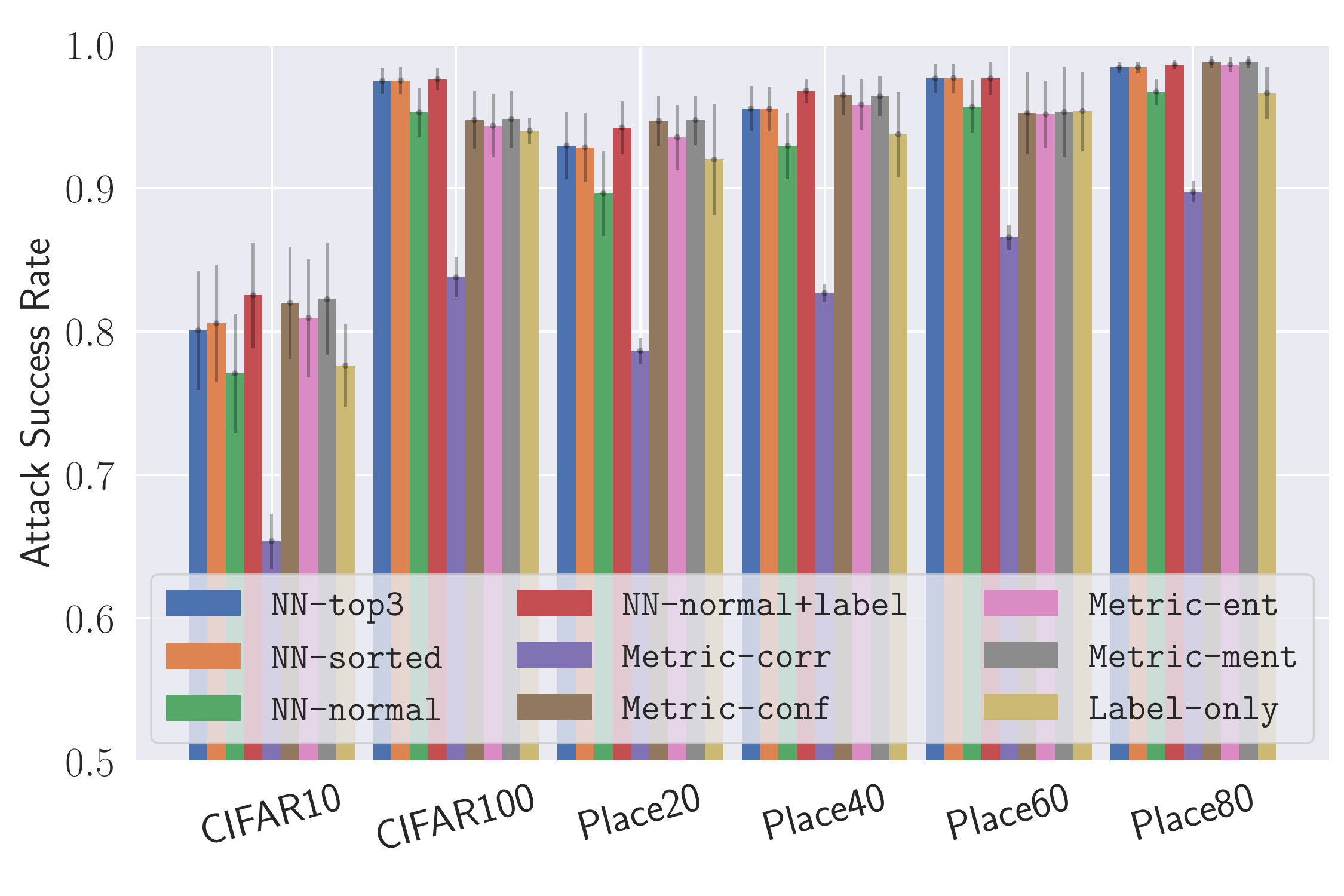}
\caption{The performance of different membership inference attacks on 6 different datasets. The x-axis represents different datasets. The y-axis represents the membership inference attack's accuracy. Note that we average the attack performance under different model architectures and show the standard deviations as well. The results for different model architectures are shown in \Cref{figure:RQ1_attack_performance_no_diff_model} (in Appendix).}
\label{figure:RQ1_attack_performance_no}
\end{figure}

\mypara{Attack Performance}
To answer which attack performs the best, we consider the target model trained normally without any defense and evaluate the attack performance against them.
\Cref{figure:RQ1_attack_performance_no} shows the performance of different attacks on different datasets.
Note that we averaged the attack performance of target models across different architectures (VGG11, MobileNetV2, ResNet18, and ResNet34) and report the mean and standard deviation values.

In general, we observe that \texttt{NN-normal+label} performs the best while \texttt{Metric-corr} performs the worst.
For instance, on CIFAR10, the average attack accuracy is 0.825 for \texttt{NN-normal+label}, while only 0.653 for \texttt{Metric-corr}.
This is expected as \texttt{NN-normal+label} considers both posteriors and predicted label's correctness while \texttt{Metric-corr} only considers predicted label's correctness.
For NN-based attacks that only consider posteriors, we observe that the attack performance can be improved by sorting the posteriors first.
For instance, on CIFAR100, the attack performance for \texttt{NN-sorted}, \texttt{NN-top3}, and \texttt{NN-normal} is 0.975, 0.974, and 0.953, respectively.
This implies that the ranked posteriors ease the process of the attack model to differentiate members and non-members.

We also find that most metric-based attacks (except \texttt{Metric-corr}) can achieve comparable performance to the best attacks.
For instance, on Place40, the attack performance of \texttt{NN-normal+label} is 0.968, while 0.965, 0.958, and 0.964 for \texttt{Metric-conf}, \texttt{Metric-ent}, and \texttt{Metric-ment}, respectively.
Also, metric-based attacks have the lowest computation cost as they do not need to train the attack models.
Therefore, if the adversary has limited computation resources, metric-based attacks may be a better choice than NN-based attacks.

When the adversary only has the predicted labels (instead of posteriors), we find that \texttt{Label-only} outperforms \texttt{Metric-corr}.
For instance, on CIFAR100, the attack performance is 0.940 for \texttt{Label-only} while only 0.838 for \texttt{Metric-corr}.
This is due to the fact that the predicted label's correctness used by \texttt{Metric-corr} is relatively coarse as many non-members are misclassified as members if the predicted label is correct.
\texttt{Label-only} provides a finer-grained metric, i.e., the magnitude of perturbation to change the predicted label, which helps to further distinguish members and non-members.
However, \texttt{Label-only} requires larger query budgets and computation cost than other attacks as it needs to query the target model multiple times and craft the adversarial perturbation to change the predicted label.

\begin{figure}[!t]
\centering
\includegraphics[width=0.85\columnwidth]{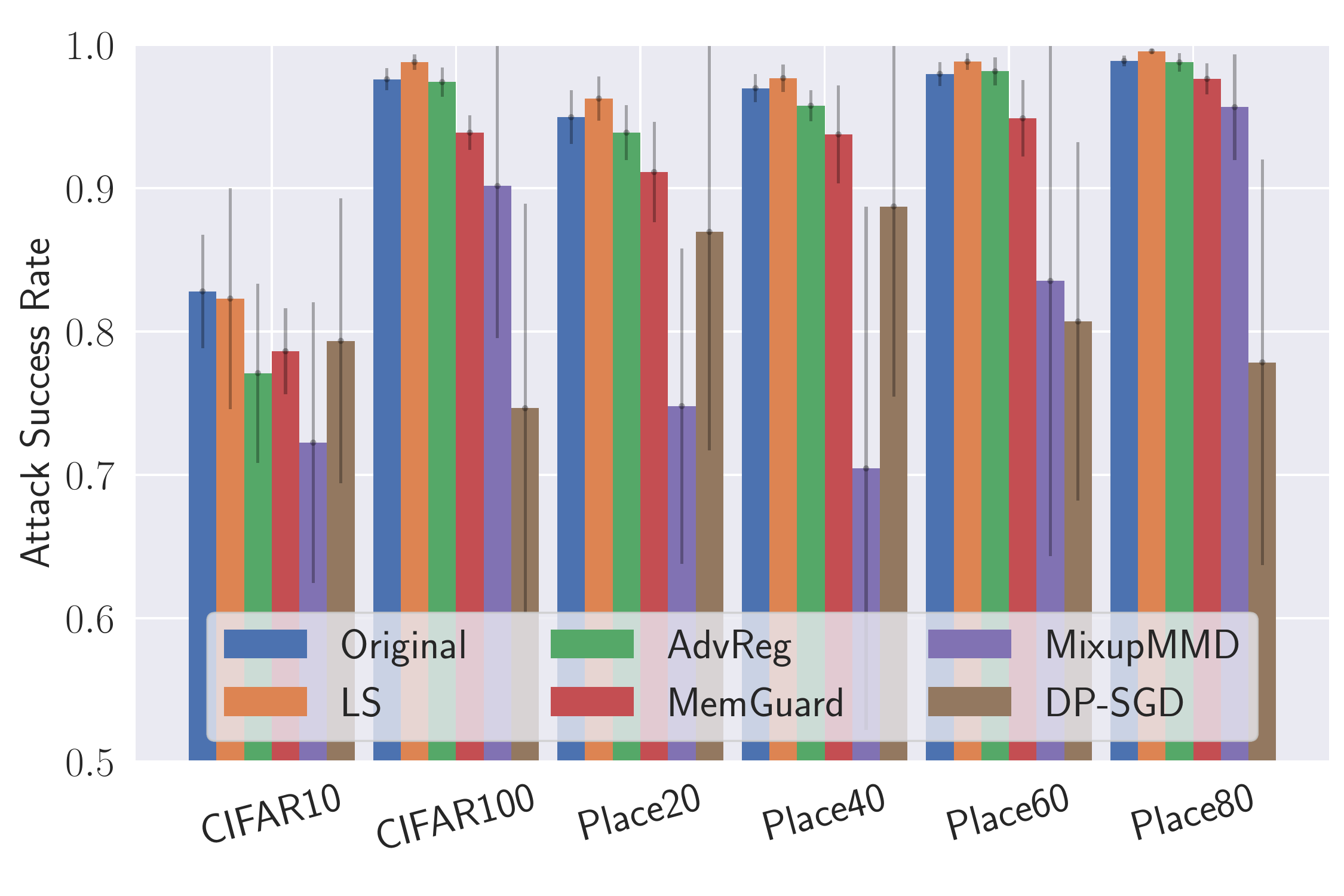}
\caption{The performance of the best membership inference attacks against the original models and models defended by different methods on 6 different datasets. The x-axis represents different datasets. The y-axis represents the membership inference attack's accuracy. Note that we average the attack performance under different model architectures and show the standard deviations as well. The results for different model architectures are shown in \Cref{figure:RQ1_defense_performance_no_diff_model} (in Appendix).}
\label{figure:RQ1_defense_performance_no}
\end{figure}

\begin{table*}[t]
\centering
\caption{The performance for original models and models defended by different methods on the original classification tasks on all the 6 datasets. Both average values and standard deviations of training accuracy and testing accuracy (in parenthesis) for different model architectures are reported.}
\resizebox{1.0\linewidth}{!} {
\begin{tabular}{l|cccccc}
\toprule
\multirow{2}{*}{\textbf{Dataset}} &
  \multicolumn{6}{c}{\textbf{Training Type}} \\
\cmidrule{2-7}
& \textbf{Original} & \textbf{LS} & \textbf{AdvReg} & \textbf{MemGuard} & \textbf{MixupMMD} & \textbf{DP-SGD} \\
\midrule
\textbf{CIFAR10} & 1.000$\pm$0.000 (0.693$\pm$0.039) & 0.984$\pm$0.026 (0.629$\pm$0.016) & 0.991$\pm$0.015 (0.658$\pm$0.035) & 1.000$\pm$0.000 (0.693$\pm$0.039) & 0.802$\pm$0.231 (0.508$\pm$0.157) & 0.891$\pm$0.174 (0.502$\pm$0.114)\\
\textbf{CIFAR100} & 1.000$\pm$0.000 (0.325$\pm$0.028) & 1.000$\pm$0.000 (0.337$\pm$0.059) & 1.000$\pm$0.000 (0.320$\pm$0.031) & 1.000$\pm$0.000 (0.325$\pm$0.028) & 0.911$\pm$0.153 (0.308$\pm$0.057) & 0.441$\pm$0.353 (0.094$\pm$0.080)\\
\textbf{Place20} & 1.000$\pm$0.000 (0.425$\pm$0.018) & 1.000$\pm$0.000 (0.400$\pm$0.041) & 1.000$\pm$0.000 (0.413$\pm$0.039) & 1.000$\pm$0.000 (0.425$\pm$0.018) & 0.710$\pm$0.220 (0.317$\pm$0.061) & 0.781$\pm$0.367 (0.216$\pm$0.082)\\
\textbf{Place40} & 1.000$\pm$0.000 (0.346$\pm$0.013) & 1.000$\pm$0.000 (0.329$\pm$0.036) & 1.000$\pm$0.000 (0.332$\pm$0.046) & 1.000$\pm$0.000 (0.346$\pm$0.013) & 0.626$\pm$0.363 (0.276$\pm$0.078) & 0.760$\pm$0.408 (0.143$\pm$0.067)\\
\textbf{Place60} & 1.000$\pm$0.000 (0.269$\pm$0.018) & 1.000$\pm$0.000 (0.256$\pm$0.037) & 1.000$\pm$0.000 (0.266$\pm$0.021) & 1.000$\pm$0.000 (0.269$\pm$0.018) & 0.785$\pm$0.359 (0.225$\pm$0.055) & 0.588$\pm$0.348 (0.084$\pm$0.047)\\
\textbf{Place80} & 1.000$\pm$0.000 (0.205$\pm$0.016) & 1.000$\pm$0.000 (0.197$\pm$0.043) & 1.000$\pm$0.000 (0.202$\pm$0.016) & 1.000$\pm$0.000 (0.205$\pm$0.016) & 0.981$\pm$0.031 (0.188$\pm$0.013) & 0.421$\pm$0.357 (0.054$\pm$0.032)\\
\bottomrule
\end{tabular}
}
\label{table:target_performance_no_aug}
\end{table*}

\mypara{Defense Performance}
To evaluate the defense performance, we consider two dimensions: defense effectiveness and the utility of the target model.
For defense effectiveness, we consider a powerful adversary that can perform all attacks and select the best one.
For utility, we consider the target model's testing accuracy on the original classification task.
We report the performance of best attacks under different defense mechanisms for each dataset as summarized in \Cref{figure:RQ1_defense_performance_no}.

Compared to the original models (blue), we find that the attack performance increases against models trained with label smoothing (orange) on almost all datasets.
On the CIFAR100 dataset, the average best attack performance is 0.976 for the original models and 0.988 for models trained with label smoothing.
Different from the cross-entropy loss calculated on the one-hot ground-truth label, which only forces the probability of the correct class to be as close to 1 as possible, label smoothing not only forces the probability of the correct class to be $1-\epsilon$, but also encourage the incorrect classes to have the same probability of $\epsilon/(k-1)$ where $k$ is the total number of class~\cite{MKH19}.
Compared to non-members, the probability of the incorrect classes is more evenly distributed around $\epsilon/(k-1)$, which makes label smoothing leaks more membership information.
Kaya and Dumitras~\cite{KD21} have the similar observation and conclude that label smoothing makes the non-maximum posteriors distribution of members to become more uniform compared to non-members, which amplifies the performance of membership inference attacks.

We find that AdvReg is more effective on datasets with fewer classes (e.g., CIFAR10, Place20, and Place40) and less effective on other datasets with more classes.
The best attacks drop the average accuracy by 0.057, 0.011, and 0.012 on CIFAR10, Place20, and Place40, respectively, while only -0.002, 0.001, and 0.002 on Place60, Place80, and CIFAR100, respectively.
AdvReg optimizes the target model by using an adversarial classifier (that infers membership information using the posteriors) to minimize membership leakage from the posteriors by increasing the loss of the adversarial classifier.
As the total number of simulated ``non-members'' (reference dataset) are fixed in our experiment, more classes mean each class has fewer ``non-members'', thus making it more difficult for AdvReg to learn the posteriors distribution of ``non-members'' and therefore hide the information of members through adversarial training.

MemGuard has better defense performance than AdvReg.
On CIFAR100, the average accuracy drop for the best attacks is 0.037 for MemGuard while only 0.001 for AdvReg.
MemGuard manually adds noises to the posteriors of members and non-members, which makes the attacks that contain the information of posteriors less effective.
However, attacks that consider predicted labels still threaten membership privacy as MemGuard does not affect \texttt{Metric-corr} and \texttt{Label-only} attacks because it does not change the predicted label.

Regarding MixupMMD, it has better defense performance compared to AdvReg and MemGuard.
For instance, on CIFAR10, the best attack performance is 0.771 and 0.786 for AdvReg and MemGuard, while only 0.722 for MixupMMD (see \Cref{figure:RQ1_defense_performance_no}).
MixupMMD leverages mix-up~\cite{ZCDL18} to train the target model which reduces the generalization gaps between training and testing sample.
It also uses MMD regularization to reduce the gap of posteriors distribution for members (training dataset) and non-members (reference dataset), which makes it more difficult for the adversary to separate them.

Similar to MixupMMD, DP-SGD, can also defend against membership inference attacks to a larger extent than AdvReg and MemGuard.
On CIFAR100, the best attack performance is 0.974 and 0.939 for AdvReg and MemGuard, while only 0.746 for DP-SGD.
This implies that adding noise during the training process reduces the risk of membership inference attacks because it decreases the boundary between members and non-members.
However, as pointed out by previous works~\cite{JE19,LLR21}, it is hard to achieve a privacy guarantee while maintaining the utility when applying DP method to train a large machine learning model (e.g., the privacy budget $\epsilon=4.34\times10^9$ for a ResNet18 trained on CIFAR10).

In addition to defense effectiveness against different attacks, another important factor to evaluate the defense is the utility (testing accuracy) on the original classification tasks.
We report the training/testing accuracy on the original classification tasks for original models and models defended by different methods in~\Cref{table:target_performance_no_aug}.

\begin{figure}[!t]
\centering
\begin{subfigure}{0.45\columnwidth}
\includegraphics[width=\columnwidth]{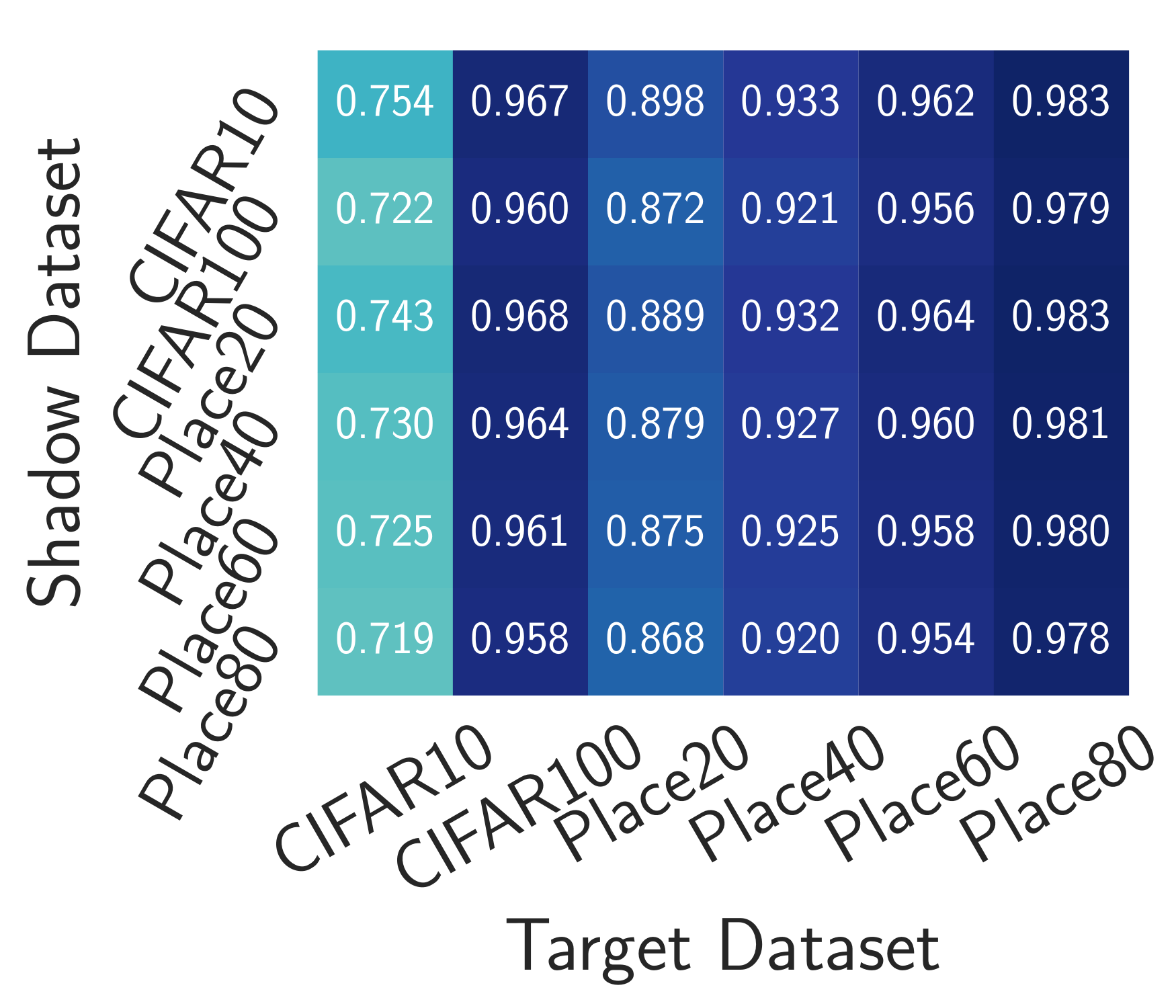}
\caption{VGG11}
\label{figure:RQ2_attack_VGG11_no_aug}
\end{subfigure}
\begin{subfigure}{0.45\columnwidth}
\includegraphics[width=\columnwidth]{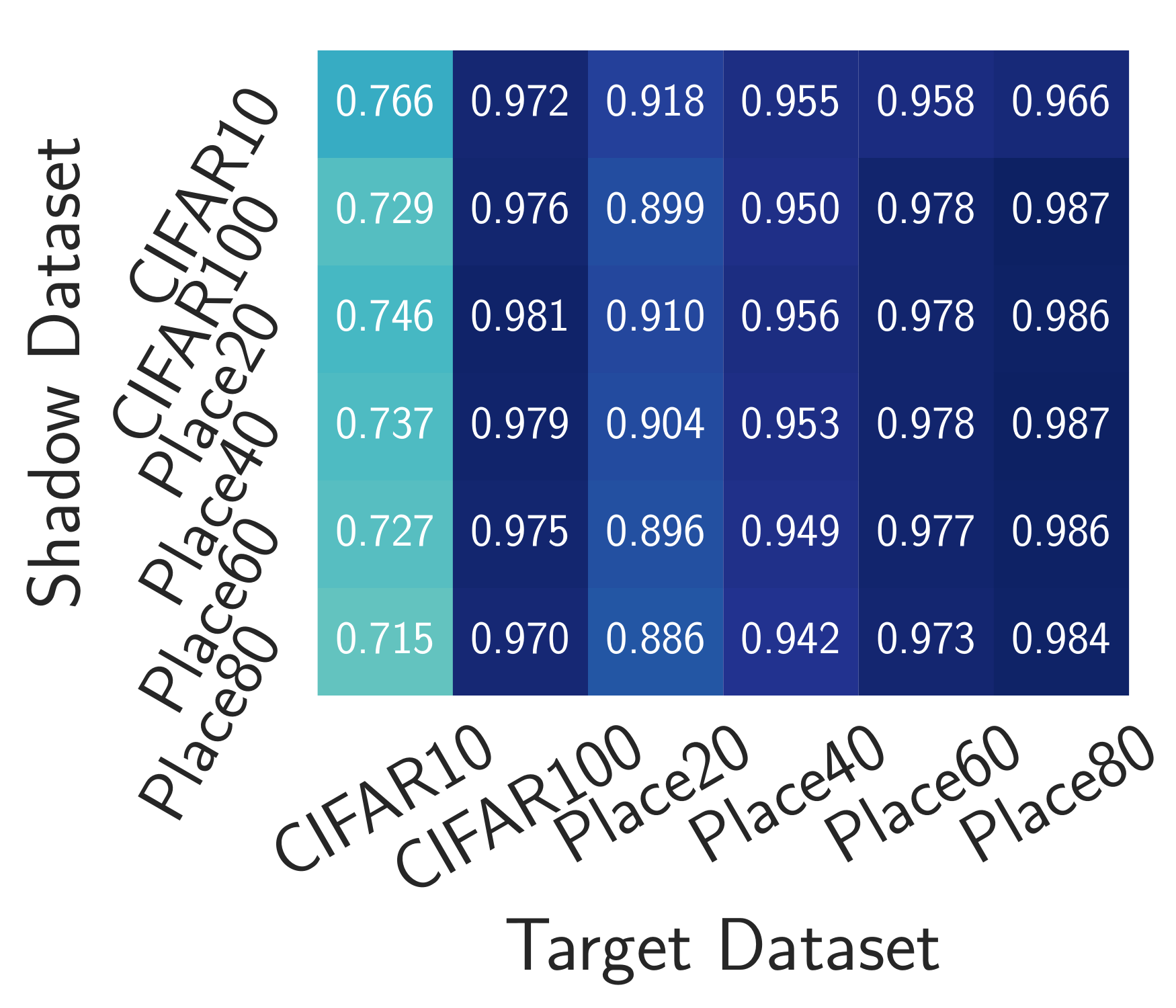}
\caption{MobileNetV2}
\label{figure:RQ2_attack_mobilenetv2_no_aug}
\end{subfigure}
\caption{The performance of membership inference attack (\texttt{NN-top3}) when the shadow dataset comes from different distributions of the target dataset.}
\label{figure:RQ2_relax_assumption_shadow_dataset_no_aug} 
\end{figure}

We find that label smoothing preserves the utility in most cases, like on Place20 where the utility drop is only 0.025.
We also observe that AdvReg can preserve the utility, where the utility drop on CIFAR100 is only 0.005.
As pointed out by Nasr et al.~\cite{NSH18}, the adversarial training procedure also serves as a regularizer and makes the model generalize to unseen samples.
As MemGuard does not temper the training process of the target model and preserve the predicted label while adding noise in the inference stage, the utility drop is 0.
For MixupMMD, the utility drop is larger, but still reasonable especially for datasets with larger numbers of classes, e.g., 0.044, 0.017, and 0.017 on Place60, Place80, and CIFAR100, respectively.
DP-SGD suffers the largest utility drop at 0.191 and 0.231 on CIFAR10 and CIFAR100, respectively.
In general, we consider MixupMMD as the best defense as it reduces the attack performance to a large extent while preserving reasonable utility.

The above defenses are evaluated in their default setting, we further investigate whether those defenses can achieve better privacy-utility trade-off if we change its default hyperparameters (see the first row of \Cref{figure:defense_tradeoff}).
We find that in most of the cases, our assumption is still hold, i.e., DP-SGD (brown) and MixupMMD (purple) can reduce the membership leakage to the largest extent but MixupMMD can better preserve the target model's utility.

\mypara{Takeaways}
Regarding the attacks, we observe that \texttt{NN-normal+label} performs best as it contains both posteriors and label information.
Most metric-based attacks achieve comparable performance to the \texttt{NN-normal+label} attack yet have less computational cost.
If the adversary only has access to predicted labels instead of the posteriors, the \texttt{Label-only} attacks provide better performance than \texttt{Metric-corr} attacks.
\texttt{Label-only} attacks leverage the magnitude of perturbation to change the prediction label as a metric that is more fine-grained to tell the difference between members and non-members.
For defenses, we observe that MixupMMD and DP-SGD provide the best defense effectiveness.
However, MixupMMD preserves more of the target model's utility.

\subsection{Implication of Different Threat Models}
\label{subsection:relax_threat_model}

\begin{figure}[!t]
\centering
\begin{subfigure}{0.45\columnwidth}
\includegraphics[width=\columnwidth]{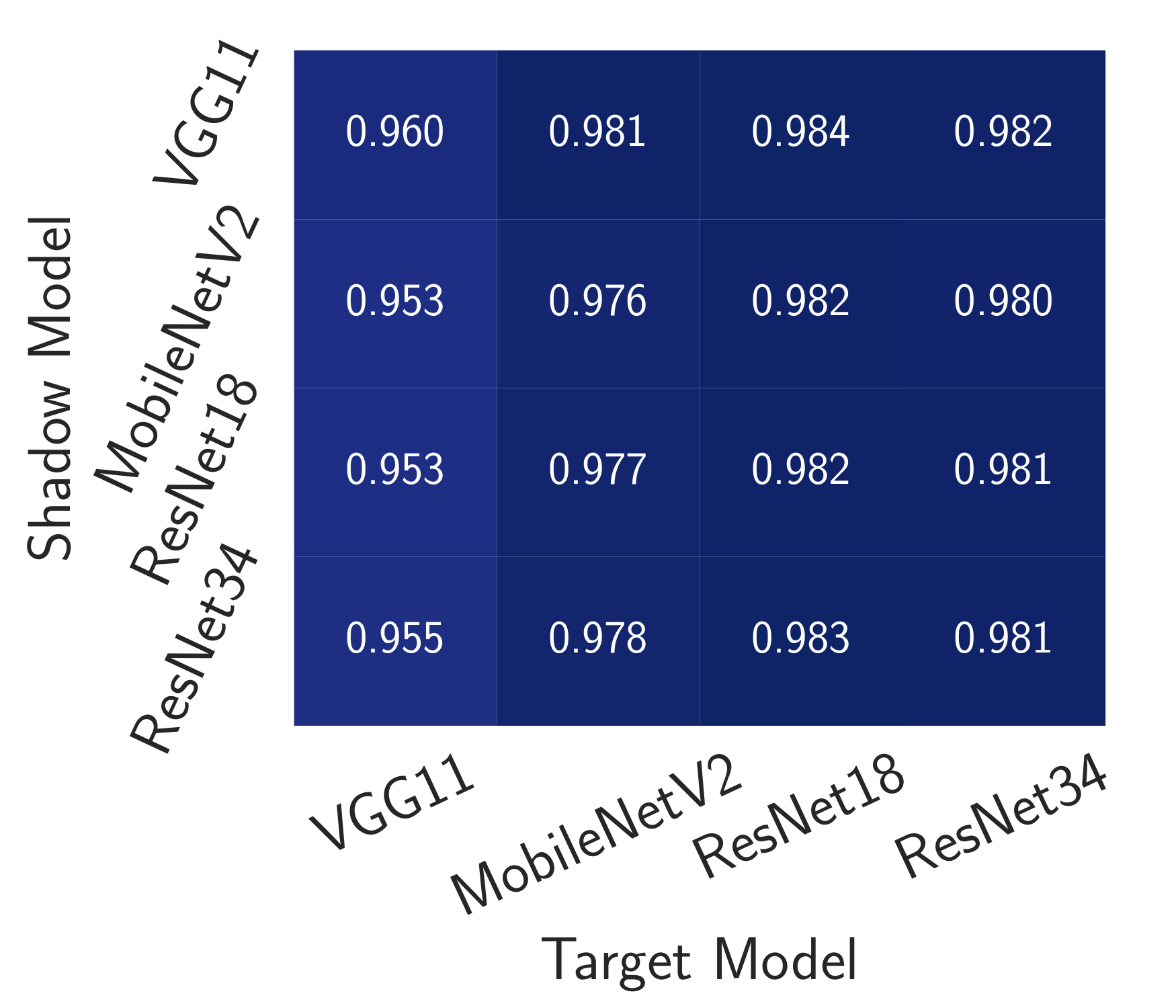}
\caption{CIFAR100}
\label{figure:RQ2_attack_CIFAR100_no_aug}
\end{subfigure}
\begin{subfigure}{0.45\columnwidth}
\includegraphics[width=\columnwidth]{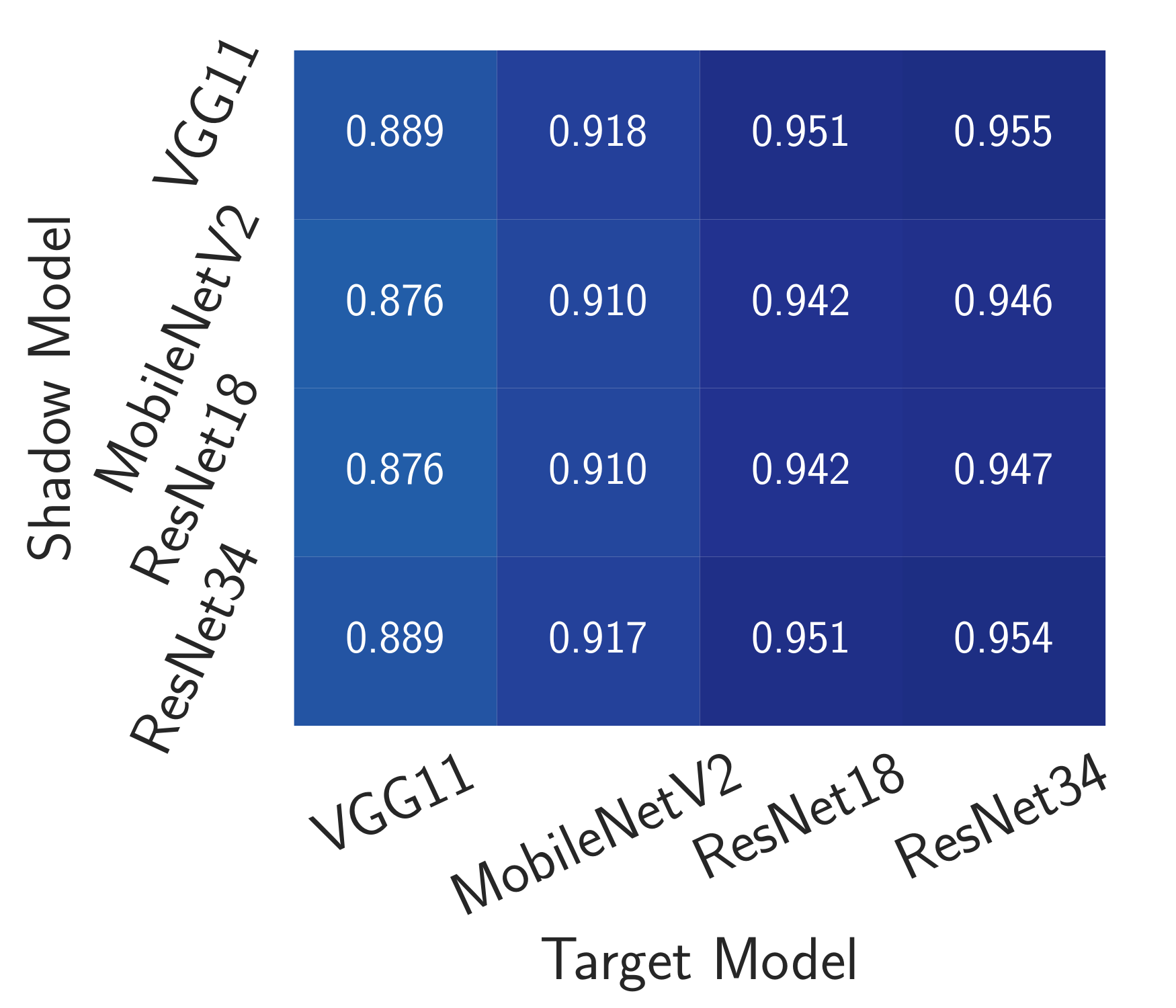}
\caption{Place20}
\label{figure:RQ2_attack_Place20_no_aug}
\end{subfigure}
\caption{The performance of membership inference attack (\texttt{NN-top3}) when the shadow model has different architecture compared to the target model.}
\label{figure:RQ2_attack_relax_assumption_shadow_model_no_aug} 
\end{figure}

Many previous works~\cite{SSSS17,SM21,LZ21} focused on the setting where the adversary trains a shadow model (of the same architecture and hyperparameters as the target model) on a shadow dataset that comes from the same distribution as the target dataset.
These works generate the target dataset and the "same distribution" shadow dataset by splitting a dataset into two parts.
We relax this assumption by creating the shadow dataset that totally collected from the Internet.
A successful attack in these cases means the threat of membership inference attacks is largely underestimated.

To compare attack performance fairly, we focus on the \texttt{NN-top3} attack, which can handle the case where the target and shadow datasets come from different distributions and have different numbers of classes.
We first investigate whether attackers can relax the assumption of a same-distribution shadow dataset and same-architecture/hyperparameters shadow model.
To do this, we evaluate whether such an attack is still effective using a shadow dataset that is totally collected from the Internet.

\mypara{Different Distribution Shadow Dataset}
\Cref{figure:RQ2_relax_assumption_shadow_dataset_no_aug} shows the attack performance when the shadow dataset distributed differently from the target dataset (see more results of other model architectures in \Cref{figure:RQ2_relax_assumption_shadow_dataset_no_aug_appendix} in Appendix).
We observe that the attack performance remains almost the same even if the shadow model is trained with different distribution shadow datasets for all model architectures.
For instance, in \Cref{figure:RQ2_attack_VGG11_no_aug} where the target dataset is CIFAR100, the attack performance is 0.960 when the shadow dataset is also CIFAR100, and the attack is still effective (0.958) when the shadow dataset is Place80.
Such observation indicates that we can relax the assumption of a same-distribution shadow dataset.

\mypara{Different Shadow Model's Architecture}
\Cref{figure:RQ2_attack_relax_assumption_shadow_model_no_aug} shows that the attacks are still effective even when target and shadow models' architectures are different (see more results of other datasets in \Cref{figure:RQ2_attack_relax_assumption_shadow_model_no_aug_appendix} in Appendix).
For instance, on Place20 (\Cref{figure:RQ2_attack_Place20_no_aug}), the attack performance is 0.954 when ResNet34 is the model architecture for both target and shadow models, and it is 0.947 when the shadow model's architecture changes to ResNet18.
Such observation hints that we can relax the assumption of a same-architecture shadow model.

\mypara{Different Hyperparameters}
We also investigate the cases when the shadow model has different hyperparameter settings compared to the target model.
In this setting, we consider the shadow dataset comes from the same distribution as the target dataset and the shadow model has the same architecture but different hyperparameters compared to the target model.
Due to the space limitation, we only show the results for the CIFAR10 dataset (shown in \Cref{table:relax_assumption_hyperparameters}) and other datasets show similar trends.
We find that the attacks are still effective even when the shadow model has a different learning rate, batch size, or training epoch compared to the target model.

\begin{figure}[!t]
\centering
\begin{subfigure}{0.45\columnwidth}
\includegraphics[width=\columnwidth]{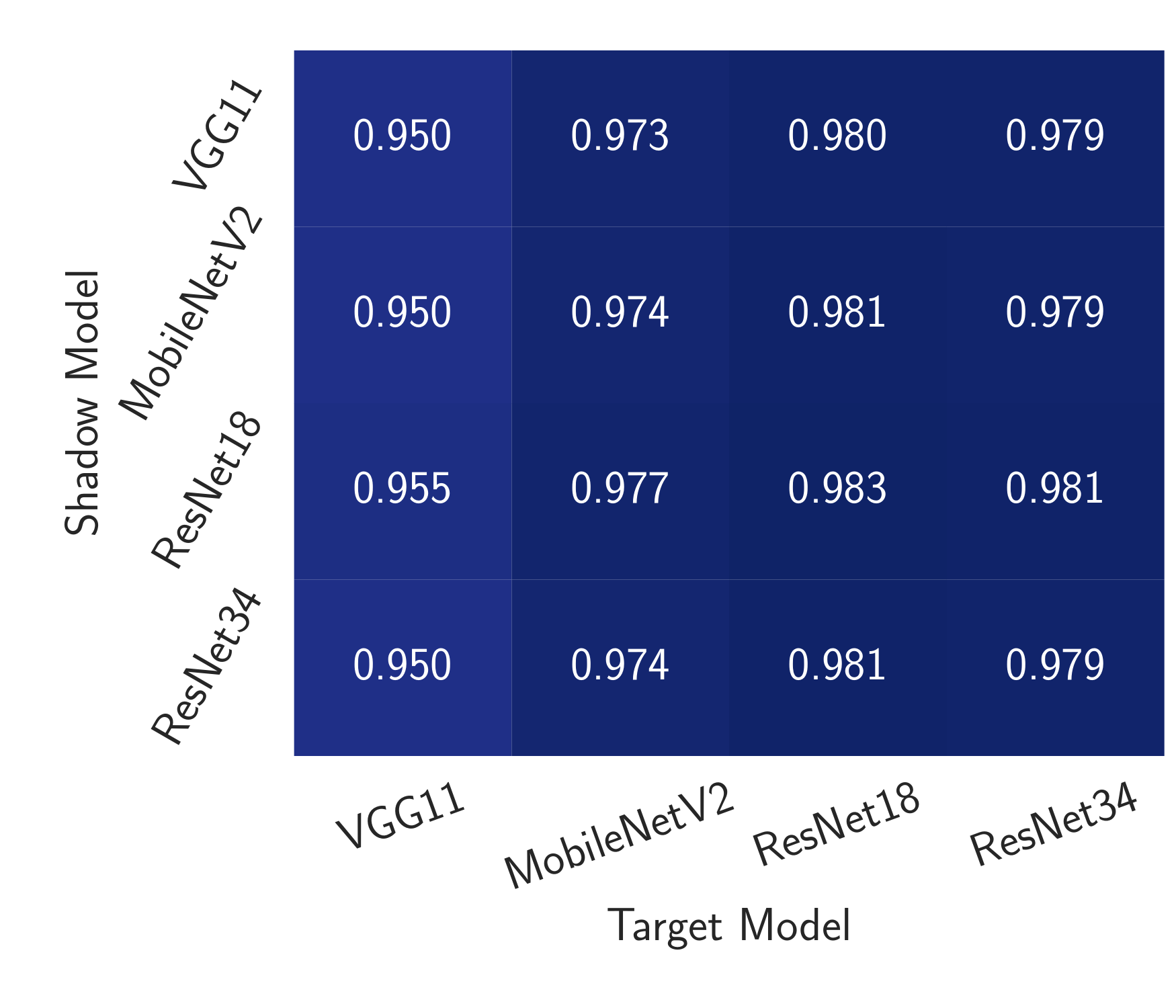}
\caption{\texttt{NN-top3}}
\label{figure:relax_assumption_wildcifar100_NN-top3}
\end{subfigure}
\begin{subfigure}{0.45\columnwidth}
\includegraphics[width=\columnwidth]{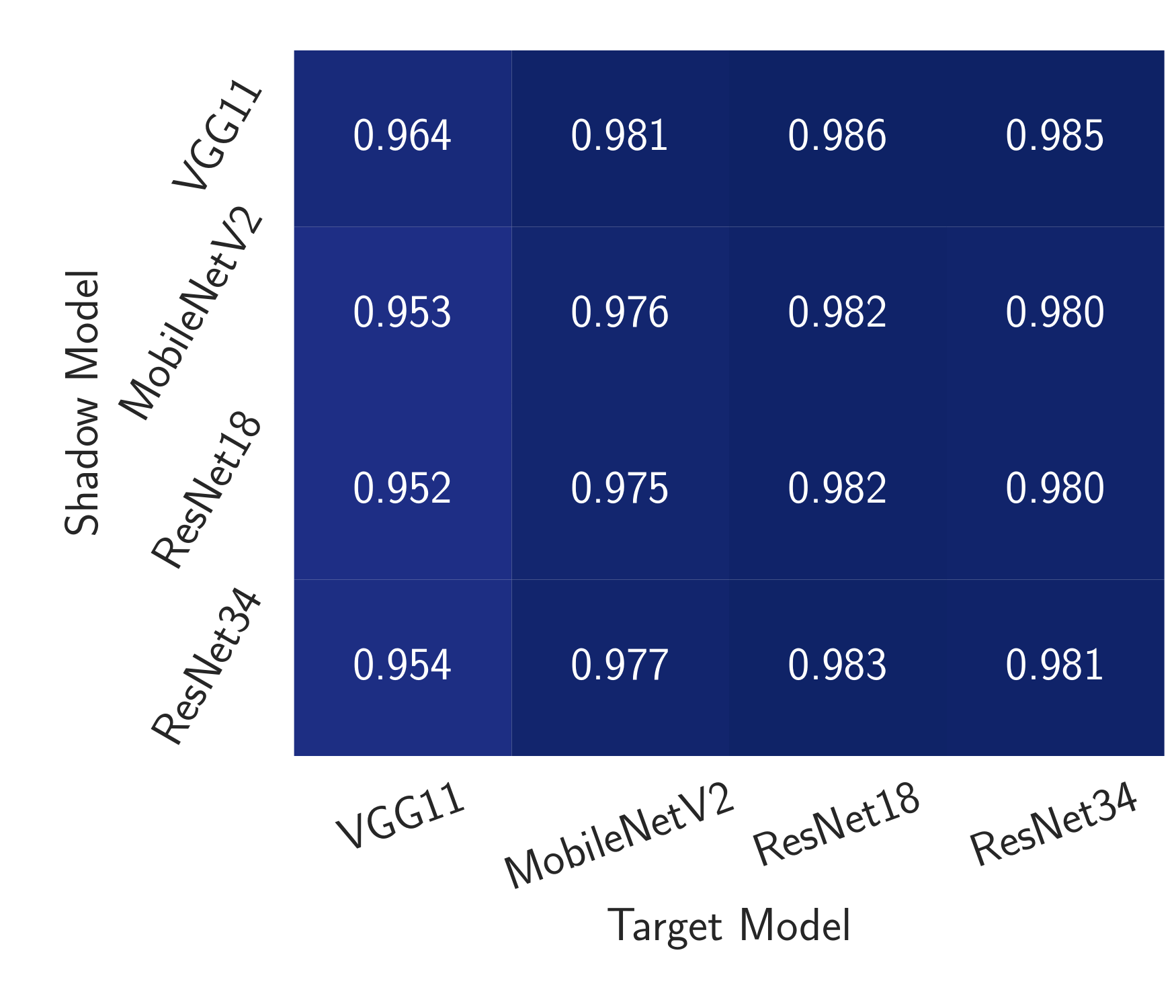}
\caption{\texttt{NN-sorted}}
\label{figure:relax_assumption_wildcifar100_NN-sorted}
\end{subfigure}
\caption{The performance of membership inference attack (\texttt{NN-top3} and \texttt{NN-sorted}) when the shadow model has different architecture compared to the target model. Note that here the target dataset is CIFAR100 and the shadow dataset is WildCIFAR100.}
\label{figure:RQ2_attack_WildCIFAR100_no_aug}
\end{figure}

\mypara{Shadow Dataset in The Wild}
So far, we consider the ``same distribution'' shadow dataset based on the images sampled from the same dataset following previous works~\cite{SSSS17,SZHBFB19,SM21}.
However, in reality, to construct a shadow dataset, the adversary can also collect the dataset from the Internet as well.
Taking CIFAR100 as a case study, to construct the shadow dataset, we first gather the 100 classes from CIFAR100 and collect images from the Internet.
Concretely, for each class, we collect roughly 550 images from Naver.\footnote{\url{https://www.naver.com/}}
We then conduct the center crop for each image and resize it to 32$\times$32.
We name this dataset WildCIFAR100 and randomly split it into six equal parts (same as other datasets).
Then we follow the same way we used before to train the shadow models and attack models using the WildCIFAR100 dataset.
Once the attack model is trained, we evaluate its performance on CIFAR100 as summarized in \Cref{figure:RQ2_attack_WildCIFAR100_no_aug}.
We observe that even if the shadow dataset is collected purely from the Internet, the attacks are still effective.
For instance, all attacks reach over 0.950 accuracy when the target dataset is CIFAR100 and the shadow dataset is WildCIFAR100, even in the cases where shadow and target models have different model architectures.
To the best of our knowledge, we are the first to quantify the effectiveness of membership inference attacks with a shadow dataset that is totally collected from the Internet.
This experiment implies that the threat of membership inference attacks is underestimated.

\mypara{Takeaways}
We show that membership inference attacks can be effective even under relaxed assumptions on the shadow dataset and shadow model's architecture/hyperparameters.
By collecting images from the Internet with the same classes provided by the target model, our evaluation shows that attacks remain effective in this more realistic case, indicating that the threat of membership inference is underrated.
Such observation reveals that there might be some more intrinsic factors that lead to the success of membership inference attacks regardless of datasets and models.
We further investigate this in \Cref{subsection:evaluation_RQ3}.

\begin{table}[!t]
\caption{The performance of membership inference attack (\texttt{NN-top3}) when target model is ResNet18 trained on CIFAR10 with learning rate (LR) as 0.1, batch size (BS) as 128, and training epoch (Epoch) as 100. We only relax one shadow model's hyperparameter for each row while keeping others the same and report the corresponding attack performance.
The value for different settings are: LR: $(0.1, 1, 0.01, 0.001)$; BS: $(128, 32, 64, 256)$; Epoch: $(100, 80, 60,40)$.}
\label{table:relax_assumption_hyperparameters}
\centering
\begin{tabular}{l | c c c c}
\toprule
Name & Original & Setting1 & Setting2 & Setting3\\
\midrule
LR & 0.850 & 0.851 & 0.853 & 0.852 \\
BS & 0.850 & 0.853 & 0.848 & 0.839 \\
Epoch & 0.850 & 0.848 & 0.842 & 0.848 \\
\bottomrule
\end{tabular}
\end{table}

\subsection{What Determines the Success of Membership Inference Attacks}
\label{subsection:evaluation_RQ3}

The above results demonstrate the effectiveness of membership inference attacks.
Here, we delve more deeply into the reasons for the success of membership inference like which factors determine the vulnerability of the model to membership inference attacks.
To do this, we focus on the correlations between several factors and attack performance.
We investigate three distance measures: the overfitting level and the Jensen-Shannon distance of entropy or cross-entropy distribution.
For attacks, we use the \texttt{NN-sorted} and \texttt{NN-normal+label} attacks, as the former performs the best with only posteriors information while the latter performs the best with both posteriors and ground truth label information.

\begin{figure}[!t]
\centering
\begin{subfigure}{0.45\columnwidth}
\includegraphics[width=\columnwidth]{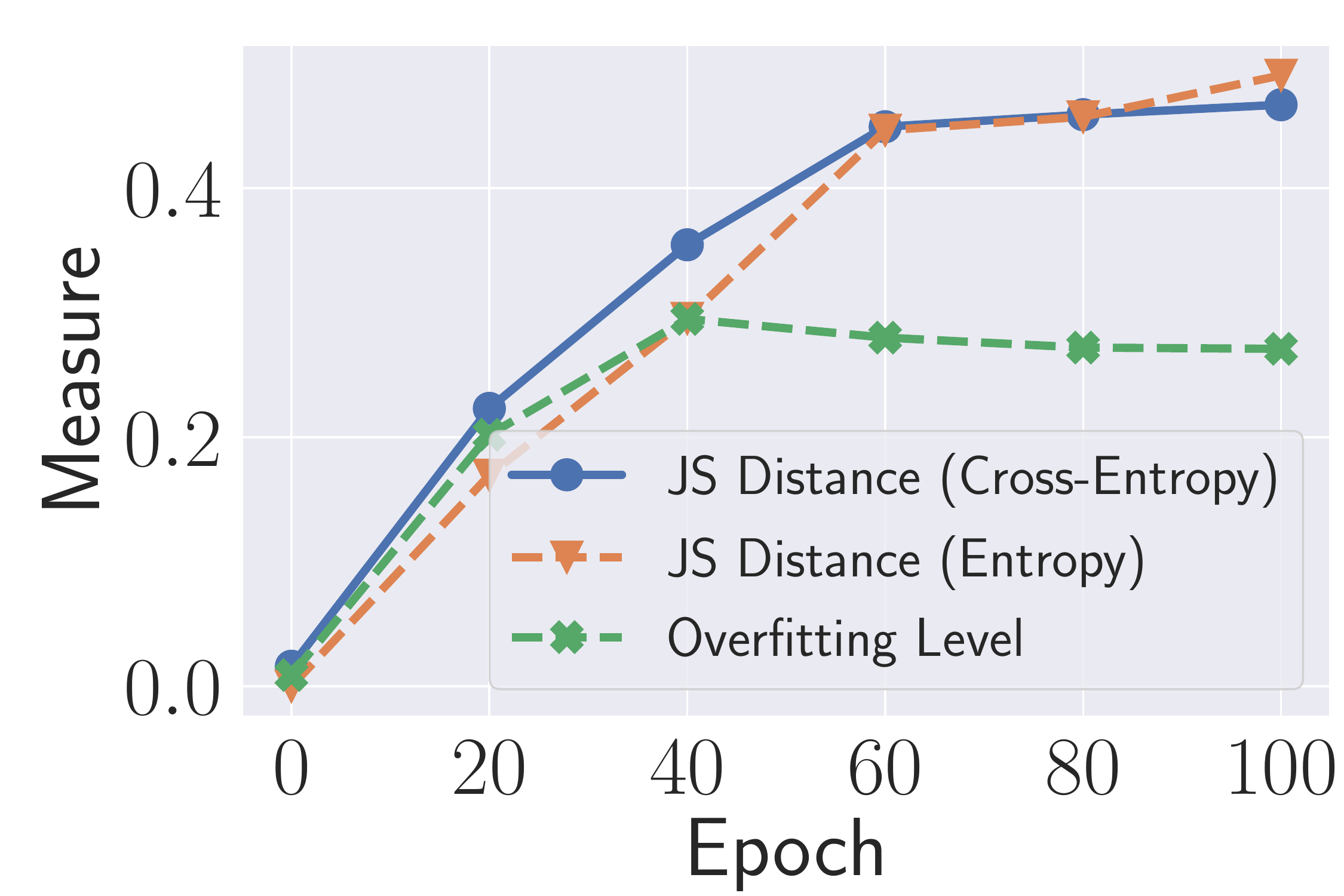}
\caption{Distance}
\label{figure:RQ3_target_mobilenetv2_CIFAR10_no_aug}
\end{subfigure}
\begin{subfigure}{0.45\columnwidth}
\includegraphics[width=\columnwidth]{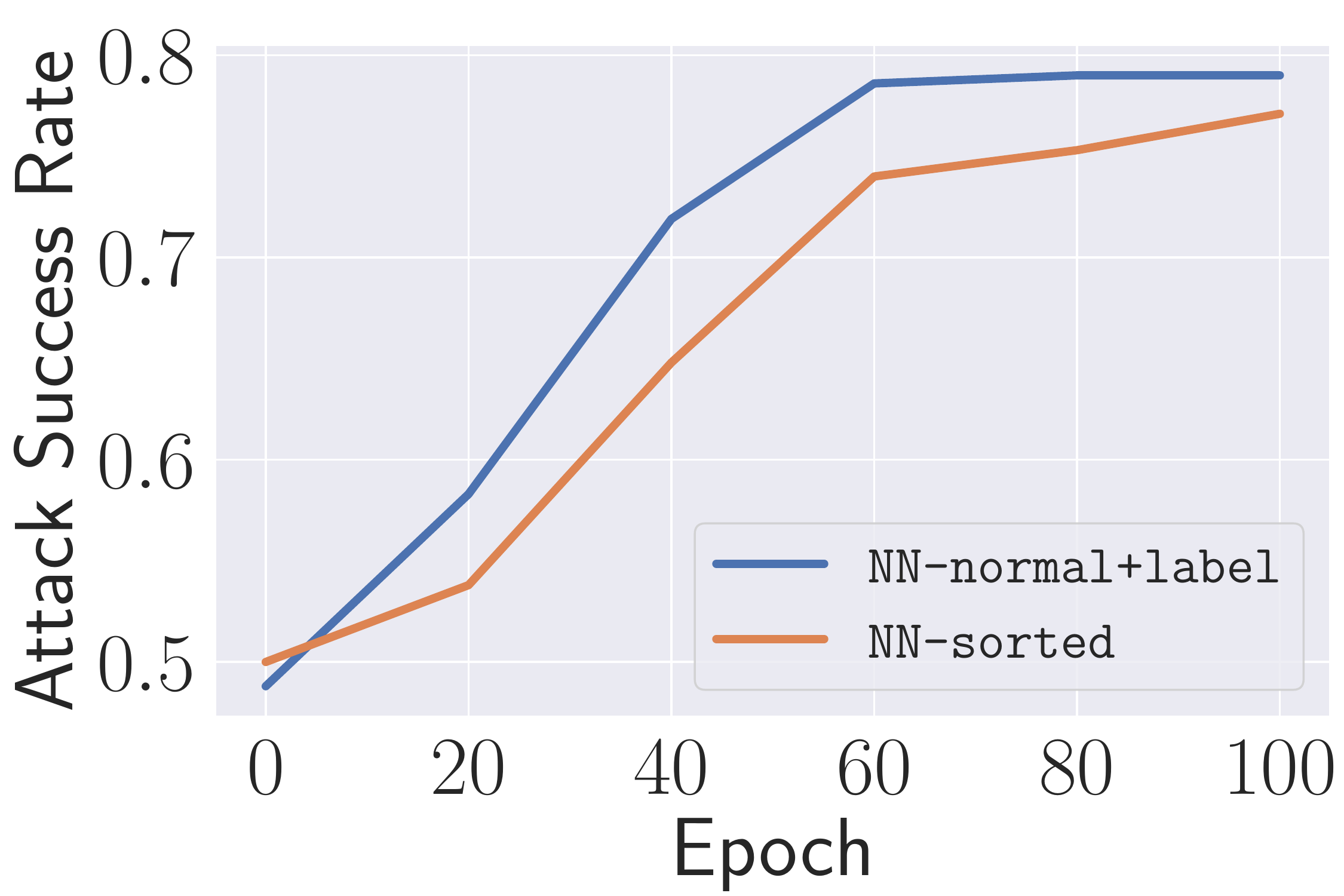}
\caption{Attack Performance}
\label{figure:RQ3_attack_mobilenetv2_CIFAR10_no_aug}
\end{subfigure}
\caption{The distance and attack performance against MobileNetV2 on CIFAR10 under different numbers of epochs for model training.}
\label{figure:RQ3_mobilenetv2_CIFAR10_no_aug} 
\end{figure}

\mypara{Overfitting} 
We notice that the attack performance varies on different models and different datasets.
Similar to previous works~\cite{SSSS17,YGFJ18,SZHBFB19,HZ21}, we also relate this to the different overfitting levels of ML models.
The overfitting level of a given model is measured by calculating the difference between its training accuracy and testing accuracy, i.e., subtracting testing accuracy from training accuracy.
In \Cref{figure:RQ3_mobilenetv2_CIFAR10_no_aug}, we can see that the overfitting level is indeed correlated with the attack performance: if a model is more overfit, it is more vulnerable to membership inference attacks. 
For example, in \Cref{figure:RQ3_target_mobilenetv2_CIFAR10_no_aug}, the overfitting level ranges from 0 to 0.202 when the target model's training epochs range from 0 to 20, and accordingly the attack success rate of \texttt{NN-normal+label} attack ranges from 0.500 to 0.583.
This observation was shown in previous works~\cite{SSSS17,YGFJ18,SZHBFB19,HZ21} and claimed that the overfitting level was the main factor contributing to the vulnerability of a model to membership inference attacks.

We also observe that the attack performance may vary across different classes due to the different overfitting levels for each class.
For instance, for VGG11 trained on CIFAR10 (\Cref{table:class_wised_performance} in Appendix), for class 3/8, the attack success rate is 0.845/0.734 with the overfitting level of 0.409/0.169.
This indicates that the membership leakage also varies across different samples from the same dataset.

However, another more interesting finding that is contradictory to the previous conclusion is that the attack performance is still increasing when the overfitting level stabilizes.
For instance, in~\Cref{figure:RQ3_mobilenetv2_CIFAR10_no_aug}, when the overfitting level is around 0.3 (epochs range from 40 to 100), the attack performance still increases rapidly with increasing training epochs.
Similar results are also observed on other datasets as shown in \Cref{appendix:relation_distance_attack} in Appendix.
This observation inspires us to rethink the relationship between overfitting levels and attack performance.
Here, different from previous works~\cite{YGFJ18,SZHBFB19,HZ21}, we argue that the overfitting level is not strongly correlated with the vulnerability of the model to membership inference attacks.

\begin{figure}[!t]
\centering
\begin{subfigure}{0.45\columnwidth}
\includegraphics[width=\columnwidth]{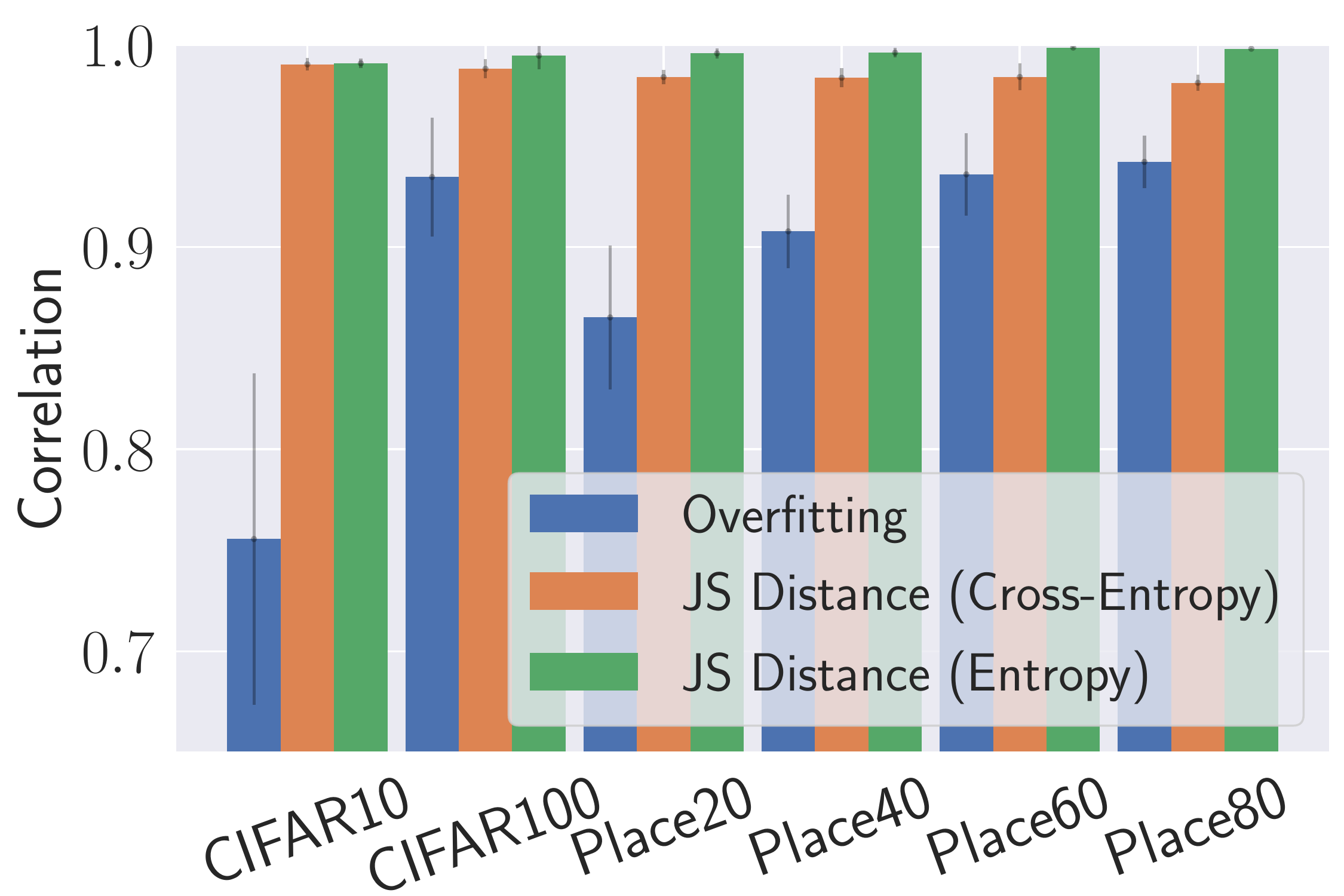}
\caption{\texttt{NN-sorted}}
\label{figure:RQ3_correlation_sorted}
\end{subfigure}
\begin{subfigure}{0.45\columnwidth}
\includegraphics[width=\columnwidth]{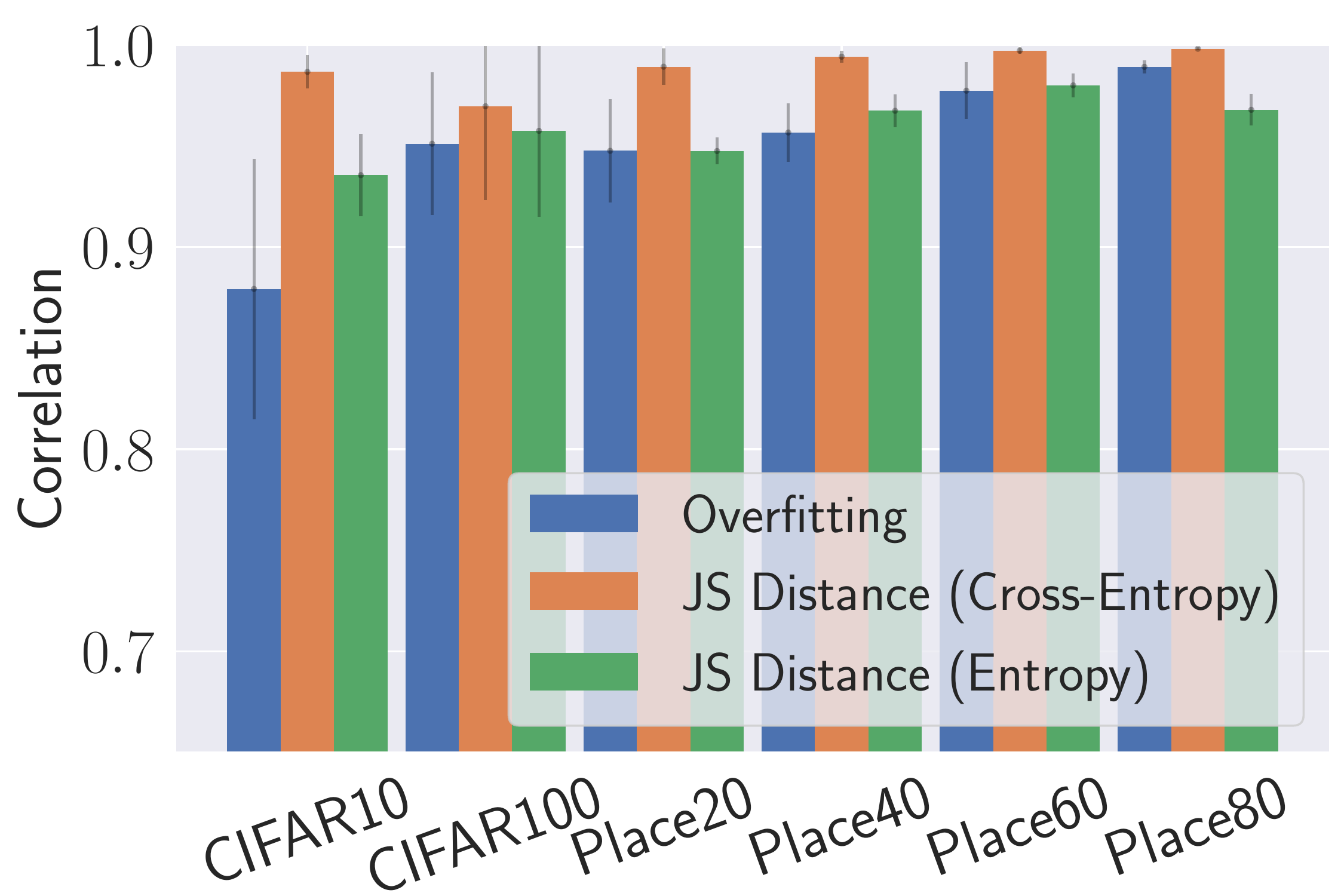}
\caption{\texttt{NN-normal+label}}
\label{figure:RQ3_correlation_normal+label}
\end{subfigure}
\caption{The Correlation between different measures and the final attack performance on 6 different datasets. Note that we average the correlation under different model architectures and show the standard deviations as well.}
\label{figure:RQ3_correlation} 
\end{figure}

\mypara{Jensen-Shannon Distance of Entropy/Cross-Entropy}
In addition, we investigate the correlation between the distance of entropy and cross-entropy distributions for members and non-members and the vulnerability of the model to membership inference attacks.
Note that here we use JS distance to measure the distance as it’s symmetric and a widely used metric to measure the distance of two probability distributions~\cite{GPMXWOCB14}.

In particular, we feed all members and non-members to the target model and calculate the entropy/cross-entropy of each sample.
We can then obtain the entropy/cross-entropy distributions for members and non-members, and normalize them into probability distributions.
Finally, we measure the JS distance of the normalized entropy/cross-entropy distributions between members and non-members.

\Cref{figure:RQ3_target_mobilenetv2_CIFAR10_no_aug} shows the JS distance of entropy/cross-entropy distributions and the overfitting level under the target model trained with different epochs when the target model is MobileNetV2 trained on CIFAR10.
We can see that the JS distance of entropy/cross-entropy is highly correlated with the attack performance.
For example, in \Cref{figure:RQ3_target_mobilenetv2_CIFAR10_no_aug}, the JS distance of entropy/cross-entropy of target model ranges from 0.02 to 0.44 when the epochs range from 0 to 60, and accordingly the attack success rate of \texttt{NN-normal+label} ranges from 0.5 to 0.783.
More interestingly, from \Cref{figure:RQ3_target_mobilenetv2_CIFAR10_no_aug} and \Cref{figure:RQ3_attack_mobilenetv2_CIFAR10_no_aug}, we can also see there is a clear turning point when the training epoch is 60, and after 60 epochs, both JS distance and attack performance become stable.
These results convincingly demonstrate that, compared to the overfitting level, the JS distance of members' and non-member's entropy/cross-entropy has a higher correlation to the attack performance.

\mypara{Comparison of Different Factors} 
To better quantify these factors, we calculated the Pearson correlation between the JS distance/overfitting level and \texttt{NN-sorted}'s attack performance as shown in \Cref{figure:RQ3_correlation_sorted}.
JS distance of entropy has the highest correlation with attack performance, followed by the JS distance of cross-entropy, while the overfitting level has a lower correlation with attack performance.
On CIFAR10, the average correlation is 0.991 for JS distance of entropy, 0.990 for JS distance of cross-entropy, and 0.756 for overfitting.
\Cref{figure:RQ3_correlation_normal+label} shows the correlation between different factors with \texttt{NN-normal+label}'s attack performance.
JS distance of cross-entropy has a higher correlation with attack performance, while the JS distance of entropy and overfitting level has a relatively lower correlation with attack performance.
On CIFAR10, the average correlation is 0.986 for JS distance of cross-entropy, 0.936 for JS distance of entropy, and 0.879 for overfitting.
JS distance of cross-entropy has a higher correlation with the attack performance of \texttt{NN-normal+label} because both the attack and the cross-entropy jointly consider the label and the posteriors.
In contrast, both the \texttt{NN-sorted} attack and entropy require only the posteriors, which can explain why the JS distance of entropy is more correlated with the \texttt{NN-sorted}'s attack performance.

\begin{figure}[t]
\centering
\begin{subfigure}{0.45\columnwidth}
\includegraphics[width=\columnwidth]{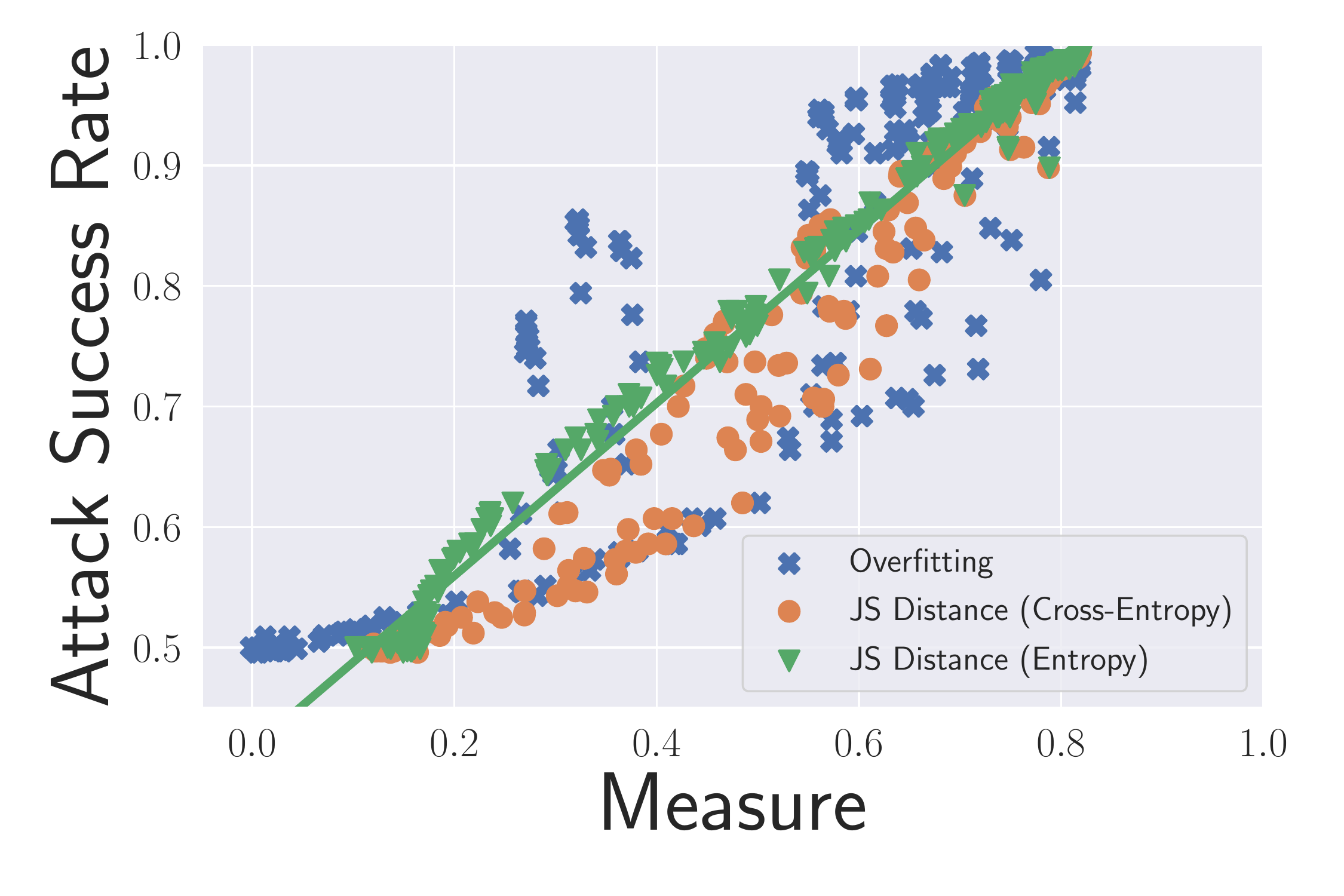}
\caption{\texttt{NN-sorted}}
\label{figure:RQ3_scatter_correlation_sorted}
\end{subfigure}
\begin{subfigure}{0.45\columnwidth}
\includegraphics[width=\columnwidth]{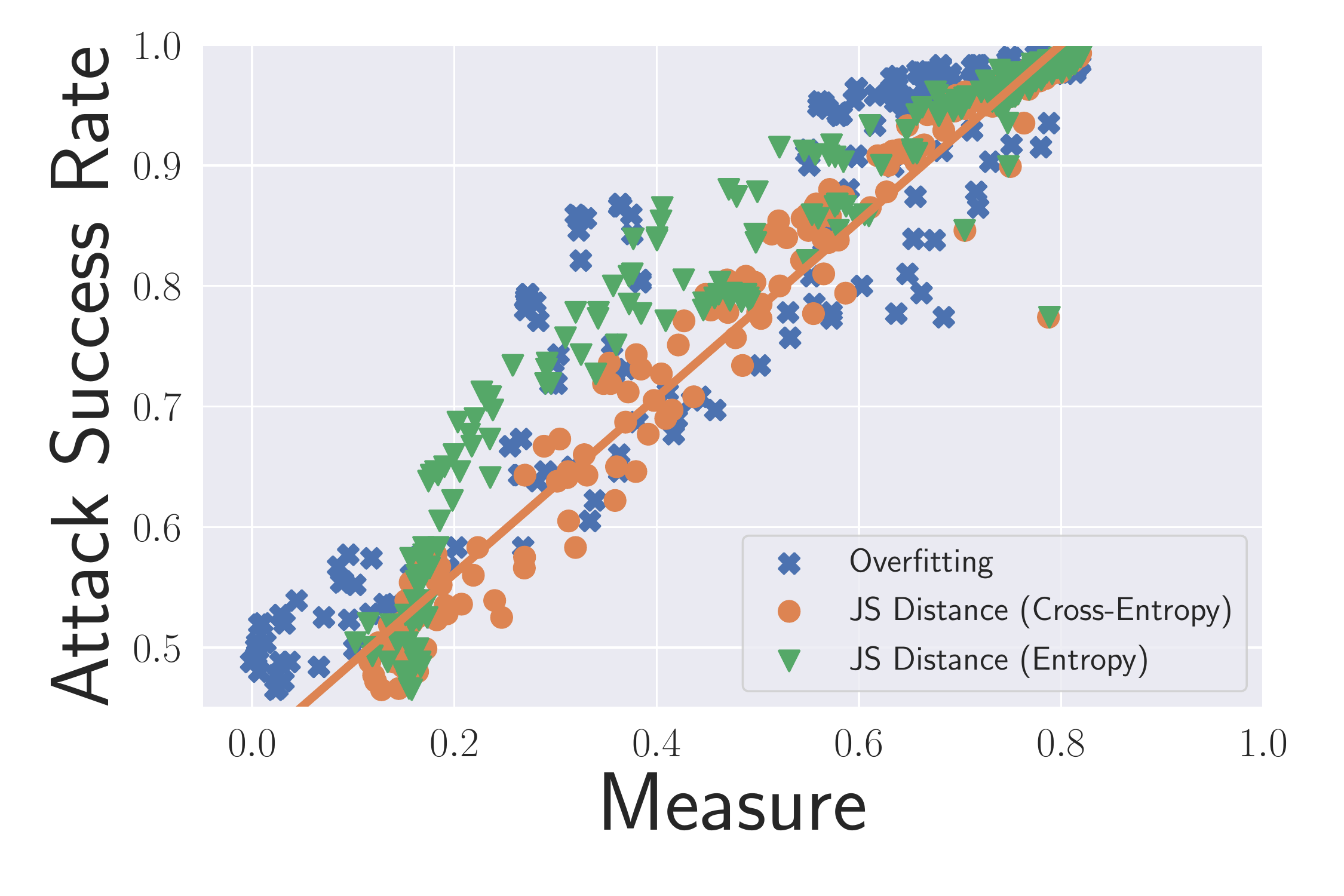}
\caption{\texttt{NN-normal+label}}
\label{figure:RQ3_scatter_correlation_normal+label}
\end{subfigure}
\caption{The relationship between different measures and \texttt{NN-normal+label} attack performance for target models with different architectures/training epochs on different datasets. The x-axis represents the value of measures and the y-axis represents the attack performance.}
\label{figure:RQ3_scatter_correlation}
\end{figure}

\mypara{Membership Inference Risks Estimation}
Given the fact that the correlation between JS distance of entropy/cross-entropy and attack performance is high for each dataset, we also investigated whether such result generalizes to different target models (with different architectures and training epochs) and different datasets.
We visualized the relationship between different measures and attack performance calculated from target models with different architectures/training epochs on different datasets in \Cref{figure:RQ3_scatter_correlation}.
For the \texttt{NN-sorted} attack in \Cref{figure:RQ3_scatter_correlation_sorted}, there is a significant linear correlation between JS distance of entropy and attack performance where the slope and intercept are 0.716 and 0.416 with a correlation coefficient of 0.996.
We can see a similar result with the \texttt{NN-normal+label} attack in \Cref{figure:RQ3_scatter_correlation_normal+label}, where the slope and intercept are 0.731 and 0.415 with a correlation coefficient of 0.984.
Thus, if defenders measure the JS distance of entropy or cross-entropy, they can leverage the slope and intercept to directly estimate the final attack performance of \texttt{NN-sorted} or \texttt{NN-normal+label}.
For instance, on CIFAR10, when the target model is VGG11 trained for 100 epochs, the JS distance of entropy is 0.492, so we estimate the attack performance of \texttt{NN-sorted} to be 0.768, which is very close to ground truth 0.760 with less than 1\% difference.
These results convincingly show that we can estimate the attack performance without conducting the attack.

\mypara{Takeaways} 
We investigated what determines the performance of membership inference attacks and show that compared to the overfitting level, the JS distance of entropy/cross-entropy distributions between members and non-members are more correlated with the attack performance.
This correlation is model and dataset agnostic and gives us a new way to accurately estimate membership inference risks even without running the attack.

\subsection{Training Machine Learning Models in Practice}
\label{subsection:evaluation_RQ4}

The previous evaluations were conducted using models trained without any data augmentation.
However, data augmentation is a broadly applied method for generating extra training samples that improve the model's performance~\cite{KSH12,ZK16,ZCDL18,CZSL20}.
We investigated how data augmentation affects the performance of membership inference attacks and whether data augmentation can facilitate the target model performance on its original classification task.

When training the target models, we considered three cases:
\begin{itemize}
    \item \textbf{No Augmentation (No)}: We directly fed the original training sample into the target model without any transformation.
    \item \textbf{Simple Augmentation (Simple)}: Following ~\cite{KSH12,ZK16}, we applied random crop with padding of 4 and horizontal flips to the original training samples.
    \item \textbf{Random Augmentation (Randaug)}: Following Cubuk et al.~\cite{CZSL20}, we applied random augmentation to the training samples, which consisted of a group of advanced augmentation operations. Concretely, we set $N=2$ and $M=10$ where $N$ denotes the number of transformations to a given sample and $M$ represents the magnitude of global distortion.
\end{itemize}

\begin{figure}[!t]
\centering
\begin{subfigure}{0.45\columnwidth}
\includegraphics[width=\columnwidth]{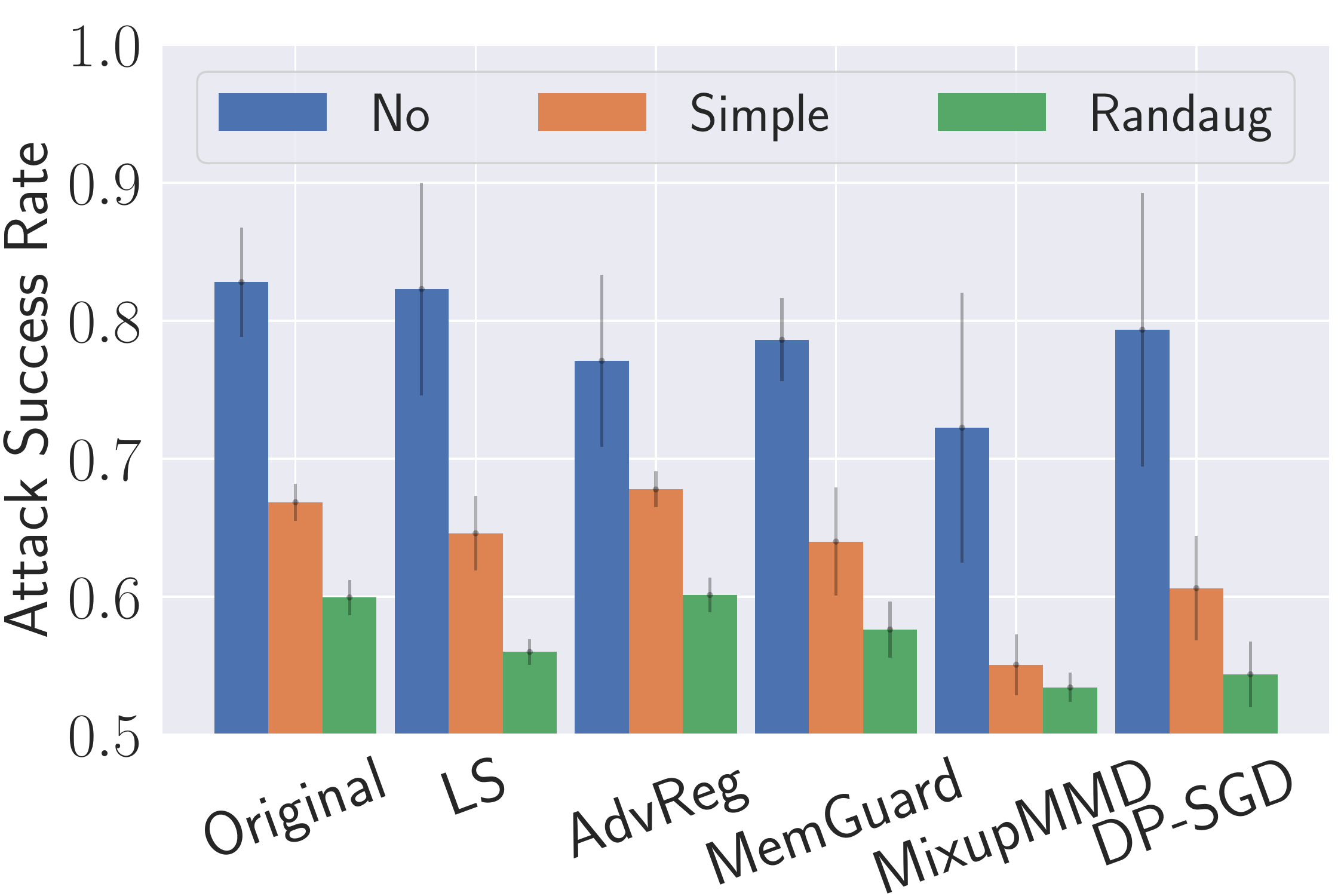}
\caption{Defense Effectiveness}
\label{figure:RQ4_CIFAR10_attack}
\end{subfigure}
\begin{subfigure}{0.45\columnwidth}
\includegraphics[width=\columnwidth]{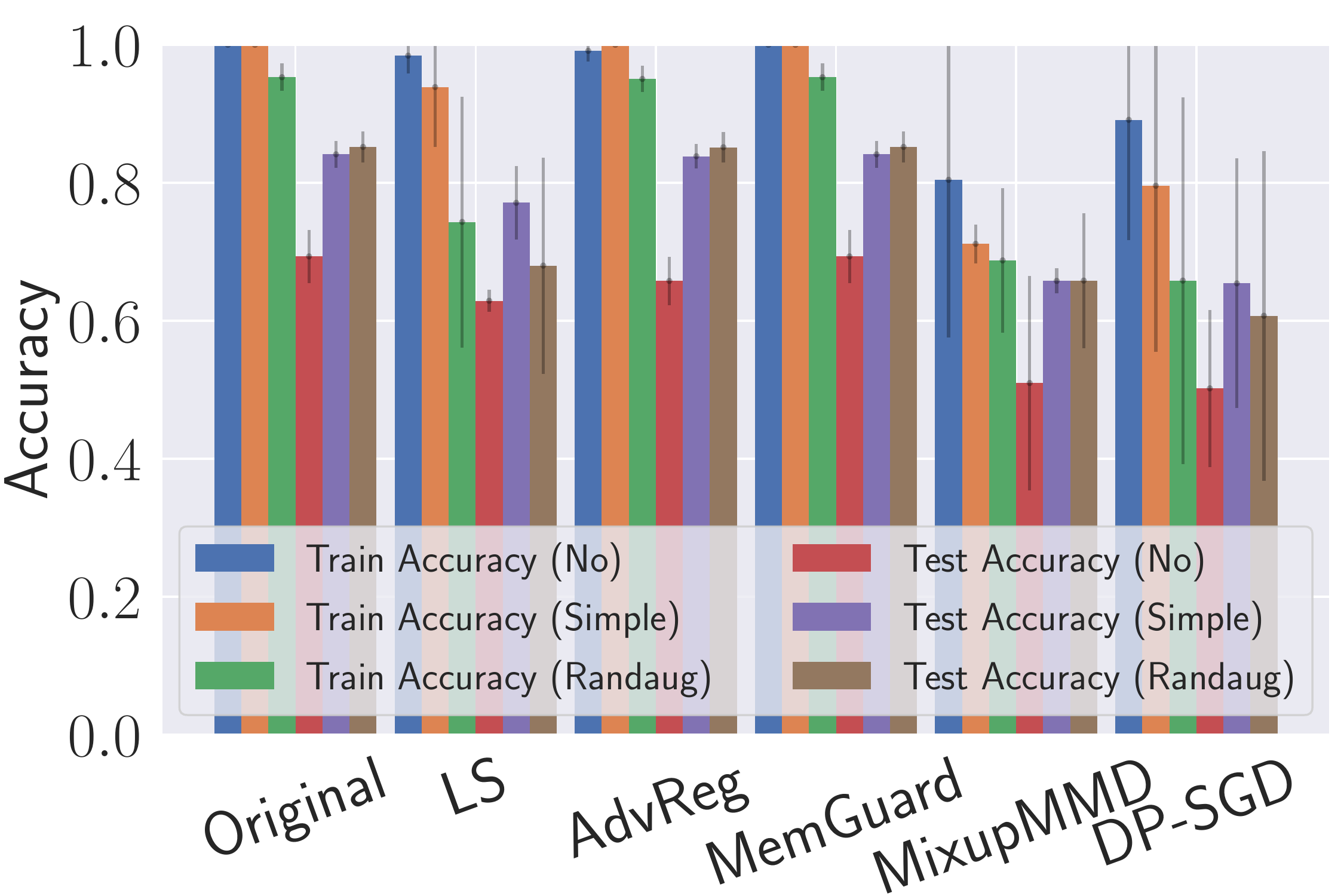}
\caption{Utility}
\label{figure:RQ4_CIFAR10_target}
\end{subfigure}
\caption{Defense effectiveness against the best attacks and utility in the original classification tasks on the CIFAR10 dataset. Note that we average the performance for different model architectures and report the standard deviations as well.}
\label{figure:RQ4_defense_effectiveness_and_utility_CIFAR10} 
\end{figure}

\begin{figure*}[!ht]
\centering
\begin{subfigure}{0.32\columnwidth}
\includegraphics[width=\columnwidth]{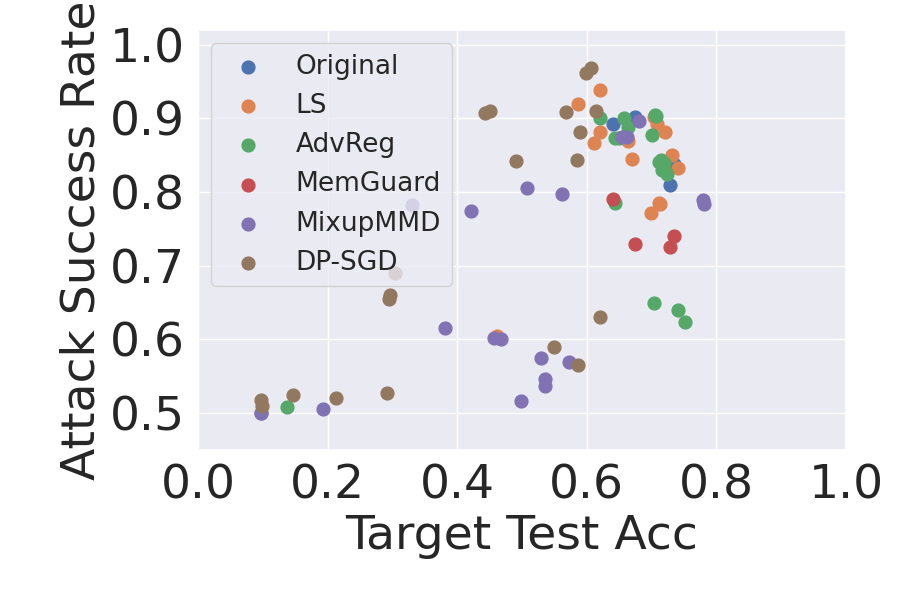}
\caption{CIFAR10}
\label{figure:defense_tradeoff_CIFAR10_no}
\end{subfigure}
\begin{subfigure}{0.32\columnwidth}
\includegraphics[width=\columnwidth]{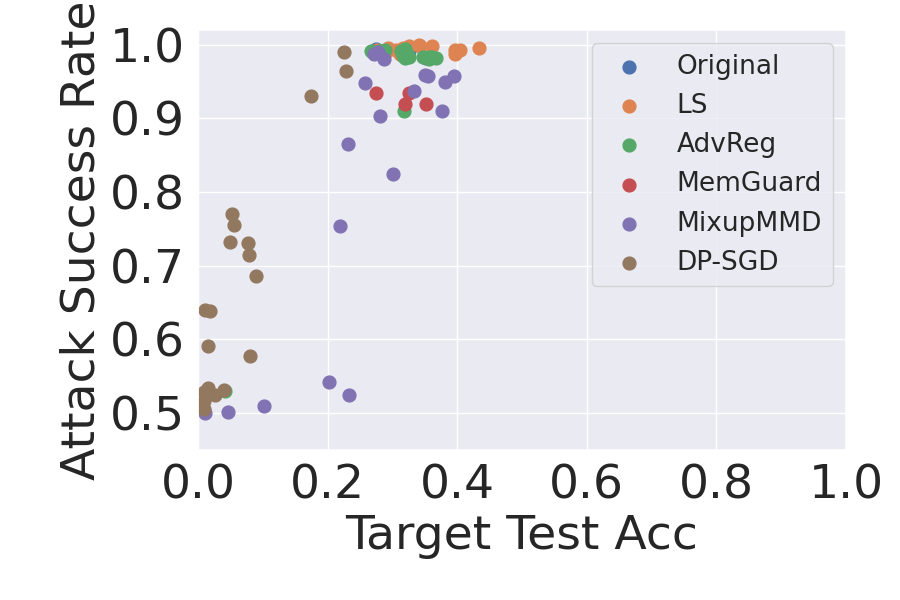}
\caption{CIFAR100}
\label{figure:defense_tradeoff_CIFAR100_no}
\end{subfigure}
\begin{subfigure}{0.32\columnwidth}
\includegraphics[width=\columnwidth]{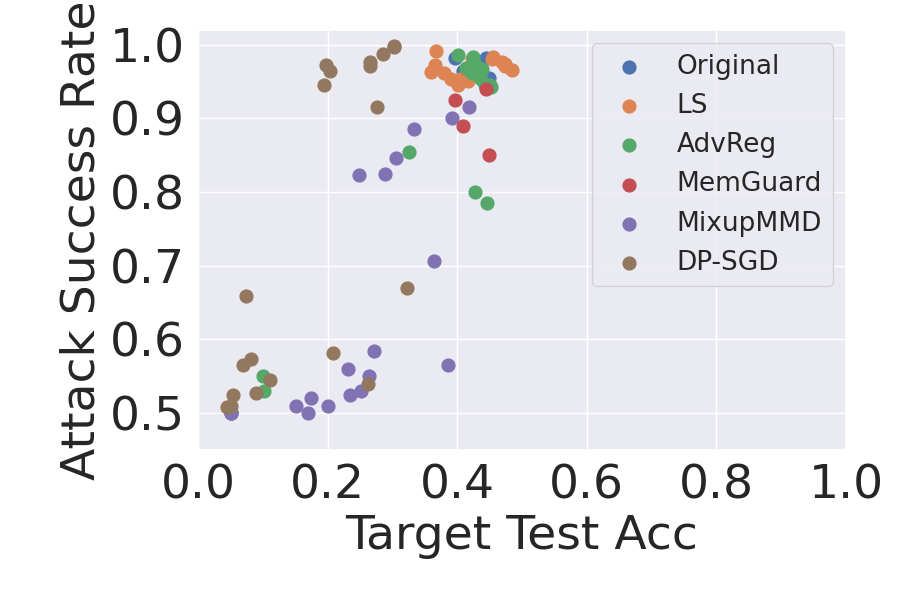}
\caption{Place20}
\label{figure:defense_tradeoff_Place20_no}
\end{subfigure}
\begin{subfigure}{0.32\columnwidth}
\includegraphics[width=\columnwidth]{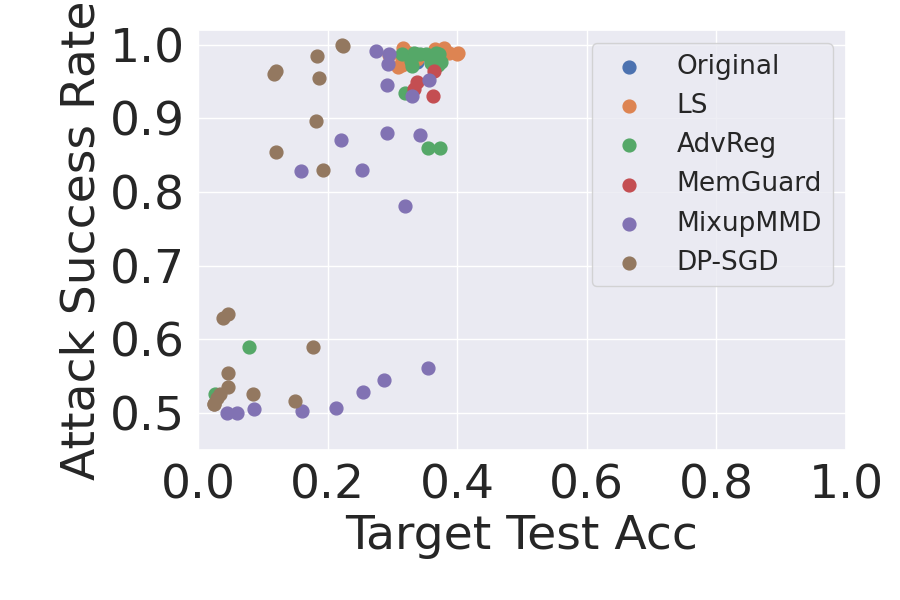}
\caption{Place40}
\label{figure:defense_tradeoff_Place40_no}
\end{subfigure}
\begin{subfigure}{0.32\columnwidth}
\includegraphics[width=\columnwidth]{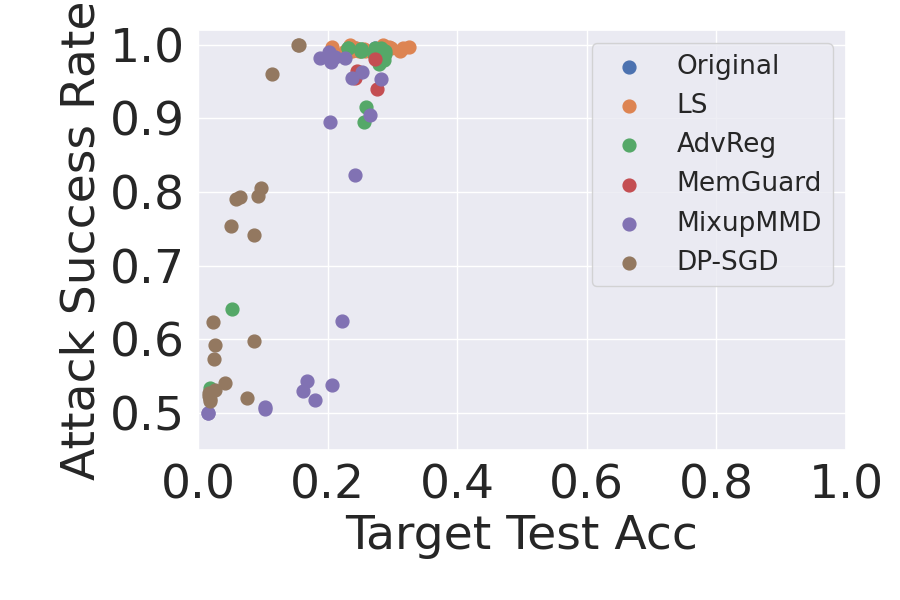}
\caption{Place60}
\label{figure:defense_tradeoff_Place60_no}
\end{subfigure}
\begin{subfigure}{0.32\columnwidth}
\includegraphics[width=\columnwidth]{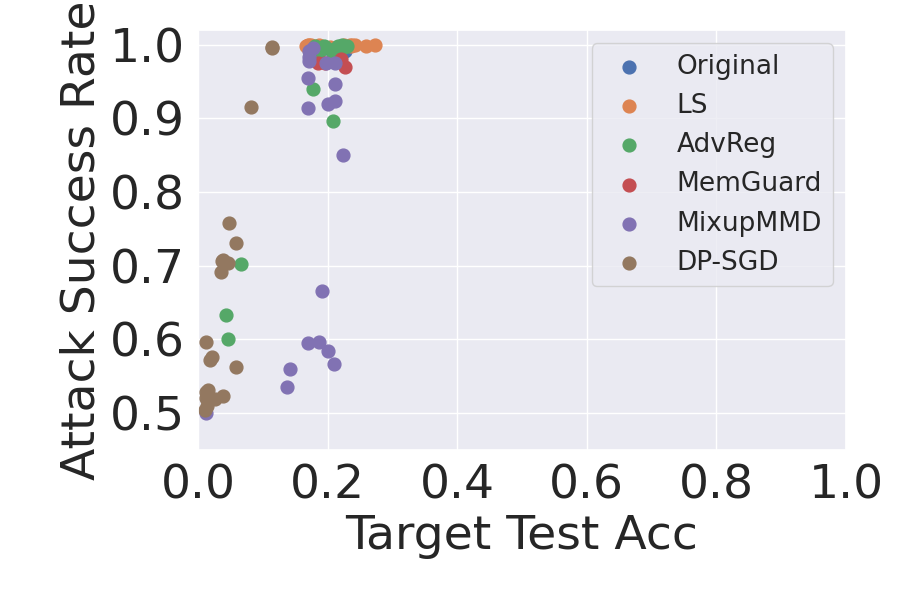}
\caption{Place80}
\label{figure:defense_tradeoff_Place80_no}
\end{subfigure}
\begin{subfigure}{0.32\columnwidth}
\includegraphics[width=\columnwidth]{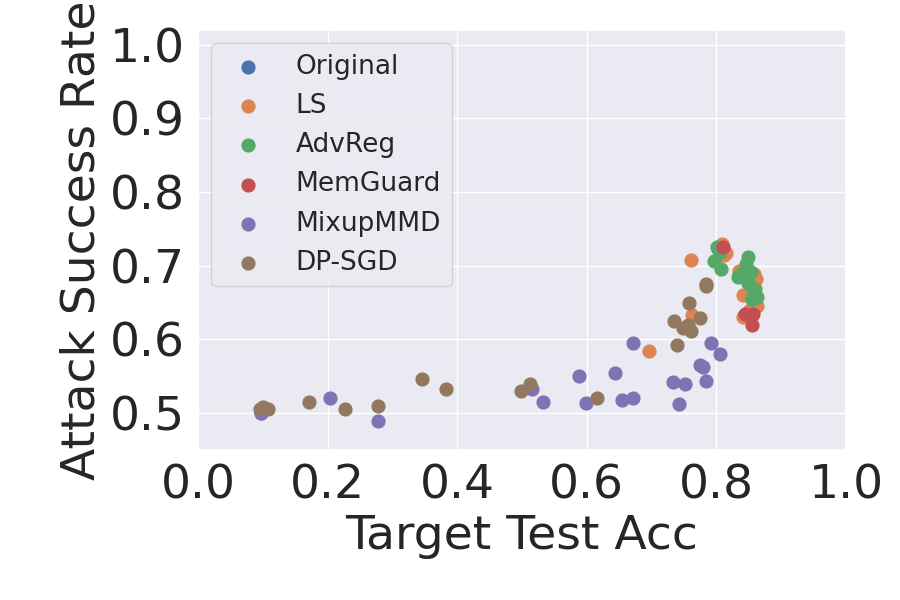}
\caption{CIFAR10}
\label{figure:defense_tradeoff_CIFAR10_simple}
\end{subfigure}
\begin{subfigure}{0.32\columnwidth}
\includegraphics[width=\columnwidth]{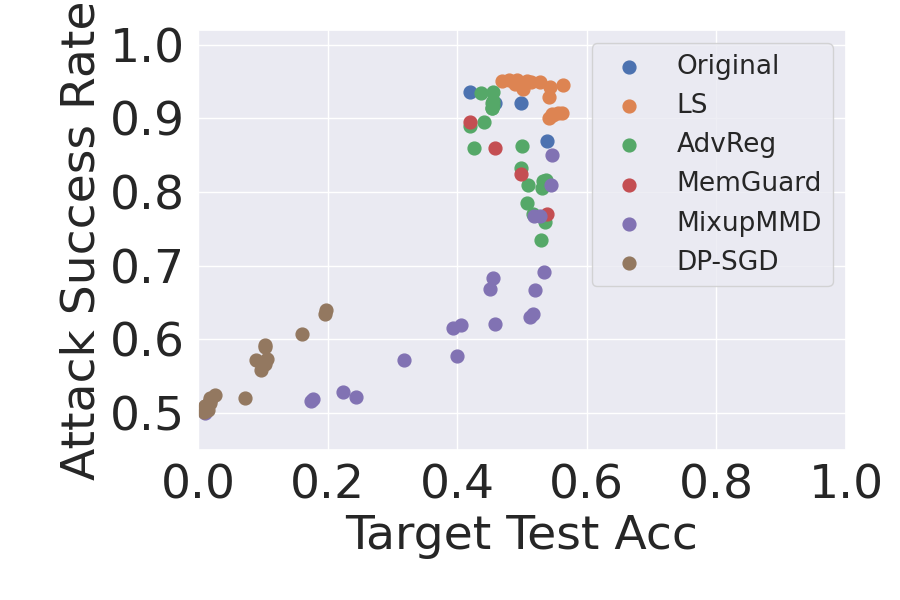}
\caption{CIFAR100}
\label{figure:defense_tradeoff_CIFAR100_simple}
\end{subfigure}
\begin{subfigure}{0.32\columnwidth}
\includegraphics[width=\columnwidth]{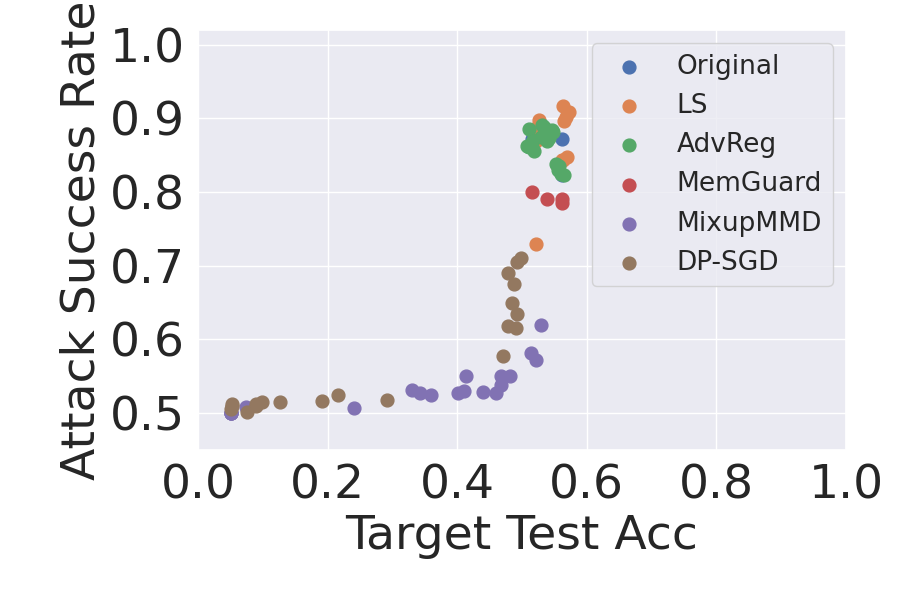}
\caption{Place20}
\label{figure:defense_tradeoff_Place20_simple}
\end{subfigure}
\begin{subfigure}{0.32\columnwidth}
\includegraphics[width=\columnwidth]{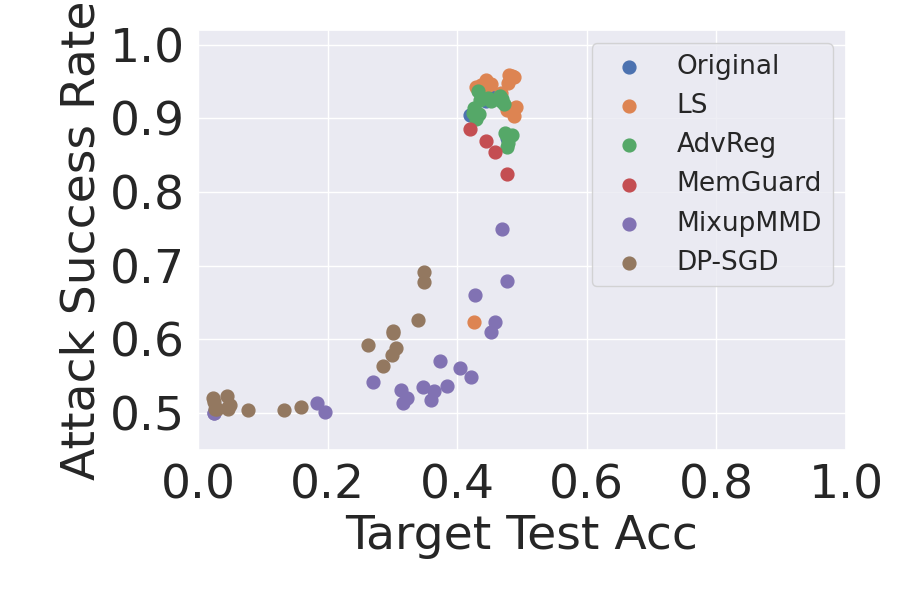}
\caption{Place40}
\label{figure:defense_tradeoff_Place40_simple}
\end{subfigure}
\begin{subfigure}{0.32\columnwidth}
\includegraphics[width=\columnwidth]{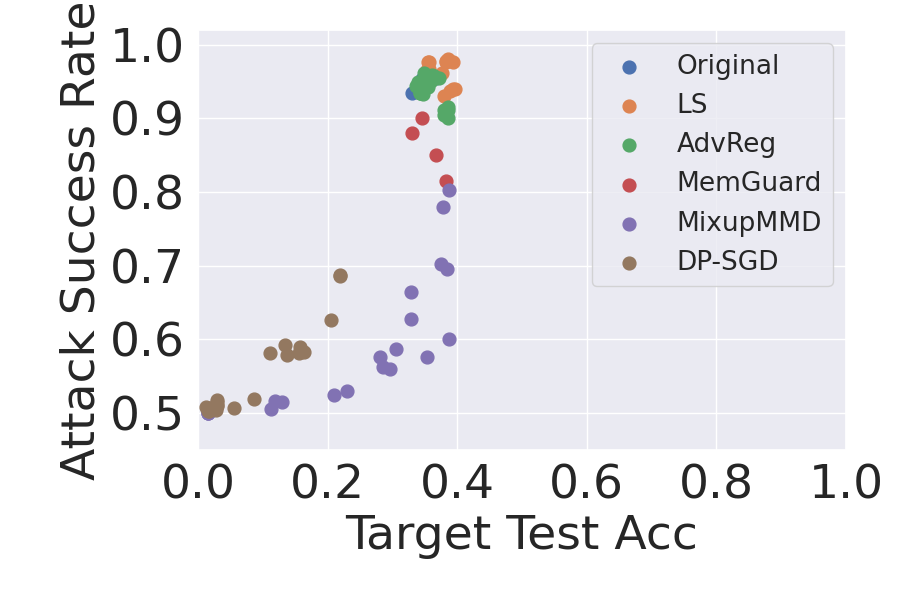}
\caption{Place60}
\label{figure:defense_tradeoff_Place60_simple}
\end{subfigure}
\begin{subfigure}{0.32\columnwidth}
\includegraphics[width=\columnwidth]{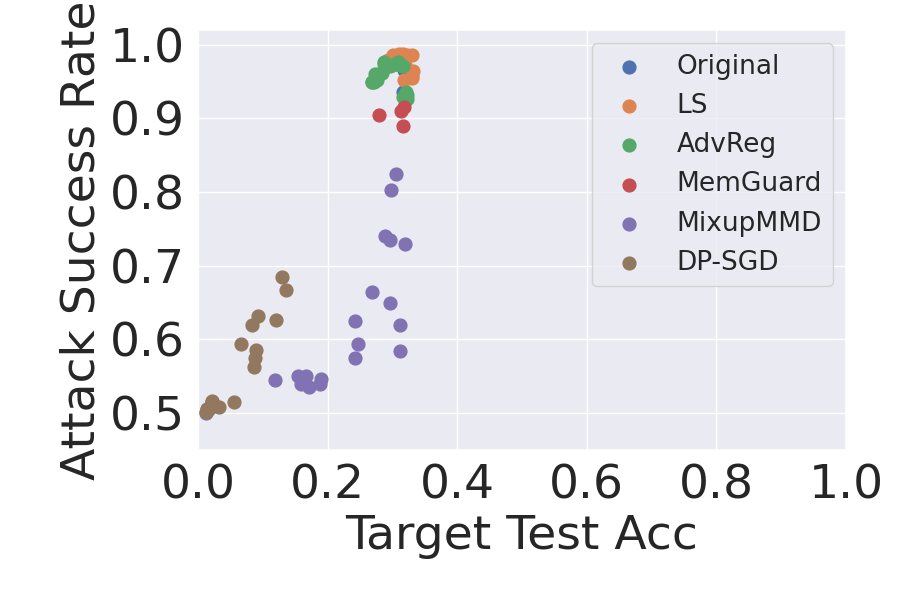}
\caption{Place80}
\label{figure:defense_tradeoff_Place80_simple}
\end{subfigure}
\begin{subfigure}{0.32\columnwidth}
\includegraphics[width=\columnwidth]{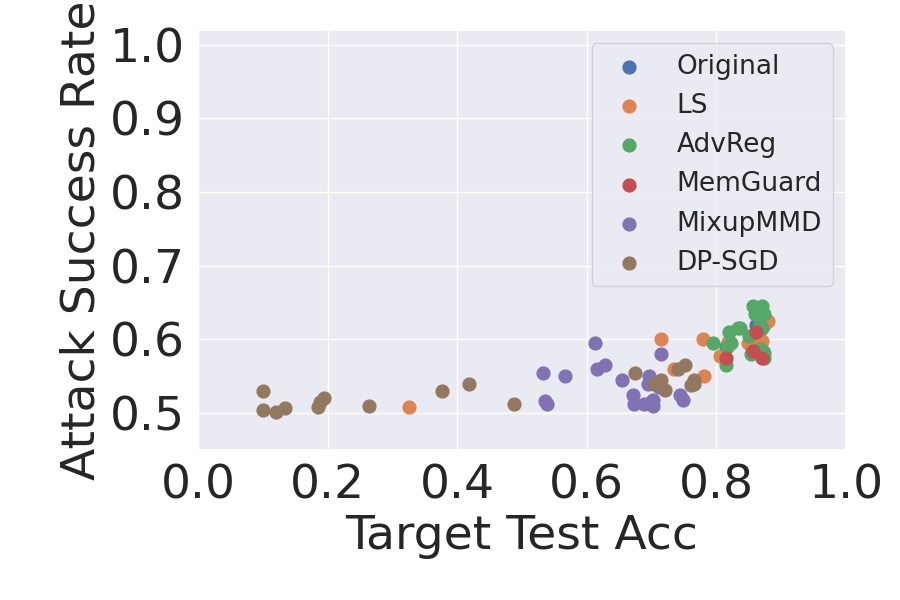}
\caption{CIFAR10}
\label{figure:defense_tradeoff_CIFAR10_randaug}
\end{subfigure}
\begin{subfigure}{0.32\columnwidth}
\includegraphics[width=\columnwidth]{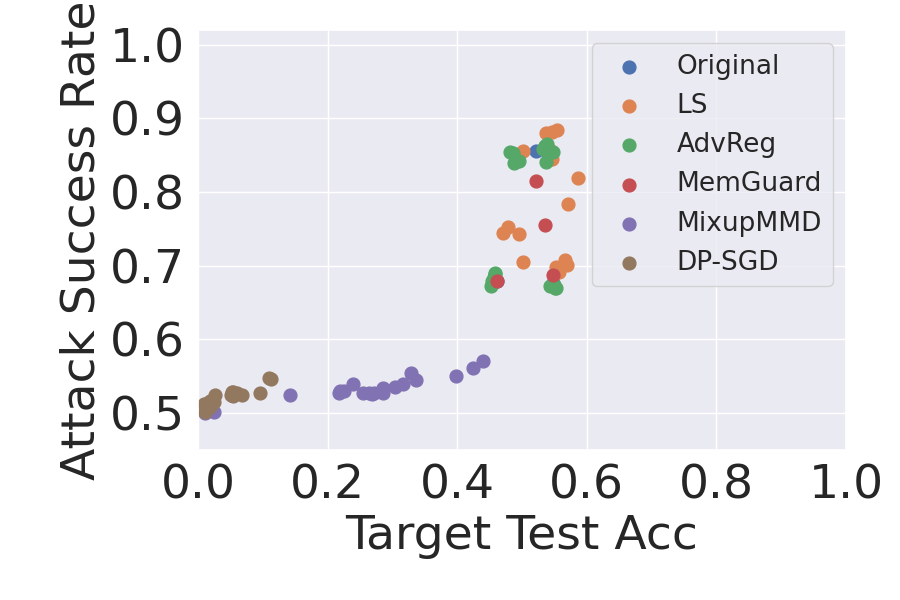}
\caption{CIFAR100}
\label{figure:defense_tradeoff_CIFAR100_randaug}
\end{subfigure}
\begin{subfigure}{0.32\columnwidth}
\includegraphics[width=\columnwidth]{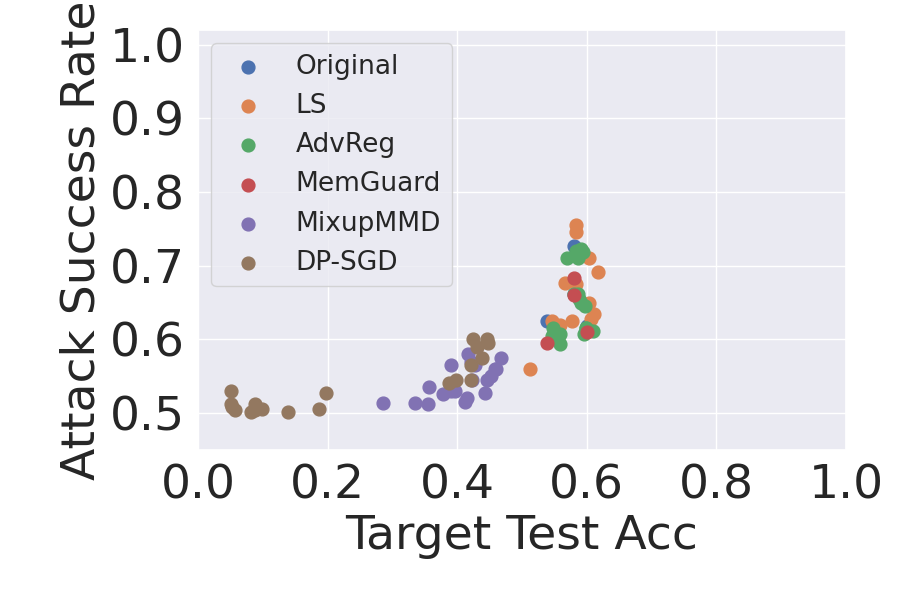}
\caption{Place20}
\label{figure:defense_tradeoff_Place20_randaug}
\end{subfigure}
\begin{subfigure}{0.32\columnwidth}
\includegraphics[width=\columnwidth]{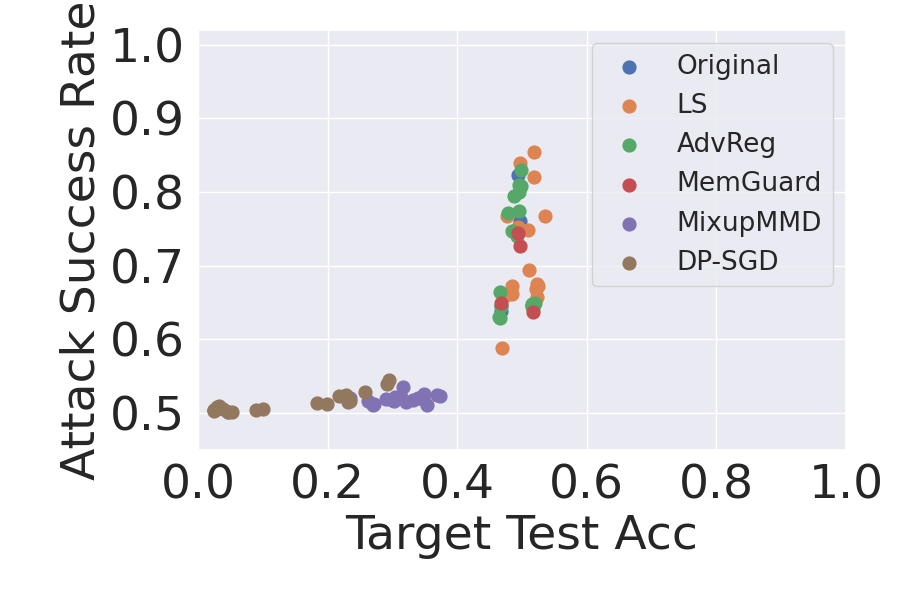}
\caption{Place40}
\label{figure:defense_tradeoff_Place40_randaug}
\end{subfigure}
\begin{subfigure}{0.32\columnwidth}
\includegraphics[width=\columnwidth]{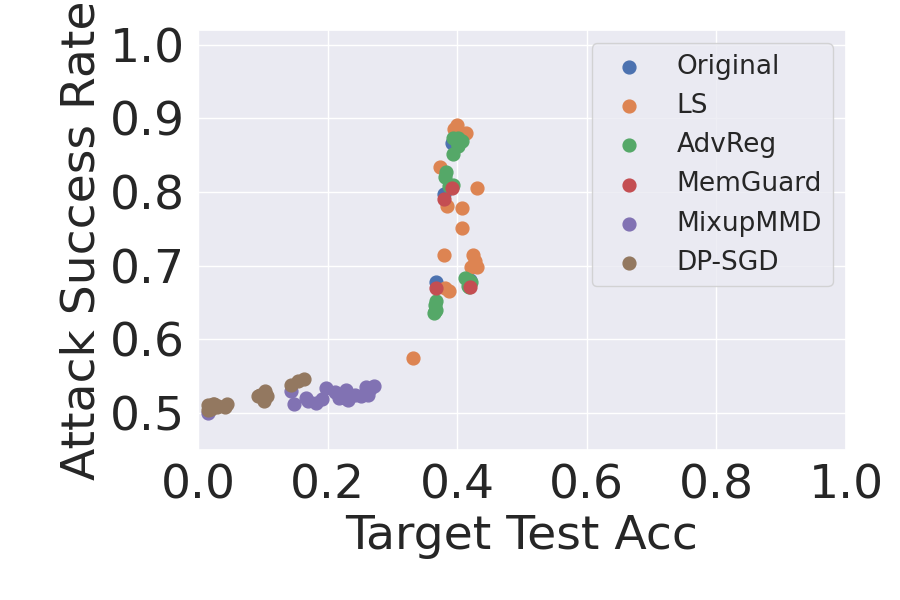}
\caption{Place60}
\label{figure:defense_tradeoff_Place60_randaug}
\end{subfigure}
\begin{subfigure}{0.32\columnwidth}
\includegraphics[width=\columnwidth]{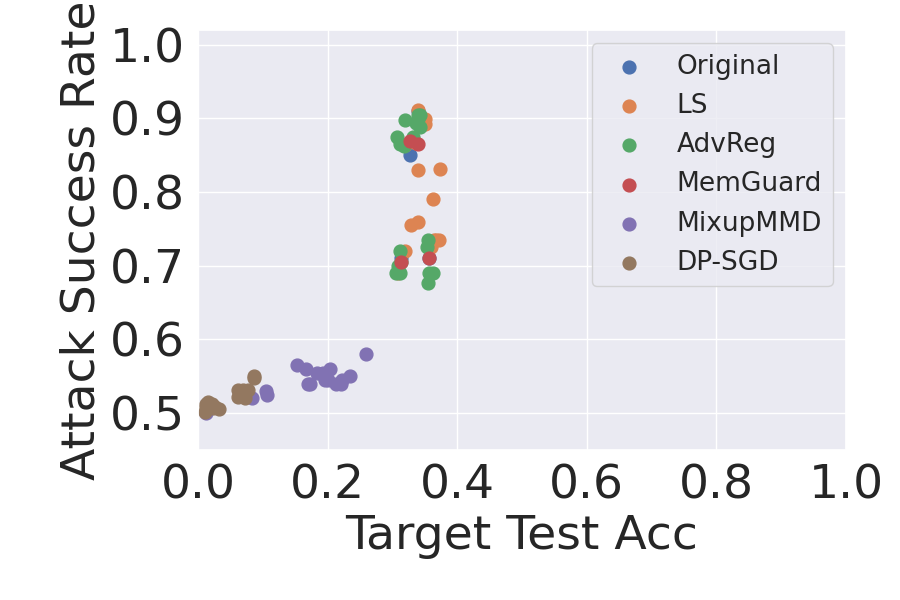}
\caption{Place80}
\label{figure:defense_tradeoff_Place80_randaug}
\end{subfigure}
\caption{Defense effectiveness against the best attacks and utility in the original classification tasks. Note that for LS, AdvReg, MixupMMD ,and DP-SGD, we varies the hyperparameters to see the privacy-utility trade-off. The first/second/third row of subfigures denotes the models trained with no/simple/random augmentation.}
\label{figure:defense_tradeoff}
\end{figure*}

The defense effectiveness against the best attacks and the corresponding utility in the original classification task under different augmentations on CIFAR10 are summarized in \Cref{figure:RQ4_defense_effectiveness_and_utility_CIFAR10}.
The corresponding results for other datasets are summarized in \Cref{appendix:subsection_defense}.

Compared to the models trained with no augmentation, models trained with augmentations better defend against membership inference attacks.
Random augmentation, a more advanced augmentation method compared to simple augmentation, reduces the threat of membership inference attacks to a larger extent.
On CIFAR10, the best average attack accuracy for the original model (trained with no augmentation) was 0.828 while only 0.669 and 0.599 with simple and random augmentations, respectively.
The corresponding attack performance drop was 0.159 and 0.229 with simple and random augmentations, which are both larger than the drop for the best defense we investigated previously (MixupMMD, which had a 0.106 attack accuracy drop on CIFAR10 as shown in \Cref{figure:RQ1_defense_performance_no}).
Existing defenses benefit from data augmentations and achieve better defense effectiveness.
On CIFAR10, the best average attack performance under MixupMMD was 0.722 with no augmentation and decreased to 0.551 and 0.534 with simple and random augmentation.

In terms of target model utility, using data augmentation improves a target model's generalization in all cases (i.e., for different datasets, model architectures, and defense mechanisms).
For instance, on CIFAR10, the average testing accuracy for normal training was 0.693 with no augmentation, while it increased to 0.841 and 0.852 with simple and random augmentations.
Data augmentation is an effective defense, and defenders can also easily use it with existing defenses to further reduce the threat of membership inference attacks.
More so, it also increases the utility of target models which facilitate the original classification as well.

\Cref{figure:defense_tradeoff} shows a more general privacy-utility trade-off for models train with different defense hyperparameters and under different data augmentations.
We observe that compared to models trained with no augmentation (the first row), models trained with simple and random augmentation (the second and third rows)
can move the points toward the lower-right area, which means that they are less vulnerable to membership inference attacks and can achieve better utility.

Given the fact that data augmentations serve as a powerful defense, we examined whether an adversary can conduct more powerful attacks under data augmentation?
With data augmentations, the target model observes augmented views (instead of the original training samples) during the training process, which, we hypothesize, causes the target model to ``remember'' those augmented views.
Given this hypothesis, rather than feeding the original view to the target model, the adversary ought to be able to query the target model with multiple augmented views of it to increase attack effectiveness.

To conduct this \textit{augmented attack}, an adversary would follow this procedure:
\begin{itemize}
    \item The adversary first trains a shadow model with the same architecture and data augmentation as the target model using the same distribution shadow dataset.
    \item The adversary then queries the shadow model with $K$ different augmented views of the original sample using only simple augmentation and obtains $K$ posteriors. Then, the $K$ posteriors are sorted by their maximum value (confidence score) in a descending order, where the first (last) posteriors have the highest (lowest) confidence score given by the shadow model. 
    \item After that, the adversary selects the 3 largest values in descending order from each posteriors (\texttt{NN-top3}) and concatenates them together as a $3K$ dimensional input to train the attack model, which is an MLP.
    \item{With the attack model, for each sample, the adversary follows the same procedure to generate the attack input by querying the target model and getting the membership prediction by querying the attack model with it.}
\end{itemize}

We used the \texttt{NN-top3} as the base attack since it performed almost the same as \texttt{NN-sorted}, which was the best attack with only posteriors information (see \Cref{subsection:best_attack_and_defense} for more details).
We did not consider the attacks using label information since the label provides the same message regardless of  whether we query the target model with the original sample or its augmented view.
We set $K=10$ and keep other settings the same as \Cref{subsection:experimental_setting}.

\begin{figure}[!t]
\centering
\includegraphics[width=0.85\columnwidth]{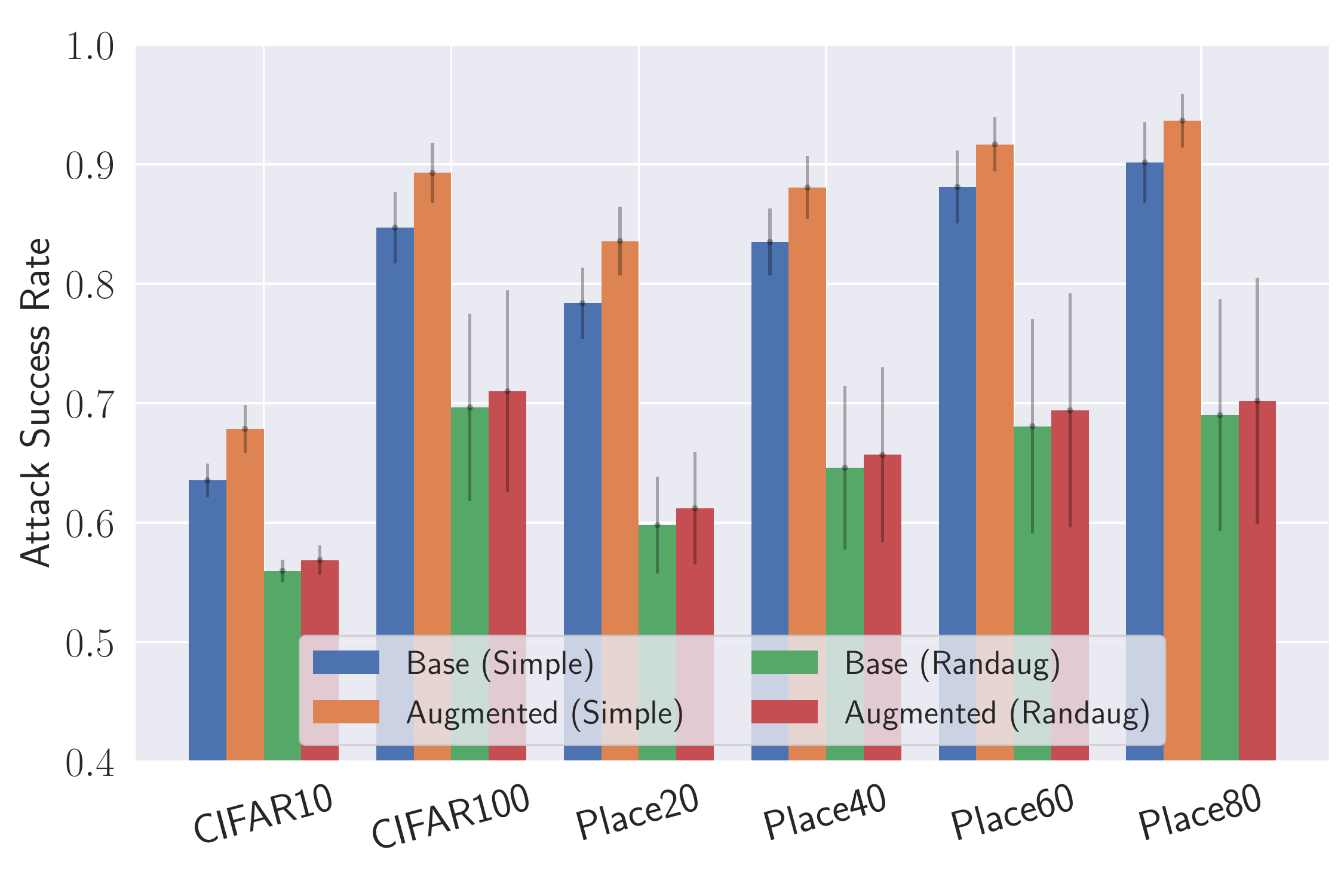}
\caption{The performance of base and augmented attacks against the models trained with simple and random augmentation on 6 different datasets. Note that we average the attack performance under different model architectures and show the standard deviations as well.}
\label{figure:RQ4_augmented_attack} 
\end{figure}

\Cref{figure:RQ4_augmented_attack} shows the performance of base and augmented attacks under simple and random augmentations.
Compared to the base attack, the augmented attack has higher attack performance.
On CIFAR10, the base attack's performance under simple and random augmentations was 0.636 and 0.560, respectively, while increasing to 0.678 and 0.569 with the augmented attack.
This augmented attack reduces the membership privacy of models trained with data augmentations.
Yu et al.~\cite{YZCYL21} made a similar observation using a set-based attack, e.g., on CIFAR10, their attack accuracy increased from 0.618 to 0.671 with the set-based attack.

Compared to attacks against target models trained with no augmentation, target models train with simple or random augmentations better preserve the membership privacy even under the augmented attacks.
For instance, on CIFAR100, the \texttt{NN-top3} attack achieved 0.975 accuracy (see \Cref{figure:RQ1_attack_performance_no}) when the target model was trained with no augmentation, while the attack accuracy was only 0.893 and 0.710 even under the augmented attacks with simple and random augmentations, respectively.
Even though this augmented attack slightly improves attack performance, using data augmentation still preserves some membership privacy.

\mypara{Takeaways}
Compared to other defense mechanisms, data augmentation is a more effective defense as it not only reduces the performance of membership inference attacks but also increases the utility of the target models.
Moreso, defenders can use it with other defense mechanisms to further reduce the risk of membership inference.
Although an augmented attack does improve the performance of membership inference attacks against models trained with data augmentations, the overall membership leakage under augmented attacks is lower than those target models trained without data augmentations.

\section{Related Work}
\label{section:related_work}

\mypara{Membership Inference Attacks}
In membership inference, the adversary's goal is to infer whether a given data sample is used to train a target model.
Currently, membership inference is one of the major methods to evaluate the privacy risks of machine learning models~\cite{SSSS17,YGFJ18,HMDC19,SZHBFB19,NSH19,SSM19,LF20,HWWBSZ21}.
Shokri et al.~\cite{SSSS17} propose the first membership inference attack against machine learning models.
They train multiple attack models using a dataset constructed from multiple shadow models, where each attack model is for each class.
These attack models take the posterior of the target sample as input and predict its membership status, i.e., member or non-member.
Then Salem et al.~\cite{SZHBFB19} propose a model and data-independent membership inference attack by gradually relaxing the assumption made by Shokri et al.~\cite{SSSS17}.
Later, Nasr et al.~\cite{NSH19} focus on the privacy risk in centralized and federated learning scenarios, and conduct extensive experiments under both black-box and white-box settings.
Song et al.~\cite{SSM19} study the relationship between adversarial examples and the privacy risk caused by membership inference attacks and find that the latter increases when model builders take measures to defend against adversarial example attacks.
Recently, researchers consider a more challenging and realistic scenario where the adversary can only receive the predicted labels.
Li and Zhang~\cite{LZ21} and Choquette-Choo et al.~\cite{CTCP21} propose the label-only membership inference attack by changing the predicted labels of the target model, then measuring the magnitude of the perturbation.
If the magnitude of the perturbation is larger than a predefined threshold, the adversary considers the data sample as a member and vice versa.

In addition to the image classification models we concentrate on in this work, there are some other works demonstrating other types of ML models, e.g., language models~\cite{SR20,SS19}, generative models~\cite{CYZF20,HMDC19}, and graph-based models ~\cite{ONK21,HWWBSZ21}, are also vulnerable to membership inference attacks.
In future work, we plan to extend our work to a broader range of ML application scenarios.

\mypara{Defenses Against Membership Inference}
Researchers have proposed to improve privacy against membership inference via different defense mechanisms.
As there exists a gap between the training set and the testing set for membership inference to be successful, multiple approaches have been proposed to reduce such a gap and improve generalization.
The standard and popular generalization enhancement techniques, such as label smoothing~\cite{SVISW16}, MixupMMD~,\cite{LLR21} and data augmentation~\cite{KSH12,ZK16,ZCDL18,CZSL20} have been shown to shrink this gap effectively.
Besides, privacy enhancement also has been applied to a wide range of applications.
Many differential privacy-based defenses~\cite{CMS11,DMNS06,INSTTW19} involve clipping and adding noise to instance-level gradients and are designed to train a model to prevent it from memorizing training data or being susceptible to membership leakage.
Shokri et al.~\cite{SSSS17} designed a differential privacy method for collaborative learning of DNNs.
Nasr et al.~\cite{NSH18} propose an adversarial training-based method to add noise to the output posteriors in the training phase, namely adversarial regularization.
Jia et al.~\cite{JSBZG19} introduce MemGuard that adds noise to the output posteriors in the inference phase, which is the first defense with formal utility-loss guarantees against membership inference.

\mypara{Other Security and Privacy Attacks}
In addition to membership inference attacks, there are many other types of attacks against ML models.
In attribute inference attacks~\cite{GL162,MSCS19,SS20} the adversary's goal is to infer a specific sensitive attribute of a data sample from its representation (generated by a target model).
Model inversion attacks~\cite{FJR15,FLJLPR14} aim to reveal the missing or hidden attributes of the training data.
Adversarial examples~\cite{PMSW18,PMGJCS17,PMJFCS16,TKPGBM17,CJW20,LXZYL20} are one of the most popular evasion attacks, where the adversary tries to mislead the target classifier by adding imperceptible noise to the data sample.
Another line of work is the backdoor attack, which embeds a backdoor into the model so that the model performs normally on clean data, but acts maliciously on trigger data.~\cite{GDG17,WYSLVZZ19,YLZZ19,SSP20,ZMZBCJ20,LMBL20}.

\section{Conclusion}
\label{section:conclusion}

In this paper, we perform the first holistic measurement of attacks and defenses in the membership inference domain against machine learning models trained on image datasets.
We systematize membership inference through the study of nine attacks and six defenses, and couple them with the most comprehensive experimental evaluation (thus far) using six benchmark datasets and four popular model architectures.
Regarding the attack, we find that \texttt{NN-normal+label} performs the best and most metric-based attacks can reach similar performance but with less computational cost.
For the defense, we observe that MixupMMD and DP-SGD have the best defense effectiveness and MixupMMD can preserve the target model's utility better.
We show  that some assumptions in the threat model, namely that the shadow dataset/model must have the same distribution/architecture as the target dataset/model, are unnecessary.
We also find that the adversary can launch more effective attacks even using a real-world dataset collected from the Internet.
Interestingly, our evaluation shows that the Jensen-Shannon distance of entropy/cross-entropy between members and non-members has a much higher correlation to the attack performance than the overfitting level, and this correlation is model and dataset agnostic.
This grants us a new way to accurately predict membership inference risks without even running the attacks.
Lastly, we find that data augmentation--a common training technique that improves utility--is a good defense as it not only mitigates existing attacks to the largest extent but also significantly increases the target model's utility.
Data augmentation can be combined with existing defenses to further reduce membership inference risks.
However, augmenting attacks to incorporate data augmentation improves attack performance against models that employ data augmentation.

Throughout the study, we derive a group of new findings that can amend and complement previous understandings of membership inference attacks and defenses and can also inspire future research in the field.
We develop a modular framework \MembershipDoctor{} that can be easily integrated with new attacks and defenses, as well as plugins for various datasets, models, etc.
In this paper, we only focus on image classification models trained in supervised manners.
We plan to integrate more ML paradigms such as contrastive learning and graph neural networks in the future.
We envision that \MembershipDoctor{} will serve as a benchmark tool to facilitate future research on membership inference attacks and defenses.

\bibliographystyle{plain}
\bibliography{normal_generated_py3}

\begin{thebibliography}{10}

\bibitem{CIFAR}
\url{https://www.cs.toronto.edu/~kriz/cifar.html}.

\bibitem{OPACUS}
\url{https://github.com/pytorch/opacus}.

\bibitem{ART}
\url{https://github.com/Trusted-AI/adversarial-robustness-toolbox}.

\bibitem{ACGMMTZ16}
Martin Abadi, Andy Chu, Ian Goodfellow, Brendan McMahan, Ilya Mironov, Kunal
  Talwar, and Li~Zhang.
\newblock {Deep Learning with Differential Privacy}.
\newblock In {\em {ACM SIGSAC Conference on Computer and Communications
  Security (CCS)}}, pages 308--318. ACM, 2016.

\bibitem{CMS11}
Kamalika Chaudhuri, Claire Monteleoni, and Anand~D Sarwate.
\newblock {Differentially Private Empirical Risk Minimization}.
\newblock {\em {Journal of Machine Learning Research}}, 2011.

\bibitem{CYZF20}
Dingfan Chen, Ning Yu, Yang Zhang, and Mario Fritz.
\newblock {GAN-Leaks: A Taxonomy of Membership Inference Attacks against
  Generative Models}.
\newblock In {\em {ACM SIGSAC Conference on Computer and Communications
  Security (CCS)}}, pages 343--362. ACM, 2020.

\bibitem{CJW20}
Jianbo Chen, Michael~I. Jordan, and Martin~J. Wainwright.
\newblock {HopSkipJumpAttack: {A} Query-Efficient Decision-Based Attack}.
\newblock In {\em {IEEE Symposium on Security and Privacy (S\&P)}}, pages
  1277--1294. IEEE, 2020.

\bibitem{CTCP21}
Christopher A.~Choquette Choo, Florian Tram{\`e}r, Nicholas Carlini, and
  Nicolas Papernot.
\newblock {Label-Only Membership Inference Attacks}.
\newblock In {\em {International Conference on Machine Learning (ICML)}}, pages
  1964--1974. PMLR, 2021.

\bibitem{CZSL20}
Ekin~Dogus Cubuk, Barret Zoph, Jon Shlens, and Quoc Le.
\newblock {RandAugment: Practical Automated Data Augmentation with a Reduced
  Search Space}.
\newblock In {\em {Annual Conference on Neural Information Processing Systems
  (NeurIPS)}}, pages 18613--18624. NeurIPS, 2020.

\bibitem{DMNS06}
Cynthia Dwork, Frank McSherry, Kobbi Nissim, and Adam Smith.
\newblock {Calibrating Noise to Sensitivity in Private Data Analysis}.
\newblock In {\em {Theory of Cryptography Conference (TCC)}}, pages 265--284.
  Springer, 2006.

\bibitem{DR14}
Cynthia Dwork and Aaron Roth.
\newblock {\em {The Algorithmic Foundations of Differential Privacy}}.
\newblock Now Publishers Inc., 2014.

\bibitem{FJR15}
Matt Fredrikson, Somesh Jha, and Thomas Ristenpart.
\newblock {Model Inversion Attacks that Exploit Confidence Information and
  Basic Countermeasures}.
\newblock In {\em {ACM SIGSAC Conference on Computer and Communications
  Security (CCS)}}, pages 1322--1333. ACM, 2015.

\bibitem{FLJLPR14}
Matt Fredrikson, Eric Lantz, Somesh Jha, Simon Lin, David Page, and Thomas
  Ristenpart.
\newblock {Privacy in Pharmacogenetics: An End-to-End Case Study of
  Personalized Warfarin Dosing}.
\newblock In {\em {USENIX Security Symposium (USENIX Security)}}, pages 17--32.
  USENIX, 2014.

\bibitem{GL162}
Neil~Zhenqiang Gong and Bin Liu.
\newblock {You are Who You Know and How You Behave: Attribute Inference Attacks
  via Users' Social Friends and Behaviors}.
\newblock In {\em {USENIX Security Symposium (USENIX Security)}}, pages
  979--995. USENIX, 2016.

\bibitem{GPMXWOCB14}
Ian Goodfellow, Jean Pouget-Abadie, Mehdi Mirza, Bing Xu, David Warde-Farley,
  Sherjil Ozair, Aaron Courville, and Yoshua Bengio.
\newblock {Generative Adversarial Nets}.
\newblock In {\em {Annual Conference on Neural Information Processing Systems
  (NIPS)}}, pages 2672--2680. NIPS, 2014.

\bibitem{GDG17}
Tianyu Gu, Brendan Dolan-Gavitt, and Siddharth Grag.
\newblock {Badnets: Identifying Vulnerabilities in the Machine Learning Model
  Supply Chain}.
\newblock {\em {CoRR abs/1708.06733}}, 2017.

\bibitem{HMDC19}
Jamie Hayes, Luca Melis, George Danezis, and Emiliano~De Cristofaro.
\newblock {LOGAN: Evaluating Privacy Leakage of Generative Models Using
  Generative Adversarial Networks}.
\newblock {\em {Privacy Enhancing Technologies Symposium}}, 2019.

\bibitem{HZRS16}
Kaiming He, Xiangyu Zhang, Shaoqing Ren, and Jian Sun.
\newblock {Deep Residual Learning for Image Recognition}.
\newblock In {\em {IEEE Conference on Computer Vision and Pattern Recognition
  (CVPR)}}, pages 770--778. IEEE, 2016.

\bibitem{HWWBSZ21}
Xinlei He, Rui Wen, Yixin Wu, Michael Backes, Yun Shen, and Yang Zhang.
\newblock {Node-Level Membership Inference Attacks Against Graph Neural
  Networks}.
\newblock {\em {CoRR abs/2102.05429}}, 2021.

\bibitem{HZ21}
Xinlei He and Yang Zhang.
\newblock {Quantifying and Mitigating Privacy Risks of Contrastive Learning}.
\newblock In {\em {ACM SIGSAC Conference on Computer and Communications
  Security (CCS)}}. ACM, 2021.

\bibitem{INSTTW19}
Roger Iyengar, Joseph~P. Near, Dawn~Xiaodong Song, Om~Dipakbhai Thakkar,
  Abhradeep Thakurta, and Lun Wang.
\newblock {Towards Practical Differentially Private Convex Optimization}.
\newblock In {\em {IEEE Symposium on Security and Privacy (S\&P)}}, pages
  299--316. IEEE, 2019.

\bibitem{JE19}
Bargav Jayaraman and David Evans.
\newblock {Evaluating Differentially Private Machine Learning in Practice}.
\newblock In {\em {USENIX Security Symposium (USENIX Security)}}, pages
  1895--1912. USENIX, 2019.

\bibitem{JSBZG19}
Jinyuan Jia, Ahmed Salem, Michael Backes, Yang Zhang, and Neil~Zhenqiang Gong.
\newblock {MemGuard: Defending against Black-Box Membership Inference Attacks
  via Adversarial Examples}.
\newblock In {\em {ACM SIGSAC Conference on Computer and Communications
  Security (CCS)}}, pages 259--274. ACM, 2019.

\bibitem{KD21}
Yigitcan Kaya and Tudor Dumitras.
\newblock {When Does Data Augmentation Help With Membership Inference Attacks?}
\newblock In {\em {International Conference on Machine Learning (ICML)}}, pages
  5345--5355. PMLR, 2021.

\bibitem{KSH12}
Alex Krizhevsky, Ilya Sutskever, and Geoffrey~E. Hinton.
\newblock {ImageNet Classification with Deep Convolutional Neural Networks}.
\newblock In {\em {Annual Conference on Neural Information Processing Systems
  (NIPS)}}, pages 1106--1114. NIPS, 2012.

\bibitem{LF20}
Klas Leino and Matt Fredrikson.
\newblock {Stolen Memories: Leveraging Model Memorization for Calibrated
  White-Box Membership Inference}.
\newblock In {\em {USENIX Security Symposium (USENIX Security)}}, pages
  1605--1622. USENIX, 2020.

\bibitem{LXZYL20}
Huichen Li, Xiaojun Xu, Xiaolu Zhang, Shuang Yang, and Bo~Li.
\newblock {{QEBA:} Query-Efficient Boundary-Based Blackbox Attack}.
\newblock In {\em {IEEE Conference on Computer Vision and Pattern Recognition
  (CVPR)}}, pages 1218--1227. IEEE, 2020.

\bibitem{LLR21}
Jiacheng Li, Ninghui Li, and Bruno Ribeiro.
\newblock {Membership Inference Attacks and Defenses in Classification Models}.
\newblock In {\em {ACM Conference on Data and Application Security and Privacy
  (CODASPY)}}, pages 5--16. ACM, 2021.

\bibitem{LLSY16}
Ninghui Li, Min Lyu, Dong Su, and Weining Yang.
\newblock {\em {Differential Privacy: From Theory to Practice}}.
\newblock Morgan \& Claypool Publishers, 2016.

\bibitem{LZ21}
Zheng Li and Yang Zhang.
\newblock {Membership Leakage in Label-Only Exposures}.
\newblock In {\em {ACM SIGSAC Conference on Computer and Communications
  Security (CCS)}}. ACM, 2021.

\bibitem{LMBL20}
Yunfei Liu, Xingjun Ma, James Bailey, and Feng Lu.
\newblock {Reflection Backdoor: A Natural Backdoor Attack on Deep Neural
  Networks}.
\newblock In {\em {European Conference on Computer Vision (ECCV)}}, pages
  182--199. Springer, 2020.

\bibitem{MSCS19}
Luca Melis, Congzheng Song, Emiliano~De Cristofaro, and Vitaly Shmatikov.
\newblock {Exploiting Unintended Feature Leakage in Collaborative Learning}.
\newblock In {\em {IEEE Symposium on Security and Privacy (S\&P)}}, pages
  497--512. IEEE, 2019.

\bibitem{MKH19}
Rafael M{\"{u}}ller, Simon Kornblith, and Geoffrey~E. Hinton.
\newblock {When Does Label Smoothing Help?}
\newblock In {\em {Annual Conference on Neural Information Processing Systems
  (NeurIPS)}}, pages 4696--4705. NeurIPS, 2019.

\bibitem{NSH18}
Milad Nasr, Reza Shokri, and Amir Houmansadr.
\newblock {Machine Learning with Membership Privacy using Adversarial
  Regularization}.
\newblock In {\em {ACM SIGSAC Conference on Computer and Communications
  Security (CCS)}}, pages 634--646. ACM, 2018.

\bibitem{NSH19}
Milad Nasr, Reza Shokri, and Amir Houmansadr.
\newblock {Comprehensive Privacy Analysis of Deep Learning: Passive and Active
  White-box Inference Attacks against Centralized and Federated Learning}.
\newblock In {\em {IEEE Symposium on Security and Privacy (S\&P)}}, pages
  1021--1035. IEEE, 2019.

\bibitem{ONK21}
Iyiola~E. Olatunji, Wolfgang Nejdl, and Megha Khosla.
\newblock {Membership Inference Attack on Graph Neural Networks}.
\newblock {\em {CoRR abs/2101.06570}}, 2021.

\bibitem{PMSW18}
Nicolas Papernot, Patrick McDaniel, Arunesh Sinha, and Michael Wellman.
\newblock {SoK: Towards the Science of Security and Privacy in Machine
  Learning}.
\newblock In {\em {IEEE European Symposium on Security and Privacy (Euro
  S\&P)}}, pages 399--414. IEEE, 2018.

\bibitem{PMGJCS17}
Nicolas Papernot, Patrick~D. McDaniel, Ian Goodfellow, Somesh Jha, Z.~Berkay
  Celik, and Ananthram Swami.
\newblock {Practical Black-Box Attacks Against Machine Learning}.
\newblock In {\em {ACM Asia Conference on Computer and Communications Security
  (ASIACCS)}}, pages 506--519. ACM, 2017.

\bibitem{PMJFCS16}
Nicolas Papernot, Patrick~D. McDaniel, Somesh Jha, Matt Fredrikson, Z.~Berkay
  Celik, and Ananthram Swami.
\newblock {The Limitations of Deep Learning in Adversarial Settings}.
\newblock In {\em {IEEE European Symposium on Security and Privacy (Euro
  S\&P)}}, pages 372--387. IEEE, 2016.

\bibitem{SDSOJ19}
Alexandre Sablayrolles, Matthijs Douze, Cordelia Schmid, Yann Ollivier, and
  Herv{\'e} J{\'e}gou.
\newblock {White-box vs Black-box: Bayes Optimal Strategies for Membership
  Inference}.
\newblock In {\em {International Conference on Machine Learning (ICML)}}, pages
  5558--5567. PMLR, 2019.

\bibitem{SSP20}
Aniruddha Saha, Akshayvarun Subramanya, and Hamed Pirsiavash.
\newblock {Hidden Trigger Backdoor Attacks}.
\newblock In {\em {AAAI Conference on Artificial Intelligence (AAAI)}}, pages
  11957--11965. AAAI, 2020.

\bibitem{SZHBFB19}
Ahmed Salem, Yang Zhang, Mathias Humbert, Pascal Berrang, Mario Fritz, and
  Michael Backes.
\newblock {ML-Leaks: Model and Data Independent Membership Inference Attacks
  and Defenses on Machine Learning Models}.
\newblock In {\em {Network and Distributed System Security Symposium (NDSS)}}.
  Internet Society, 2019.

\bibitem{SHZZC18}
Mark Sandler, Andrew~G. Howard, Menglong Zhu, Andrey Zhmoginov, and
  Liang{-}Chieh Chen.
\newblock {MobileNetV2: Inverted Residuals and Linear Bottlenecks}.
\newblock In {\em {IEEE Conference on Computer Vision and Pattern Recognition
  (CVPR)}}, pages 4510--4520. IEEE, 2018.

\bibitem{SSSS17}
Reza Shokri, Marco Stronati, Congzheng Song, and Vitaly Shmatikov.
\newblock {Membership Inference Attacks Against Machine Learning Models}.
\newblock In {\em {IEEE Symposium on Security and Privacy (S\&P)}}, pages
  3--18. IEEE, 2017.

\bibitem{SZ15}
Karen Simonyan and Andrew Zisserman.
\newblock {Very Deep Convolutional Networks for Large-Scale Image Recognition}.
\newblock In {\em {International Conference on Learning Representations
  (ICLR)}}, 2015.

\bibitem{SR20}
Congzheng Song and Ananth Raghunathan.
\newblock {Information Leakage in Embedding Models}.
\newblock In {\em {ACM SIGSAC Conference on Computer and Communications
  Security (CCS)}}, pages 377--390. ACM, 2020.

\bibitem{SS19}
Congzheng Song and Vitaly Shmatikov.
\newblock {Auditing Data Provenance in Text-Generation Models}.
\newblock In {\em {ACM Conference on Knowledge Discovery and Data Mining
  (KDD)}}, pages 196--206. ACM, 2019.

\bibitem{SS20}
Congzheng Song and Vitaly Shmatikov.
\newblock {Overlearning Reveals Sensitive Attributes}.
\newblock In {\em {International Conference on Learning Representations
  (ICLR)}}, 2020.

\bibitem{SM21}
Liwei Song and Prateek Mittal.
\newblock {Systematic Evaluation of Privacy Risks of Machine Learning Models}.
\newblock In {\em {USENIX Security Symposium (USENIX Security)}}. USENIX, 2021.

\bibitem{SSM19}
Liwei Song, Reza Shokri, and Prateek Mittal.
\newblock {Privacy Risks of Securing Machine Learning Models against
  Adversarial Examples}.
\newblock In {\em {ACM SIGSAC Conference on Computer and Communications
  Security (CCS)}}, pages 241--257. ACM, 2019.

\bibitem{SVISW16}
Christian Szegedy, Vincent Vanhoucke, Sergey Ioffe, Jonathon Shlens, and
  Zbigniew Wojna.
\newblock {Rethinking the Inception Architecture for Computer Vision}.
\newblock In {\em {IEEE Conference on Computer Vision and Pattern Recognition
  (CVPR)}}, pages 2818--2826. IEEE, 2016.

\bibitem{TKPGBM17}
Florian Tram{\`e}r, Alexey Kurakin, Nicolas Papernot, Ian Goodfellow, Dan
  Boneh, and Patrick McDaniel.
\newblock {Ensemble Adversarial Training: Attacks and Defenses}.
\newblock In {\em {International Conference on Learning Representations
  (ICLR)}}, 2017.

\bibitem{WYSLVZZ19}
Bolun Wang, Yuanshun Yao, Shawn Shan, Huiying Li, Bimal Viswanath, Haitao
  Zheng, and Ben~Y. Zhao.
\newblock {Neural Cleanse: Identifying and Mitigating Backdoor Attacks in
  Neural Networks}.
\newblock In {\em {IEEE Symposium on Security and Privacy (S\&P)}}, pages
  707--723. IEEE, 2019.

\bibitem{YLZZ19}
Yuanshun Yao, Huiying Li, Haitao Zheng, and Ben~Y. Zhao.
\newblock {Latent Backdoor Attacks on Deep Neural Networks}.
\newblock In {\em {ACM SIGSAC Conference on Computer and Communications
  Security (CCS)}}, pages 2041--2055. ACM, 2019.

\bibitem{YGFJ18}
Samuel Yeom, Irene Giacomelli, Matt Fredrikson, and Somesh Jha.
\newblock {Privacy Risk in Machine Learning: Analyzing the Connection to
  Overfitting}.
\newblock In {\em {IEEE Computer Security Foundations Symposium (CSF)}}, pages
  268--282. IEEE, 2018.

\bibitem{YZCYL21}
Da~Yu, Huishuai Zhang, Wei Chen, Jian Yin, and Tie{-}Yan Liu.
\newblock {How Does Data Augmentation Affect Privacy in Machine Learning?}
\newblock In {\em {AAAI Conference on Artificial Intelligence (AAAI)}}, pages
  10746--10753. AAAI, 2021.

\bibitem{ZK16}
Sergey Zagoruyko and Nikos Komodakis.
\newblock {Wide Residual Networks}.
\newblock In {\em {Proceedings of the British Machine Vision Conference
  (BMVC)}}. {BMVA} Press, 2016.

\bibitem{ZCDL18}
Hongyi Zhang, Moustapha Ciss{\'{e}}, Yann~N. Dauphin, and David Lopez{-}Paz.
\newblock {mixup: Beyond Empirical Risk Minimization}.
\newblock In {\em {International Conference on Learning Representations
  (ICLR)}}, 2018.

\bibitem{ZMZBCJ20}
Shihao Zhao, Xingjun Ma, Xiang Zheng, James Bailey, Jingjing Chen, and Yu-Gang
  Jiang.
\newblock {Clean-Label Backdoor Attacks on Video Recognition Models}.
\newblock In {\em {IEEE Conference on Computer Vision and Pattern Recognition
  (CVPR)}}, pages 14443--144528. IEEE, 2020.

\bibitem{ZLKOT18}
Bolei Zhou, {\`{A}}gata Lapedriza, Aditya Khosla, Aude Oliva, and Antonio
  Torralba.
\newblock {Places: {A} 10 Million Image Database for Scene Recognition}.
\newblock {\em {IEEE Transactions on Pattern Analysis and Machine
  Intelligence}}, 2018.

\end{thebibliography}

\newpage
\appendix
\section{Appendix}
\label{section:Appendix}

\subsection{Evaluate Membership Inference Attacks and Defenses with Different Model Architectures}
\label{appendix:evaluate_diff_model}

\begin{figure*}[!ht]
\centering
\begin{subfigure}{0.48\columnwidth}
\includegraphics[width=\columnwidth]{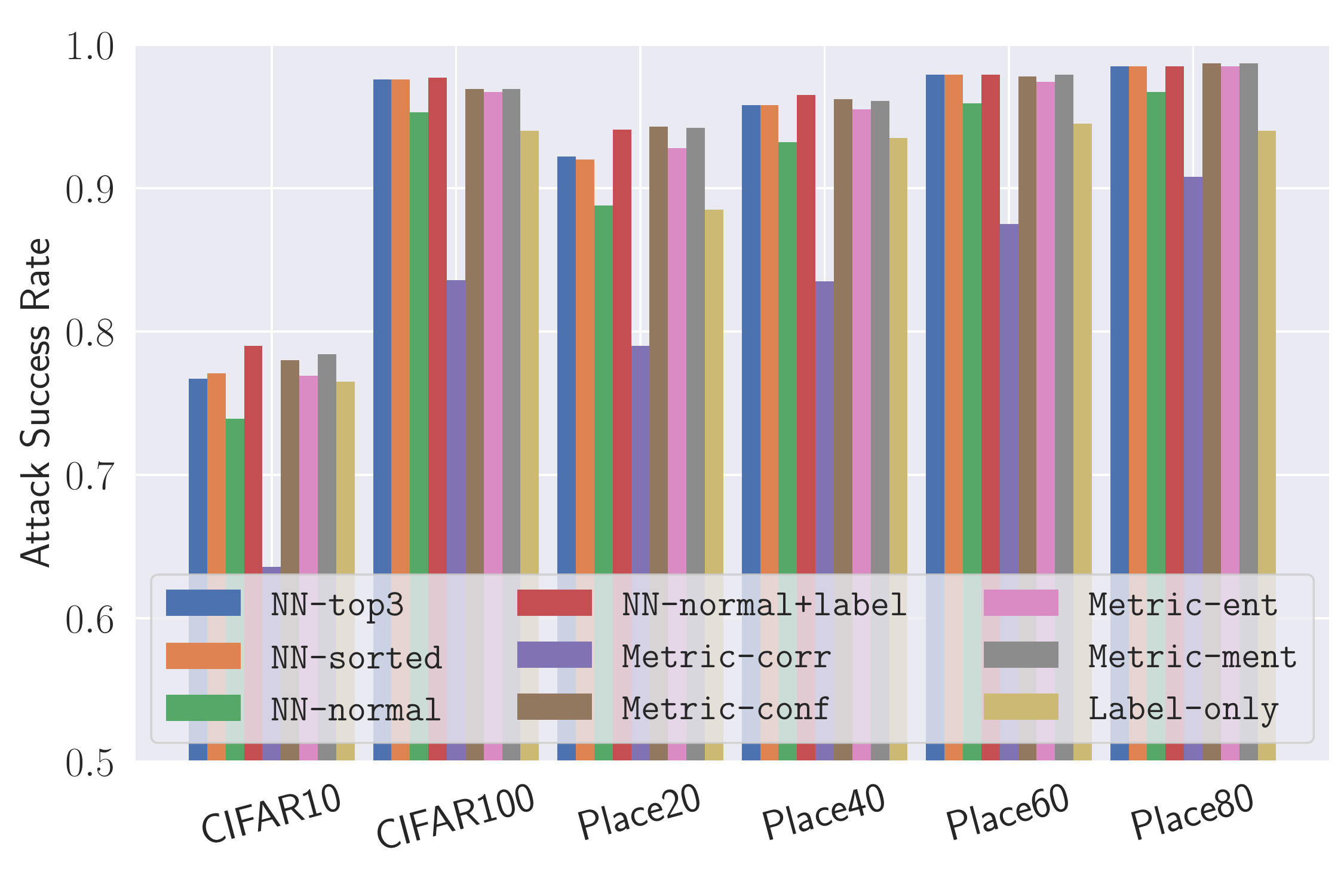}
\caption{MobileNetV2}
\label{figure:RQ1_attack_performance_no_mobilenetv2}
\end{subfigure}
\begin{subfigure}{0.48\columnwidth}
\includegraphics[width=\columnwidth]{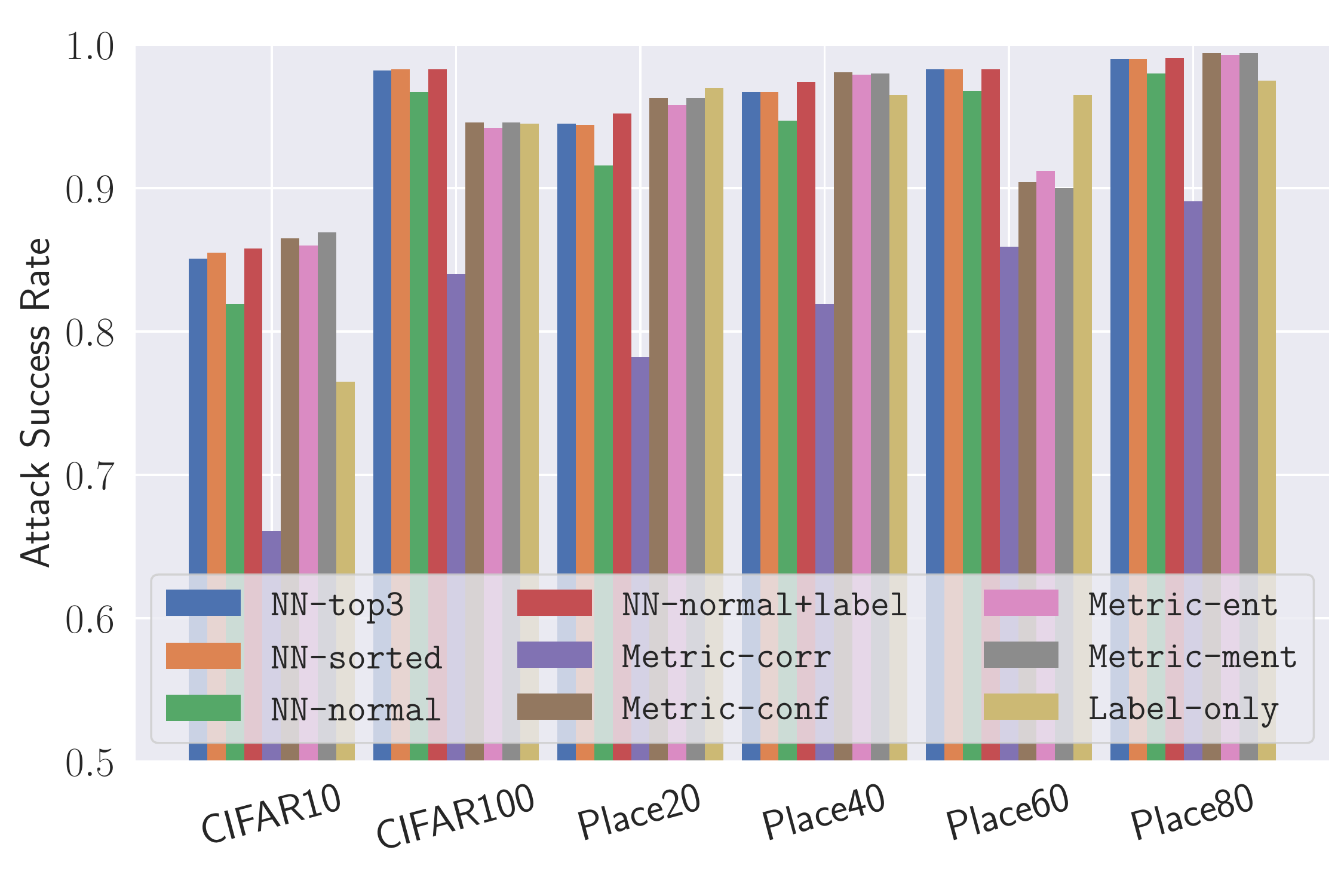}
\caption{ResNet18}
\label{figure:RQ1_attack_performance_no_resnet18}
\end{subfigure}
\begin{subfigure}{0.48\columnwidth}
\includegraphics[width=\columnwidth]{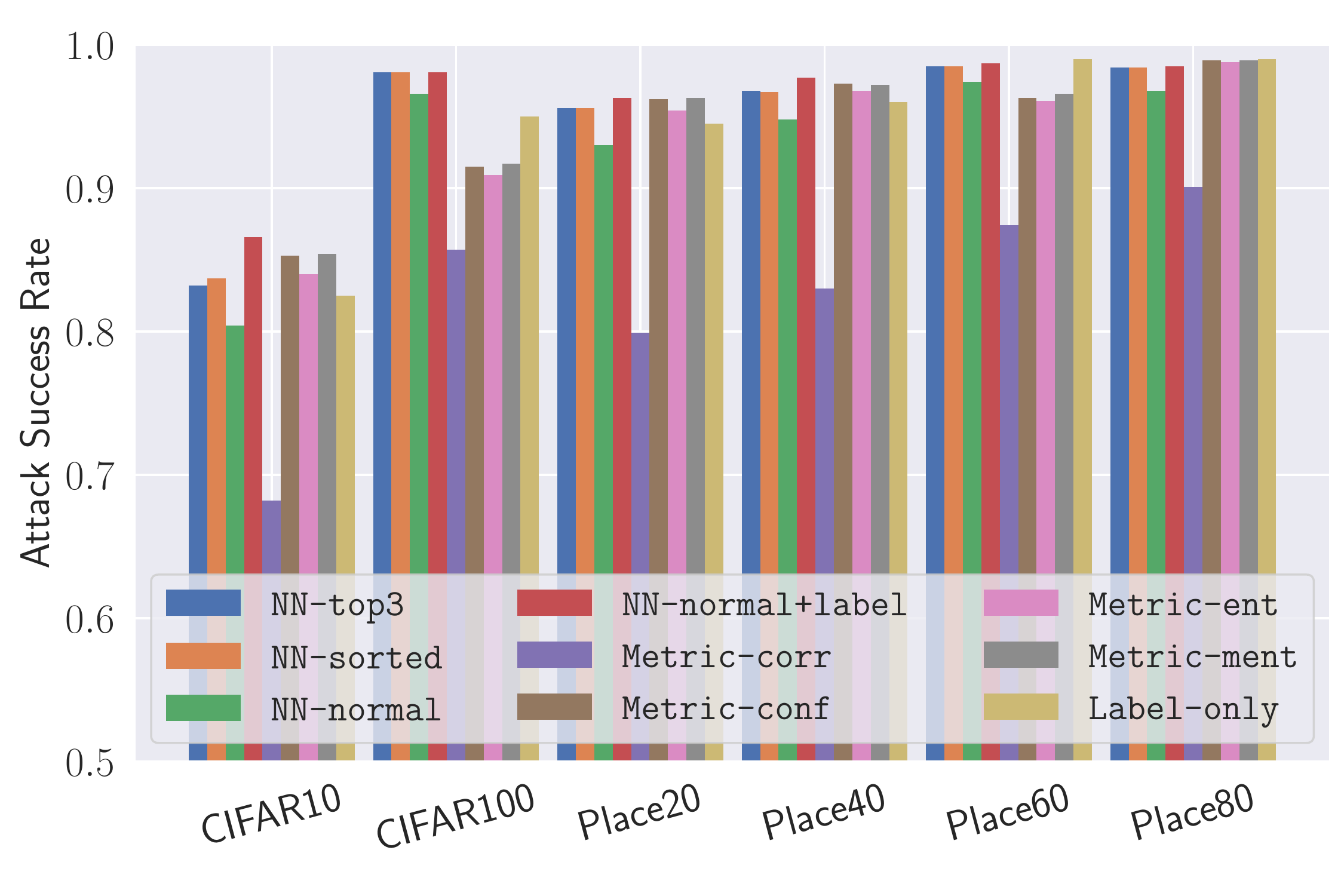}
\caption{ResNet34}
\label{figure:RQ1_attack_performance_no_resnet34}
\end{subfigure}
\begin{subfigure}{0.48\columnwidth}
\includegraphics[width=\columnwidth]{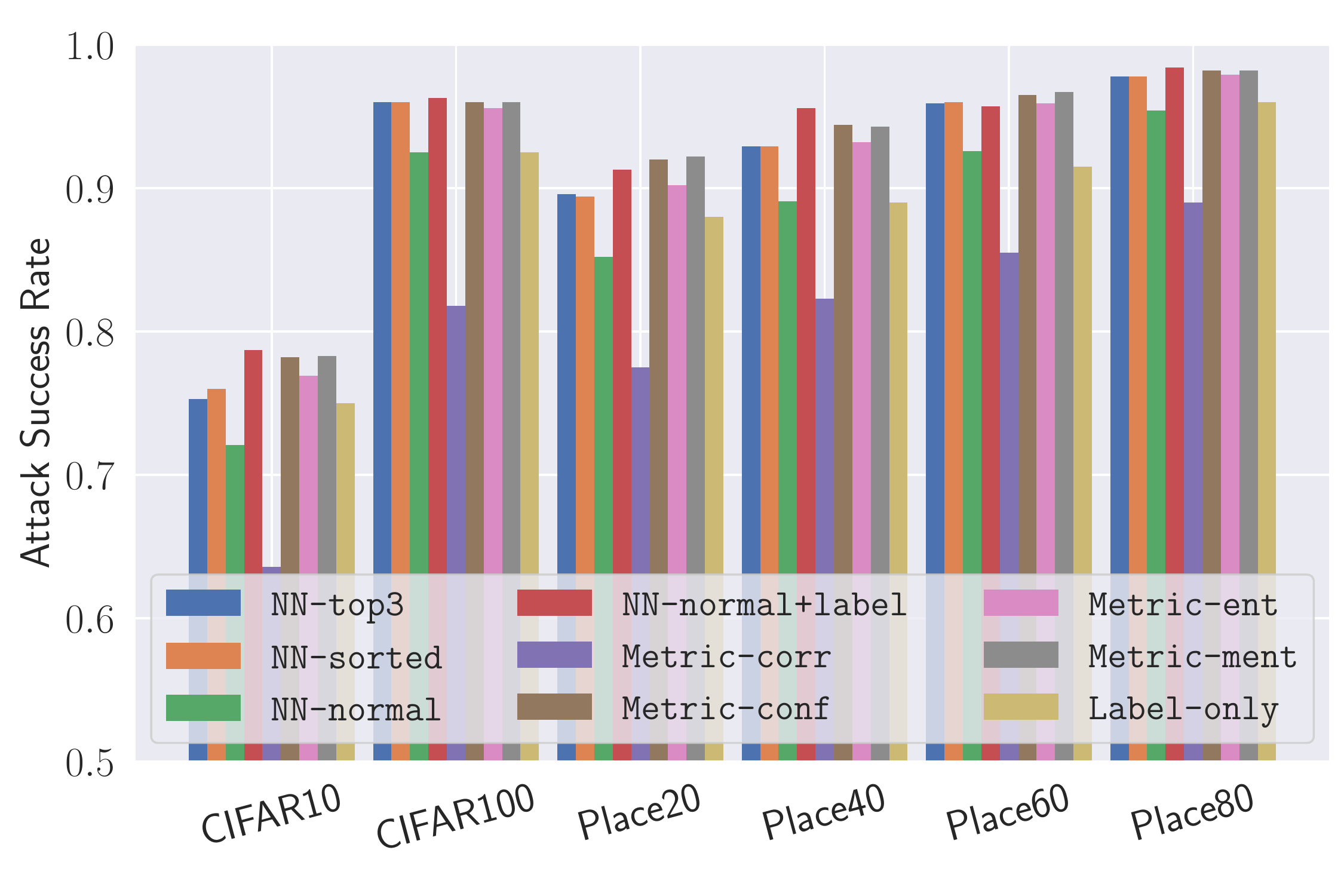}
\caption{VGG11}
\label{figure:RQ1_attack_performance_no_vgg11}
\end{subfigure}
\caption{The performance of different membership inference attacks on 6 different datasets. The x-axis represents different datasets. The y-axis represents the membership inference attack's accuracy.}
\label{figure:RQ1_attack_performance_no_diff_model}
\end{figure*}

The performance of different membership inference attacks on 6 different datasets when both target and shadow models are MobileNetV2, ResNet18, ResNet34, and VGG11 are shown in \Cref{figure:RQ1_attack_performance_no_mobilenetv2}, \Cref{figure:RQ1_attack_performance_no_resnet18}, \Cref{figure:RQ1_attack_performance_no_resnet34}, and \Cref{figure:RQ1_attack_performance_no_vgg11}, respectively.

\begin{figure*}[!ht]
\centering
\begin{subfigure}{0.48\columnwidth}
\includegraphics[width=\columnwidth]{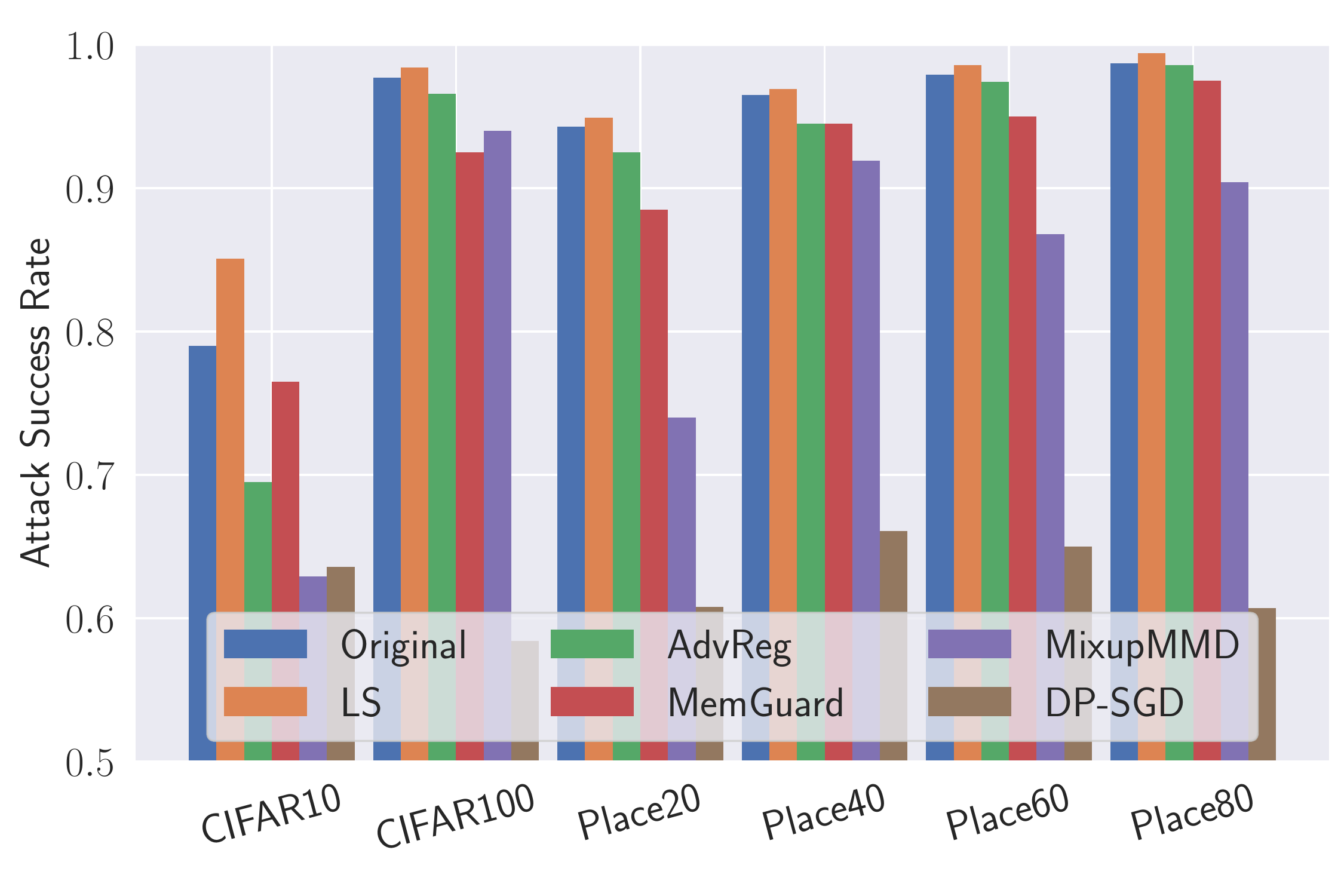}
\caption{MobileNetV2}
\label{figure:RQ1_defense_performance_no_mobilenetv2}
\end{subfigure}
\begin{subfigure}{0.48\columnwidth}
\includegraphics[width=\columnwidth]{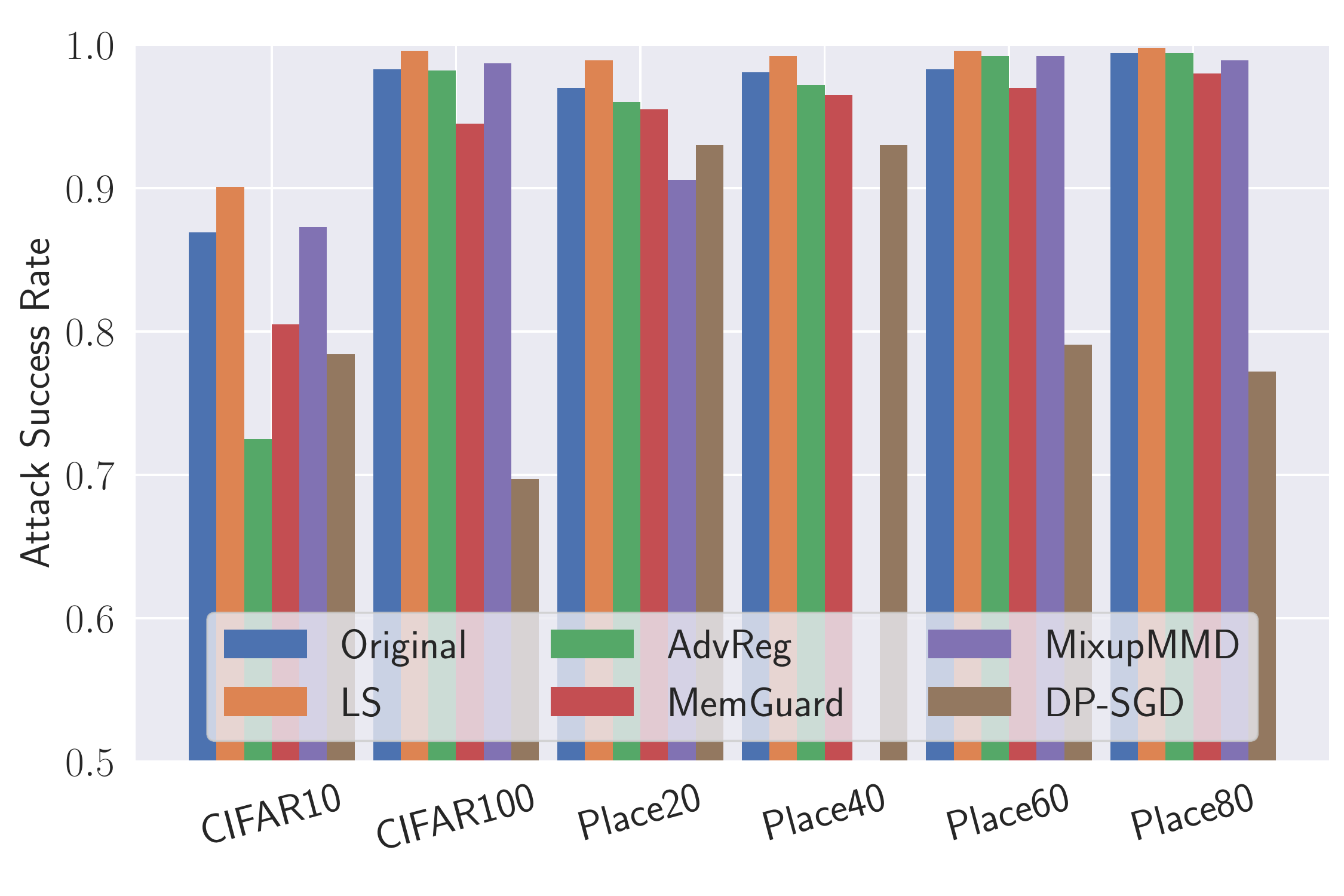}
\caption{ResNet18}
\label{figure:RQ1_defense_performance_no_resnet18}
\end{subfigure}
\begin{subfigure}{0.48\columnwidth}
\includegraphics[width=\columnwidth]{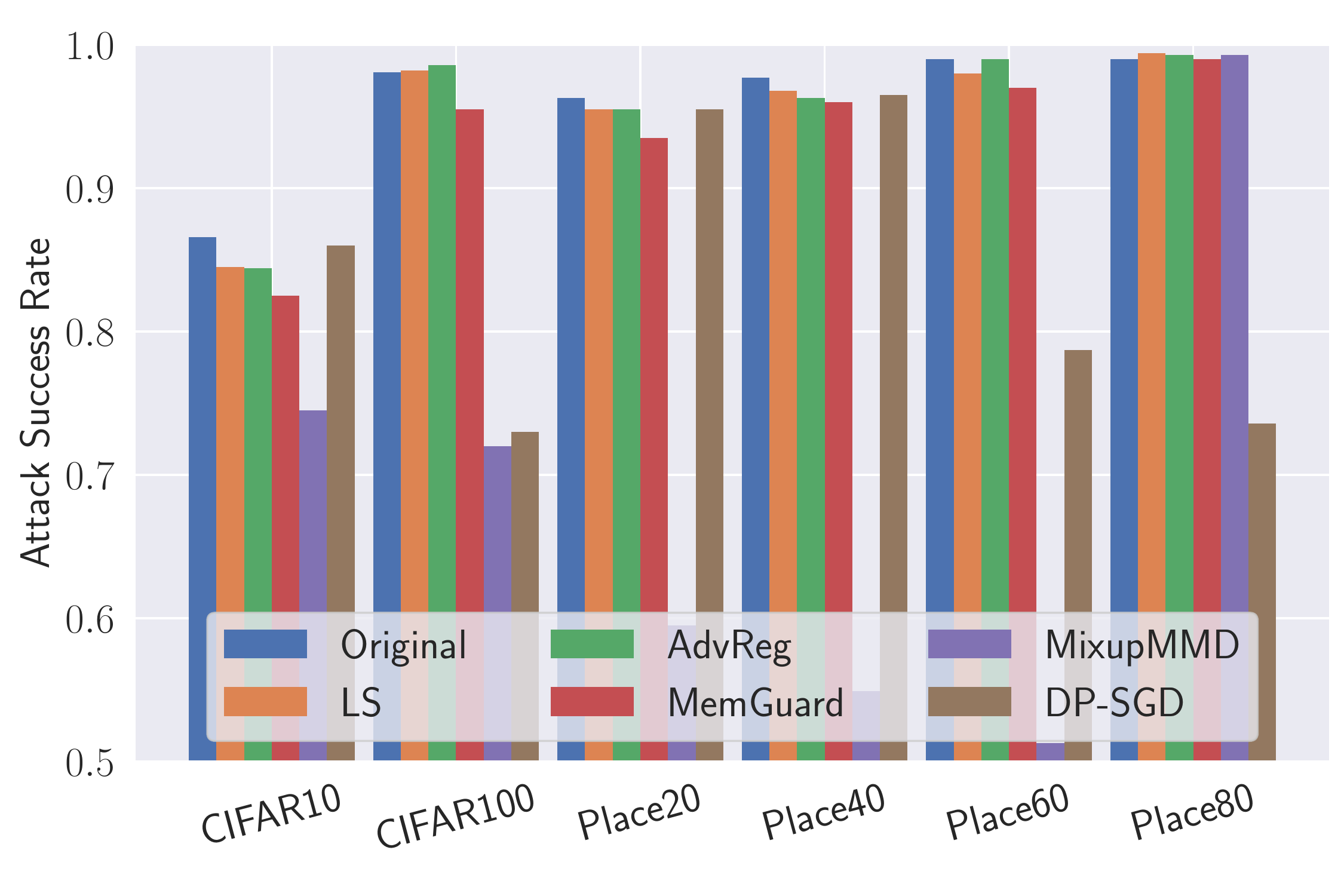}
\caption{ResNet34}
\label{figure:RQ1_defense_performance_no_resnet34}
\end{subfigure}
\begin{subfigure}{0.48\columnwidth}
\includegraphics[width=\columnwidth]{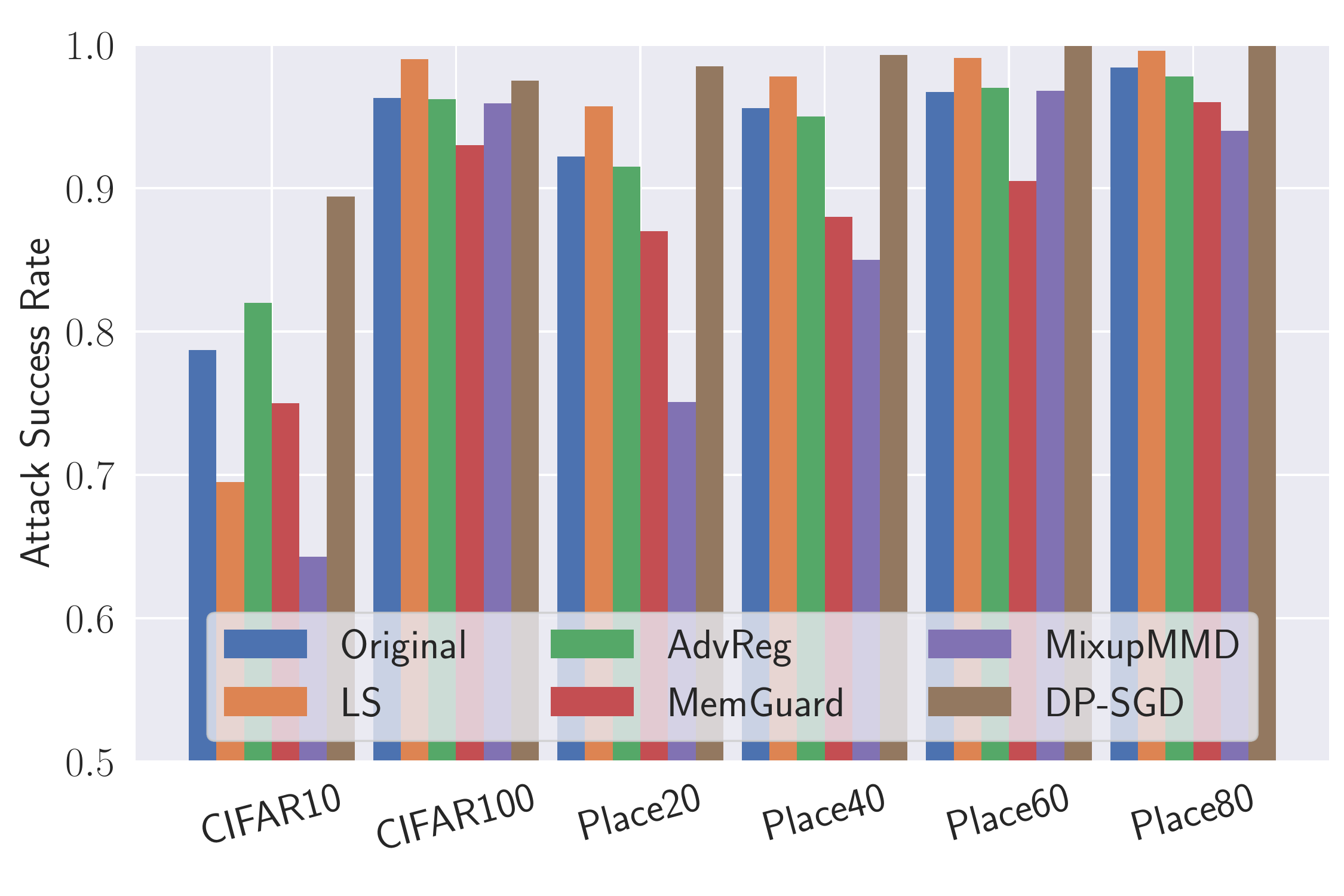}
\caption{VGG11}
\label{figure:RQ1_defense_performance_no_vgg11}
\end{subfigure}
\caption{The performance of the best membership inference attacks against the original models and models defended by different methods on 6 different datasets. The x-axis represents different datasets. The y-axis represents the membership inference attack's accuracy.}
\label{figure:RQ1_defense_performance_no_diff_model}
\end{figure*}

The performance of the best membership inference attacks against the original models and models defended by different methods on 6 different datasets when both target and shadow models are MobileNetV2, ResNet18, ResNet34, and VGG11 are shown in \Cref{figure:RQ1_defense_performance_no_mobilenetv2}, \Cref{figure:RQ1_defense_performance_no_resnet18}, \Cref{figure:RQ1_defense_performance_no_resnet34}, and \Cref{figure:RQ1_defense_performance_no_vgg11}, respectively.

\subsection{Evaluate Membership Inference Attacks and Defenses with Different Metrics}
\label{appendix:diff_evaluate_metric}

\begin{figure*}[!ht]
\centering
\begin{subfigure}{0.48\columnwidth}
\includegraphics[width=\columnwidth]{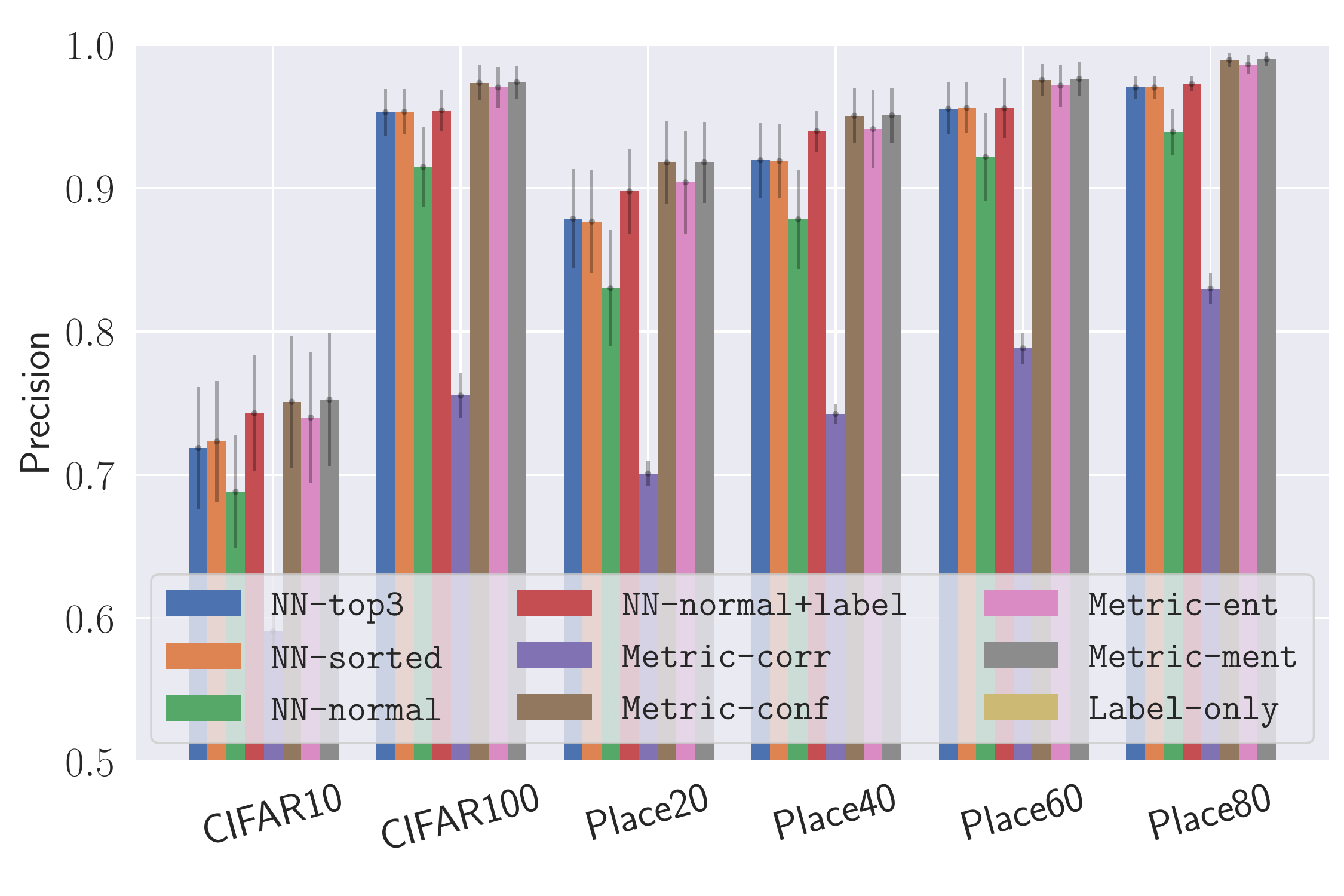}
\caption{Precision}
\label{figure:RQ1_attack_performance_precision_no}
\end{subfigure}
\begin{subfigure}{0.48\columnwidth}
\includegraphics[width=\columnwidth]{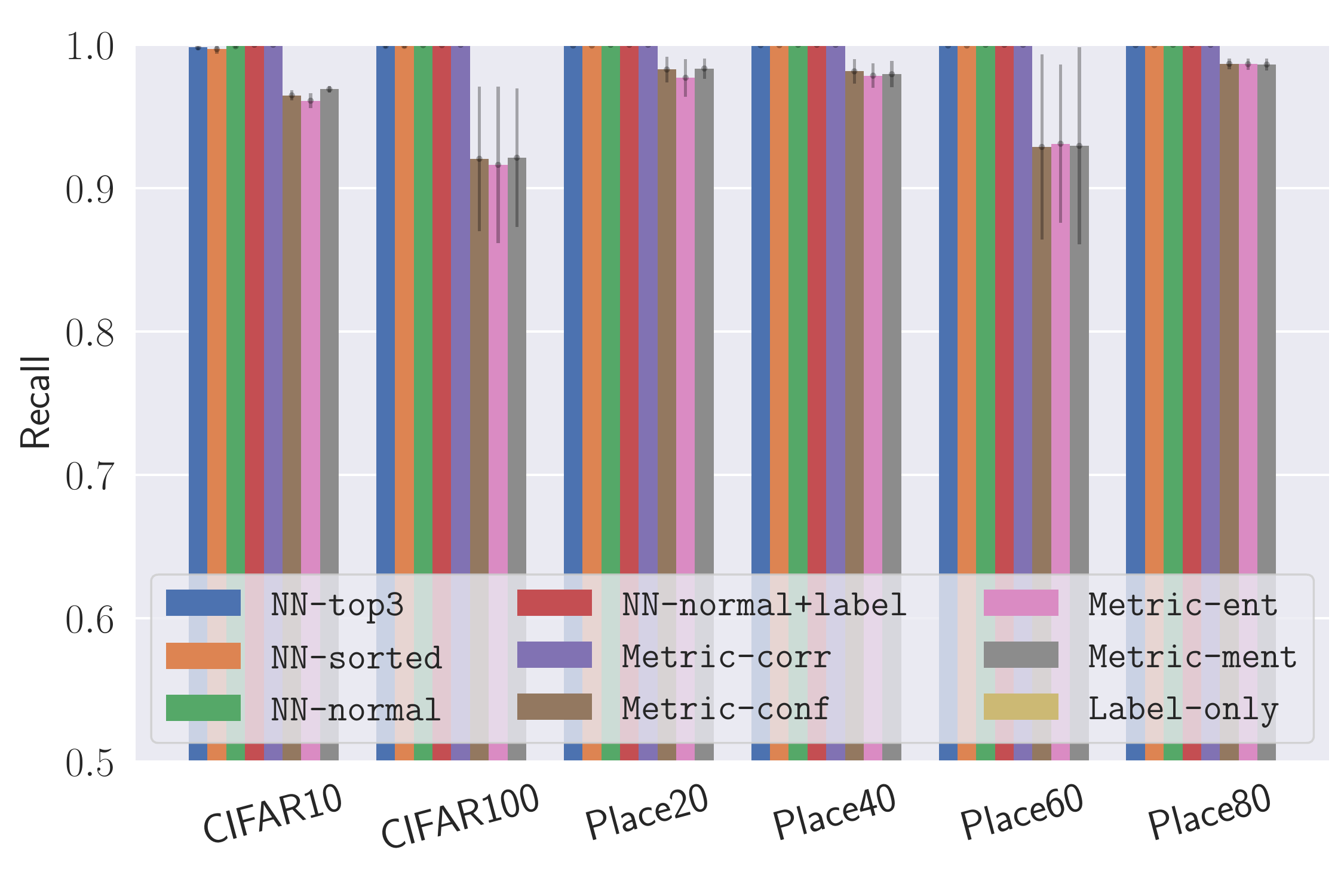}
\caption{Recall}
\label{figure:RQ1_attack_performance_recall_no}
\end{subfigure}
\label{figure:RQ1_attack_performance_precision_recall}
\begin{subfigure}{0.48\columnwidth}
\includegraphics[width=\columnwidth]{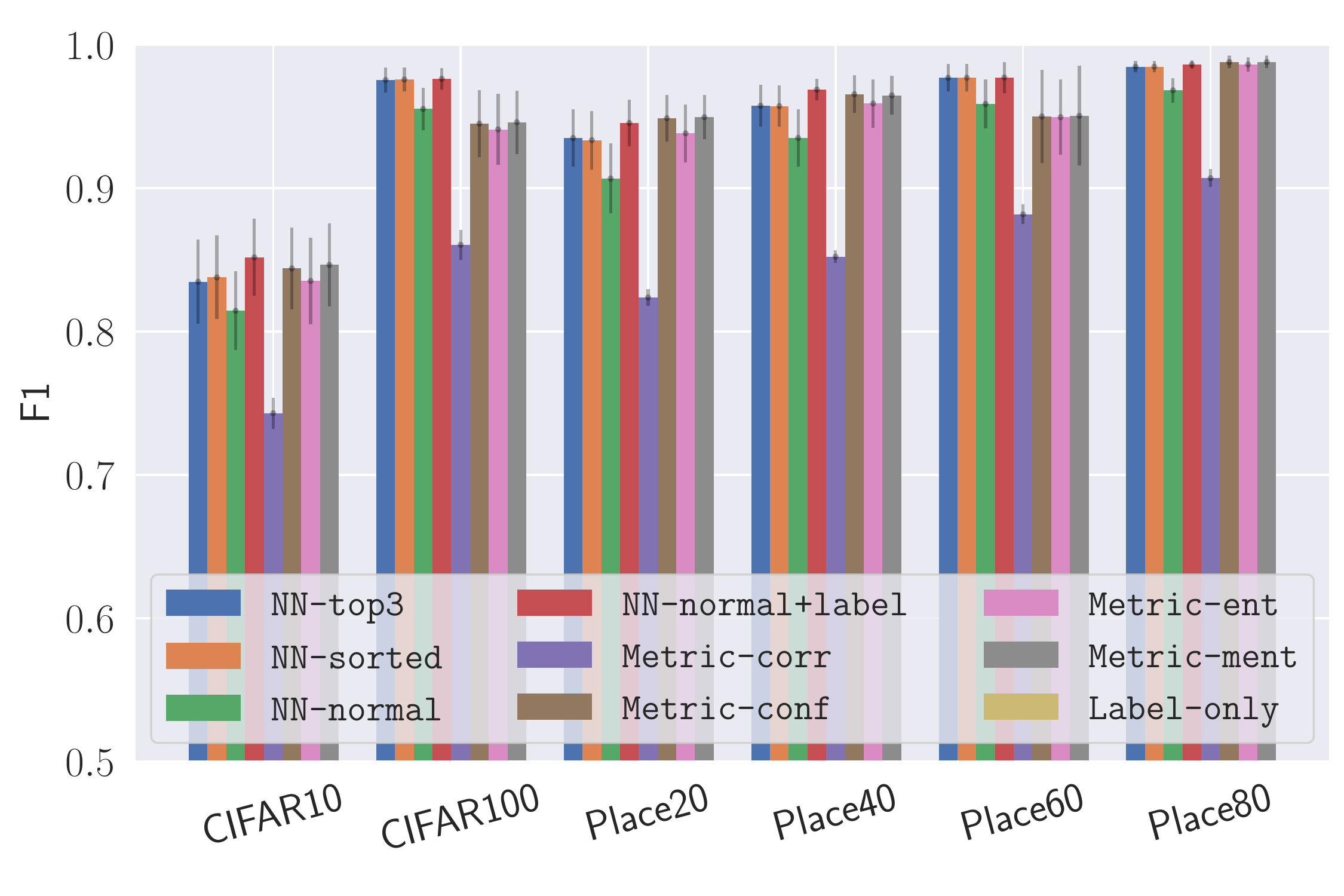}
\caption{F1}
\label{figure:RQ1_attack_performance_f1_no}
\end{subfigure}
\label{figure:RQ1_attack_performance_precision_f1}
\begin{subfigure}{0.48\columnwidth}
\includegraphics[width=\columnwidth]{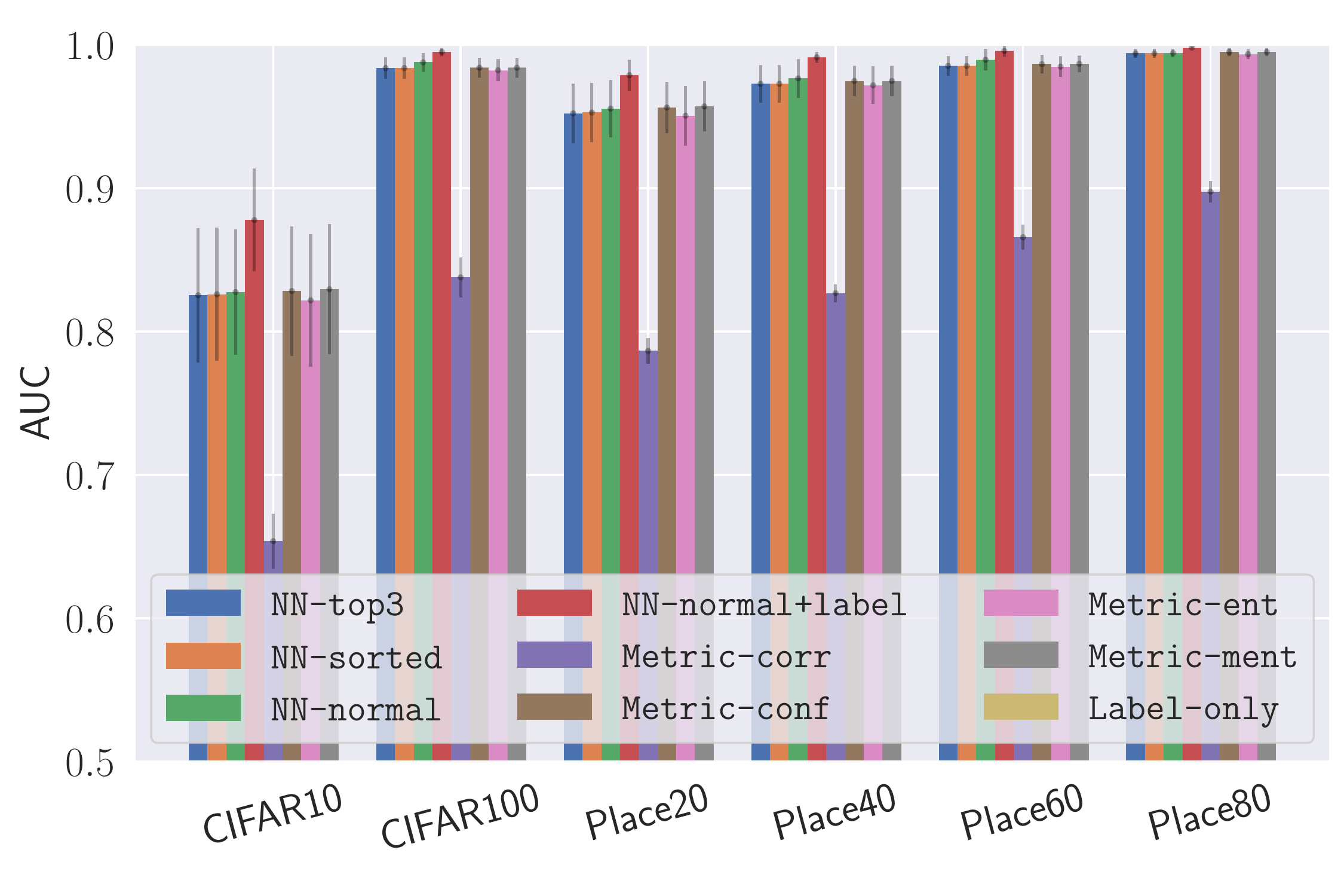}
\caption{AUC}
\label{figure:RQ1_attack_performance_auc_no}
\end{subfigure}
\caption{The performance of different membership inference attacks on 6 different datasets. The x-axis represents different datasets. The y-axis represents the membership inference attack's Precision, Recall, F1, and AUC, respectively. Note that we average the attack performance under different model architectures and show the standard deviations as well.}
\label{figure:RQ1_attack_performance_prf1auc_no}
\end{figure*}

\begin{figure*}[!ht]
\centering
\begin{subfigure}{0.48\columnwidth}
\includegraphics[width=\columnwidth]{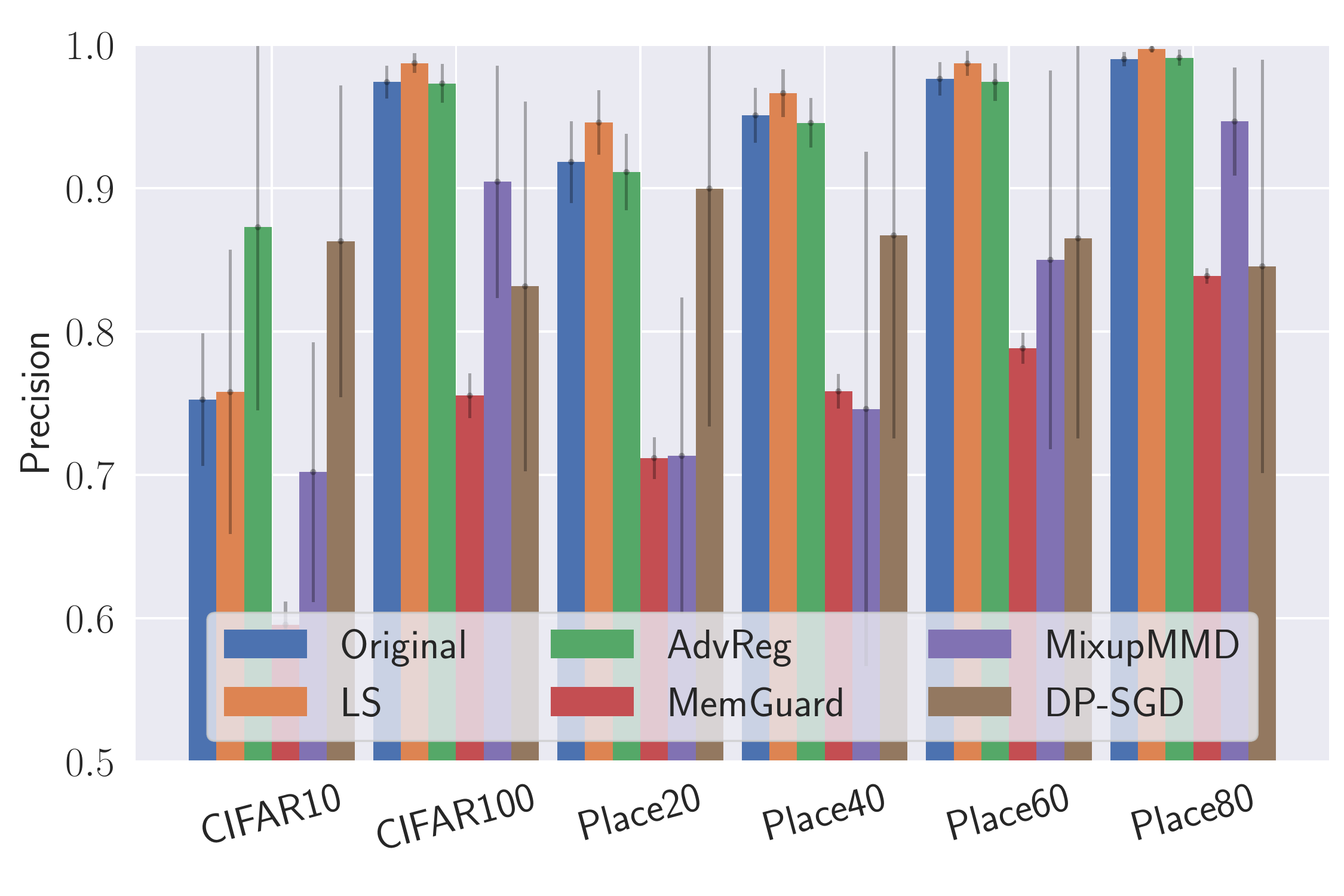}
\caption{Precision}
\label{figure:RQ1_defense_performance_precision_no}
\end{subfigure}
\begin{subfigure}{0.48\columnwidth}
\includegraphics[width=\columnwidth]{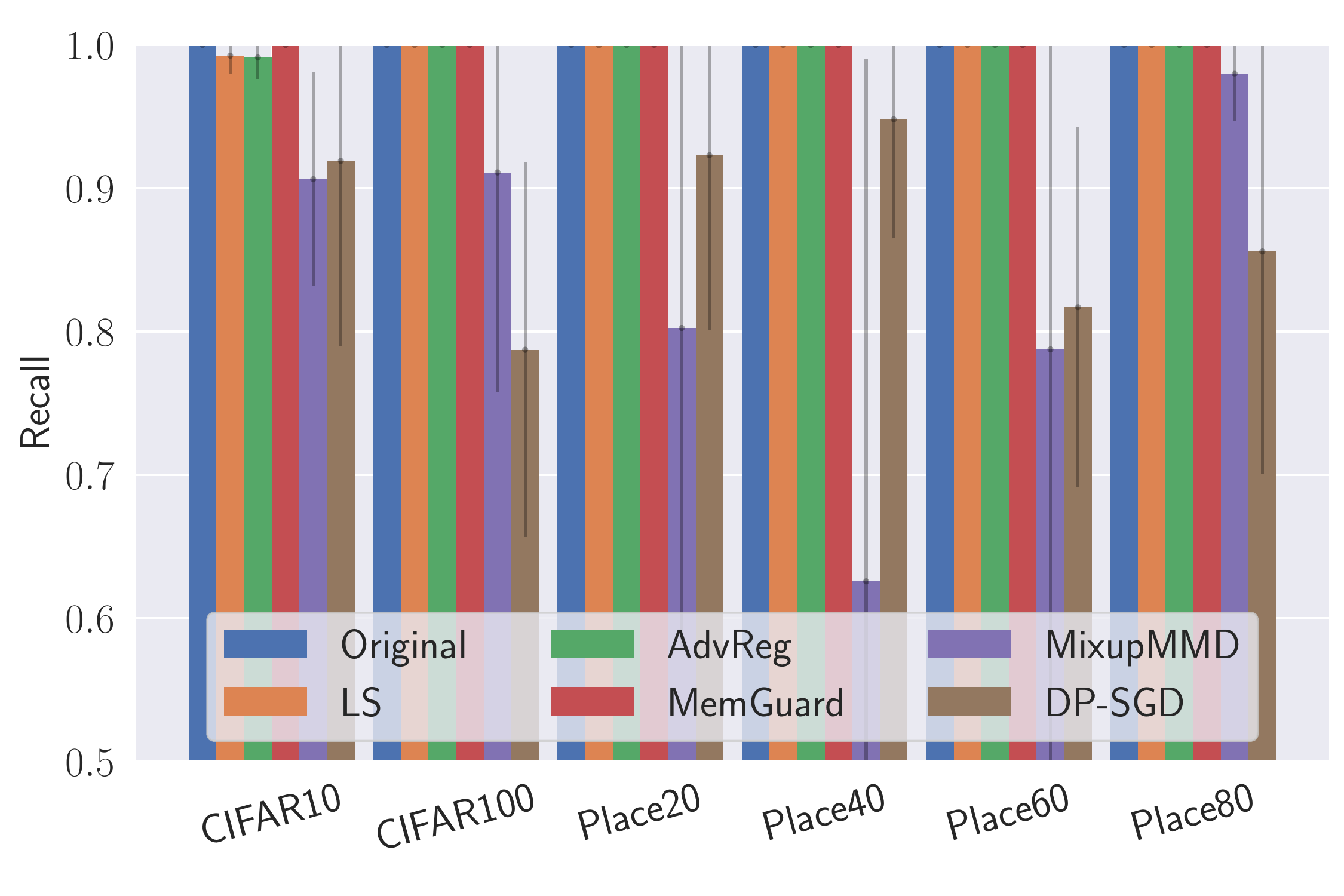}
\caption{Recall}
\label{figure:RQ1_defense_performance_recall_no}
\end{subfigure}
\begin{subfigure}{0.48\columnwidth}
\includegraphics[width=\columnwidth]{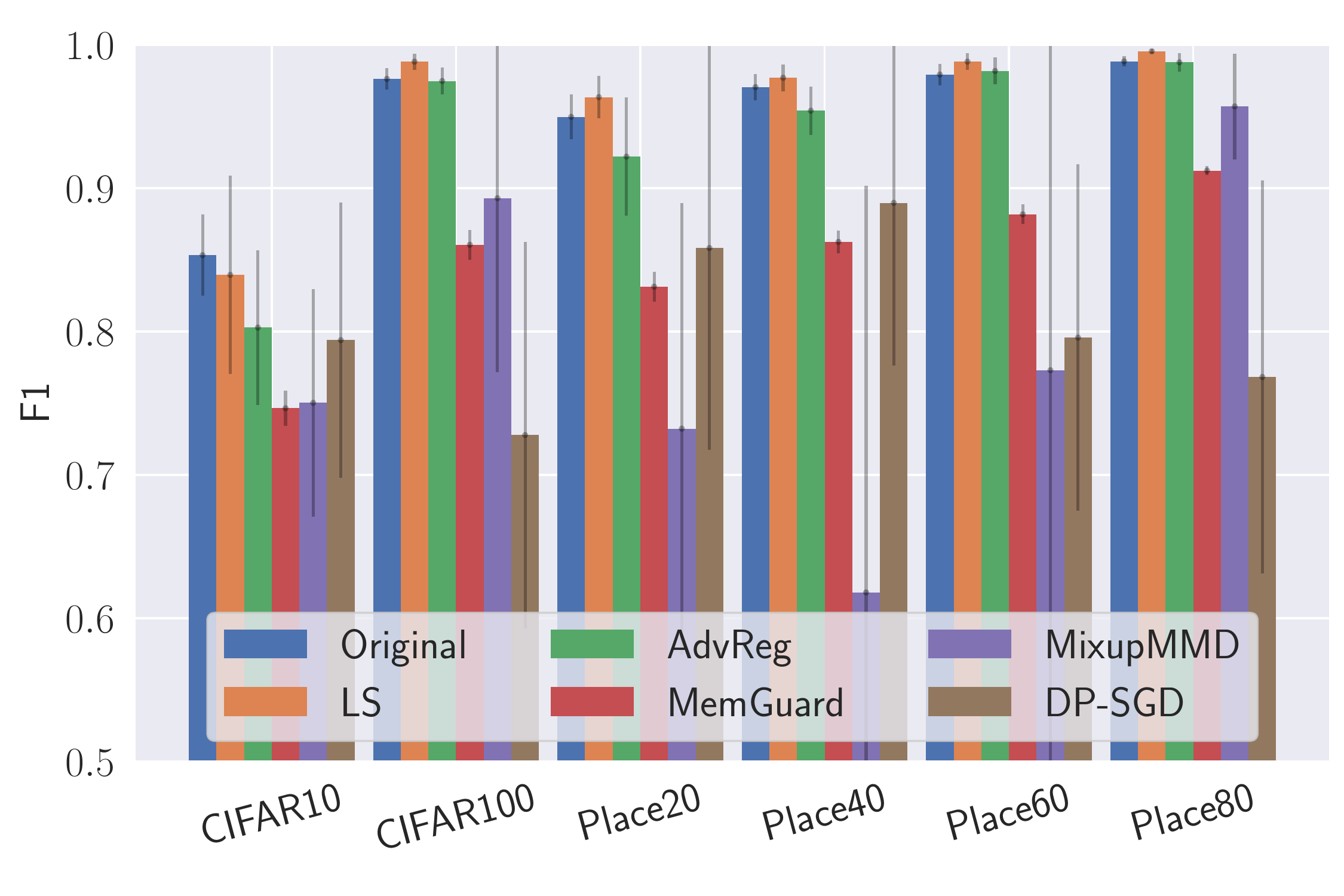}
\caption{F1}
\label{figure:RQ1_defense_performance_f1_no}
\end{subfigure}
\begin{subfigure}{0.48\columnwidth}
\includegraphics[width=\columnwidth]{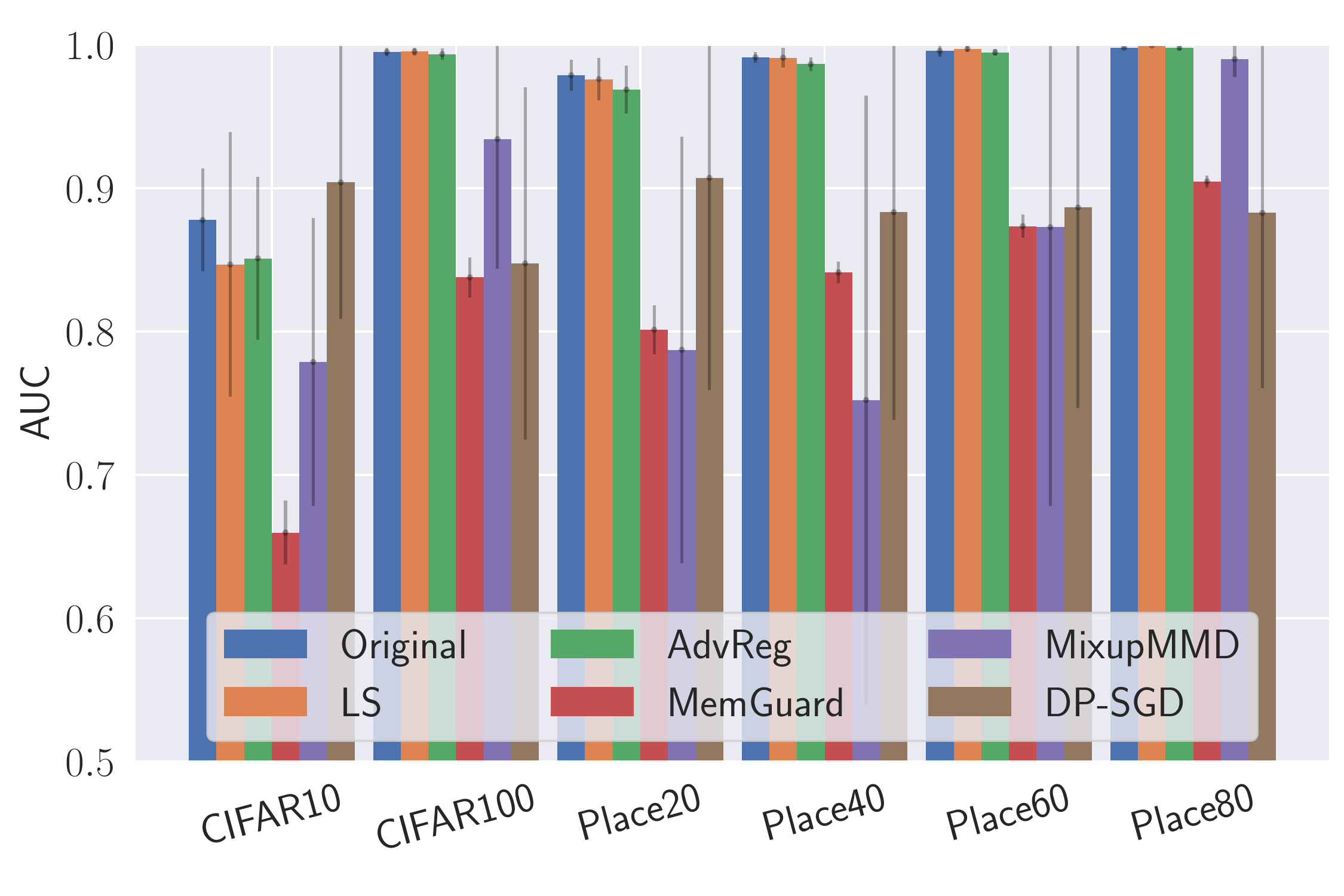}
\caption{AUC}
\label{figure:RQ1_defense_performance_auc_no}
\end{subfigure}
\caption{The performance of the best membership inference attacks against the original models and models defended by different methods on 6 different datasets. The x-axis represents different datasets. The y-axis represents the membership inference attack's Precision, Recall, F1, and AUC, respectively. Note that we average the attack performance under different model architectures and show the standard deviations as well.}
\label{figure:RQ1_defense_performance_prf1auc_no}
\end{figure*}

We summarize the attack and defense performance with different evaluation metrics in \Cref{figure:RQ1_attack_performance_prf1auc_no} and \Cref{figure:RQ1_defense_performance_prf1auc_no}, respectively.
Concretely, we evaluate the Precision, Recall, F1, and AUC.

\subsection{Relationship between Distance and Attack Performance}
\label{appendix:relation_distance_attack}

\begin{figure*}[htbp]
\centering
\begin{subfigure}{0.45\columnwidth}
\includegraphics[width=\columnwidth]{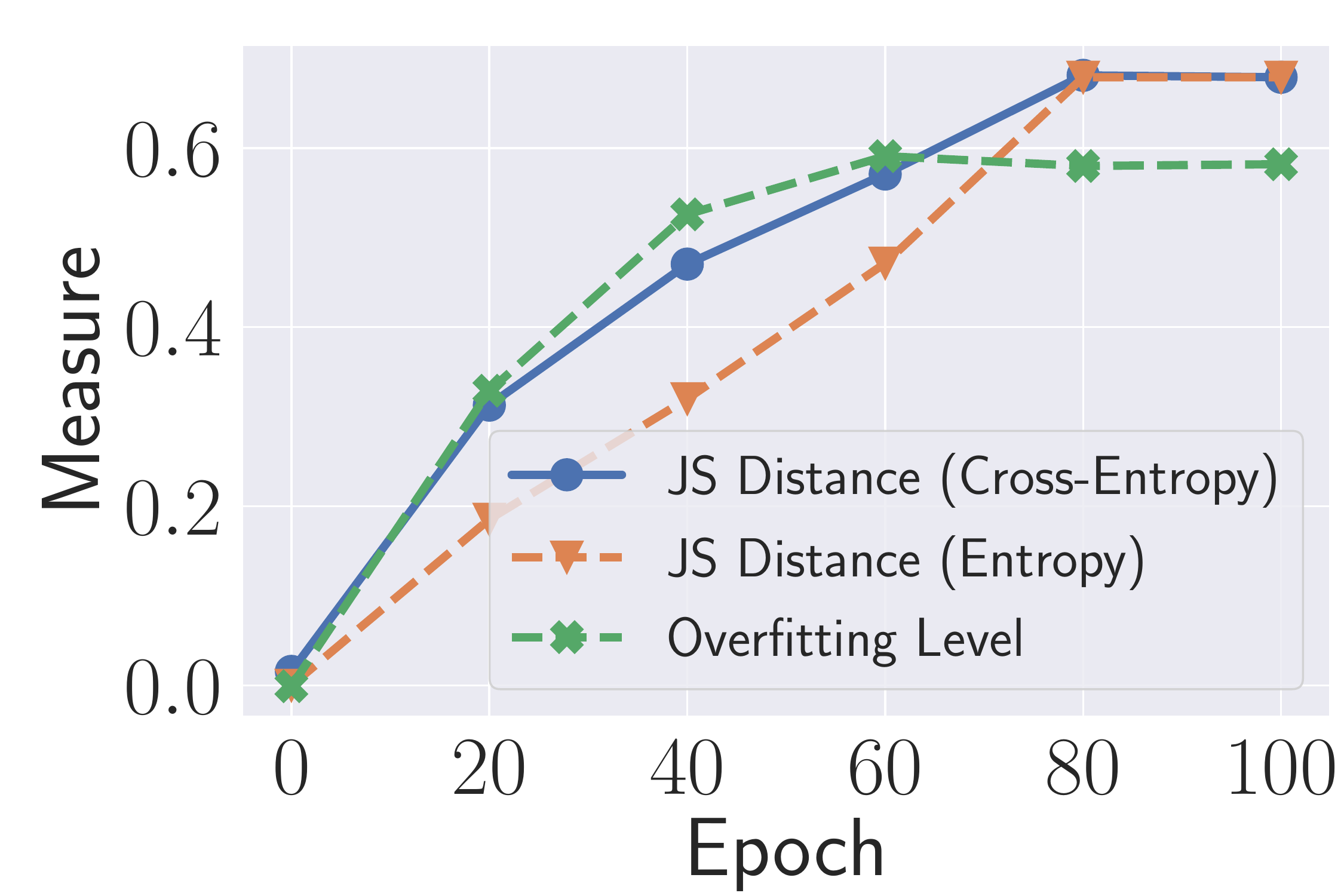}
\caption{Place20}
\label{figure:RQ3_target_mobilenetv2_Place20_no_aug}
\end{subfigure}
\begin{subfigure}{0.45\columnwidth}
\includegraphics[width=\columnwidth]{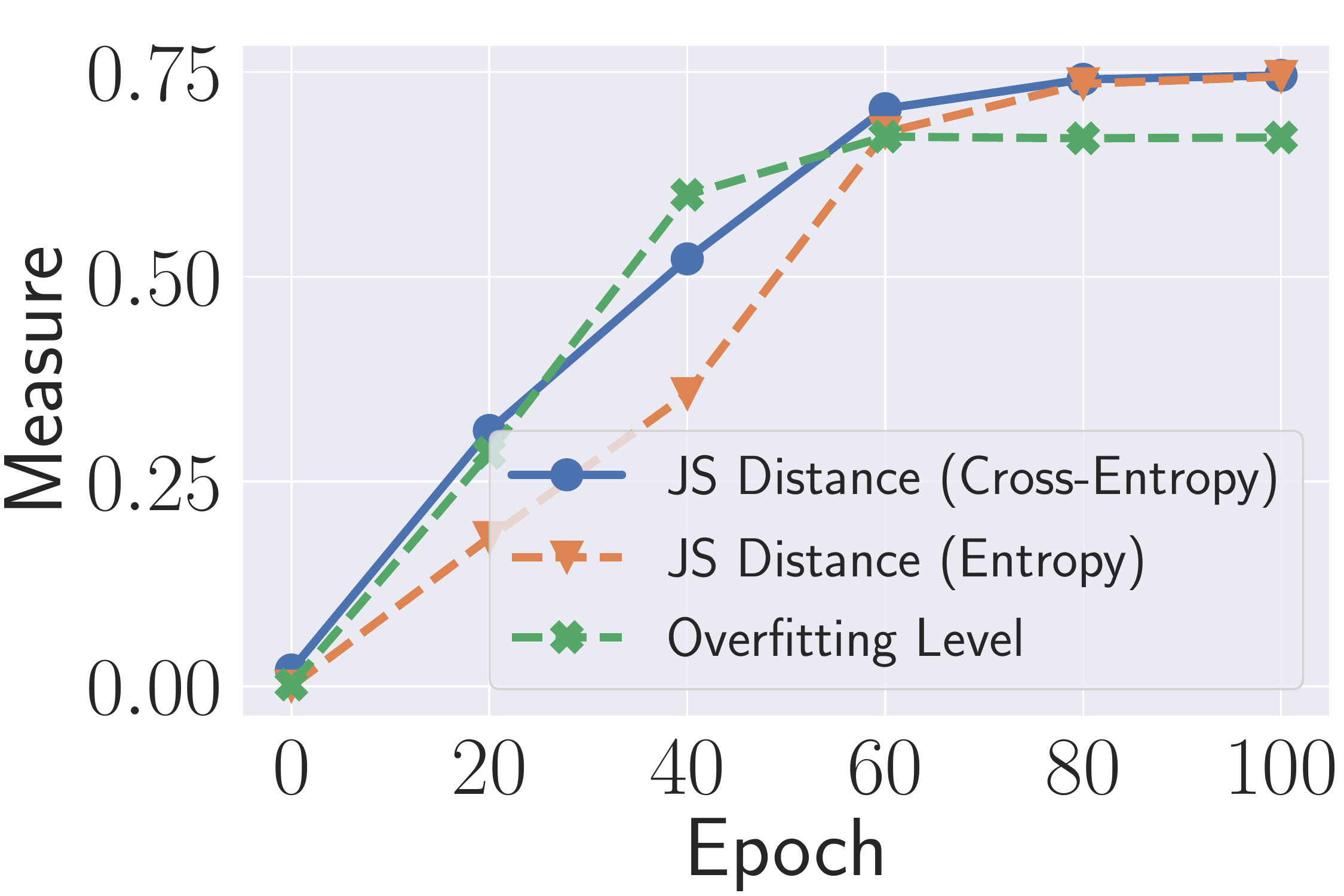}
\caption{Place40}
\label{figure:RQ3_target_mobilenetv2_Place40_no_aug}
\end{subfigure}
\centering
\begin{subfigure}{0.45\columnwidth}
\includegraphics[width=\columnwidth]{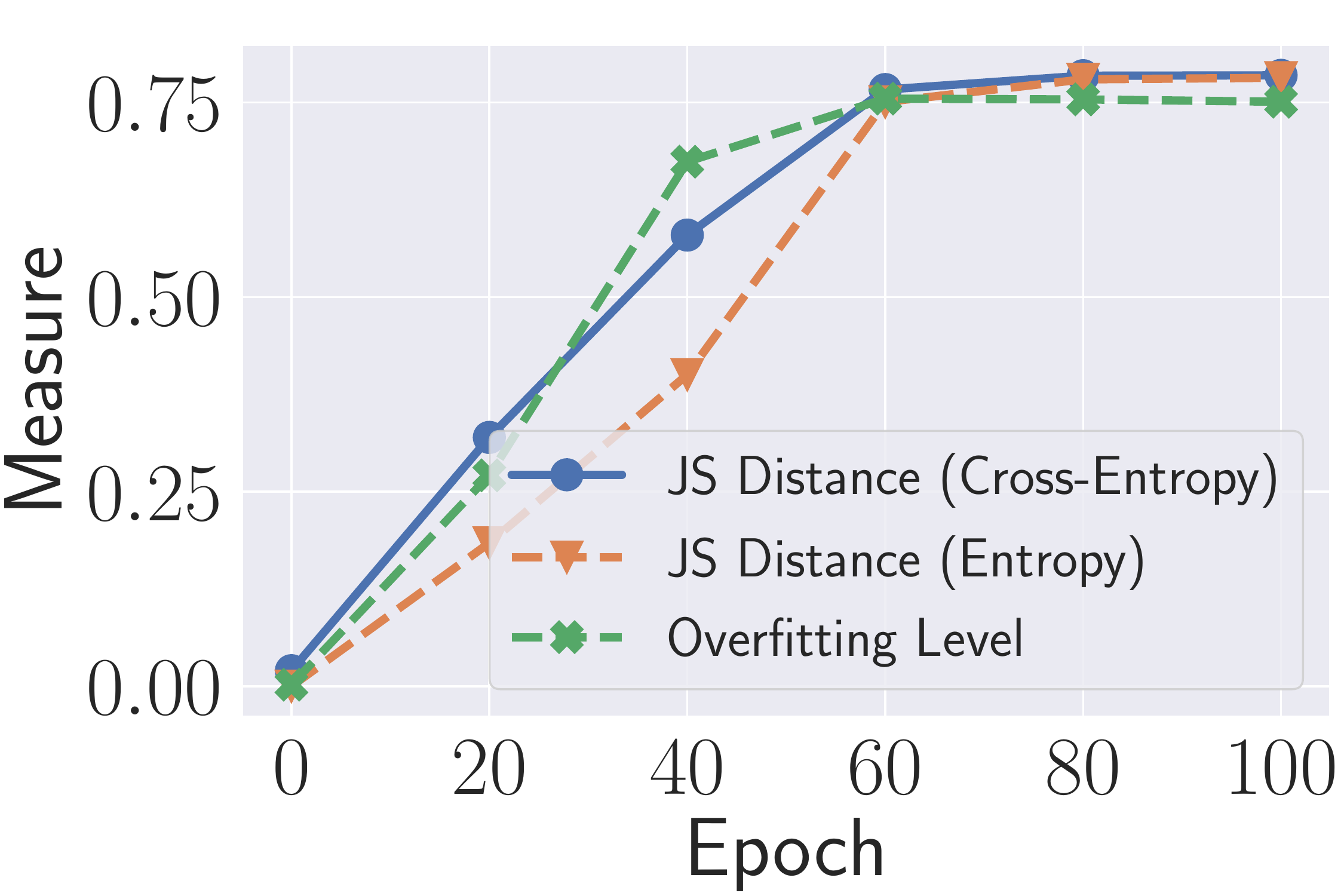}
\caption{Place60}
\label{figure:RQ3_target_mobilenetv2_Place60_no_aug}
\end{subfigure}
\begin{subfigure}{0.45\columnwidth}
\includegraphics[width=\columnwidth]{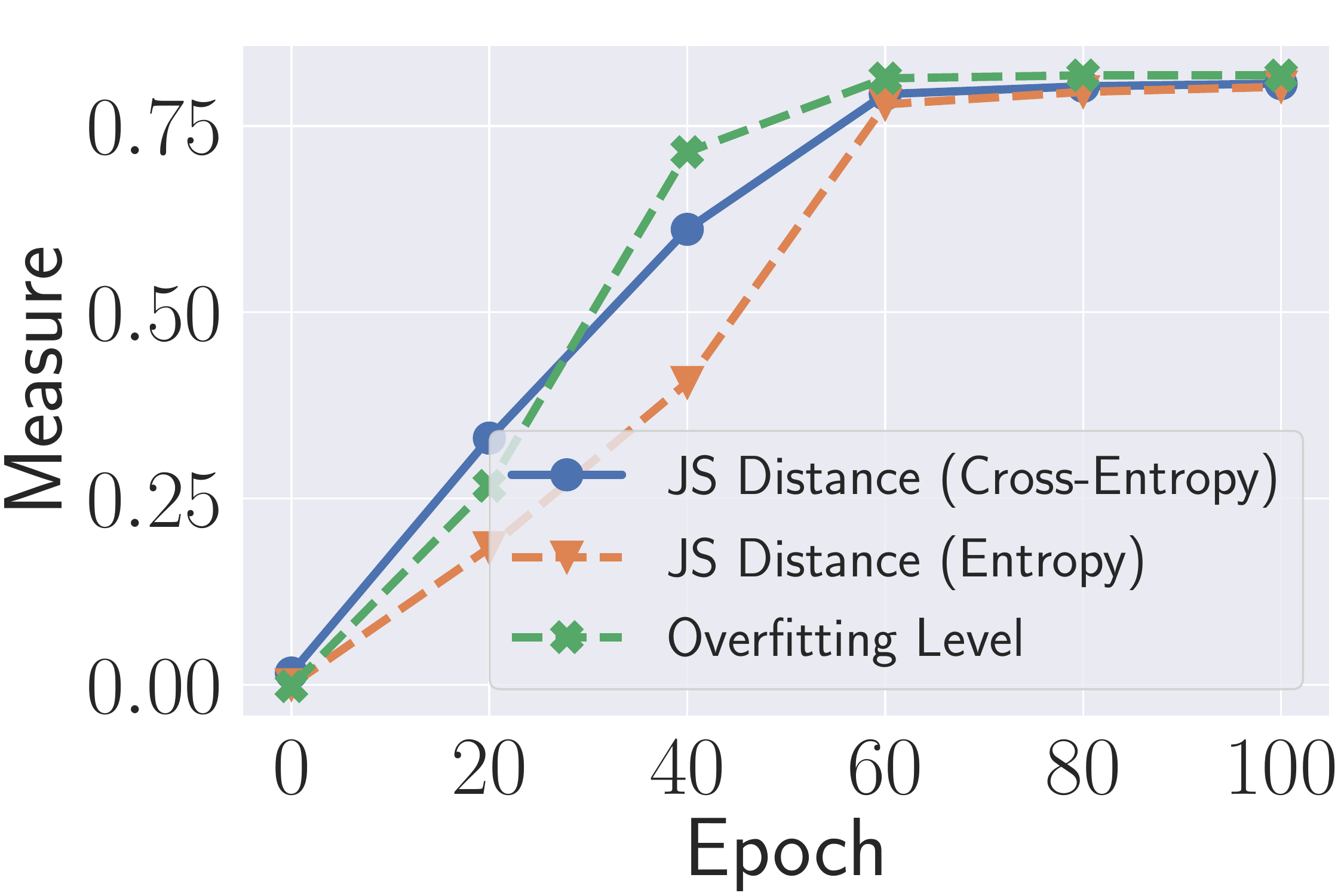}
\caption{Place80}
\label{figure:RQ3_target_mobilenetv2_Place80_no_aug}
\end{subfigure}
\begin{subfigure}{0.45\columnwidth}
\includegraphics[width=\columnwidth]{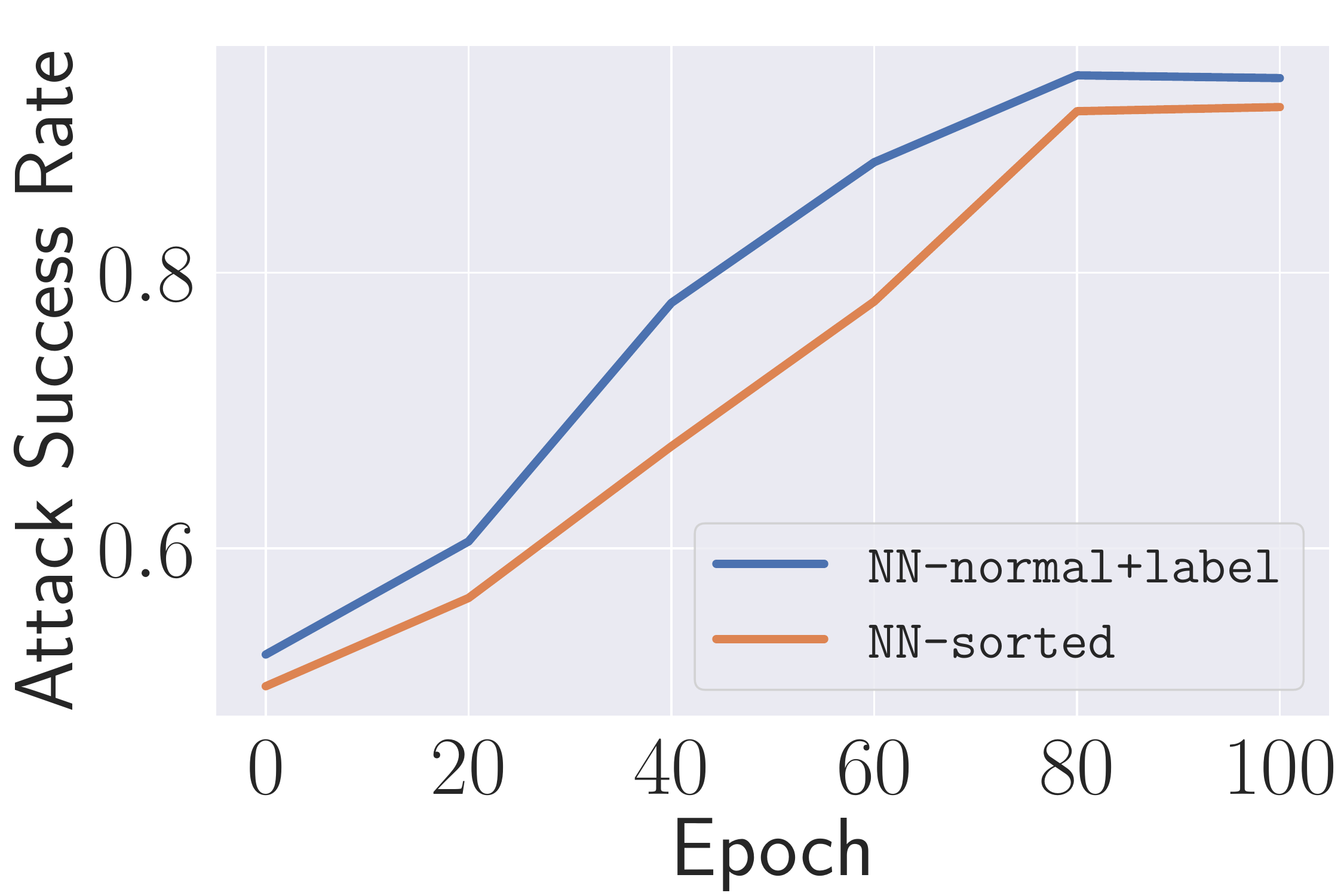}
\caption{Place20}
\label{figure:RQ3_attack_mobilenetv2_Place20_no_aug}
\end{subfigure}
\begin{subfigure}{0.45\columnwidth}
\includegraphics[width=\columnwidth]{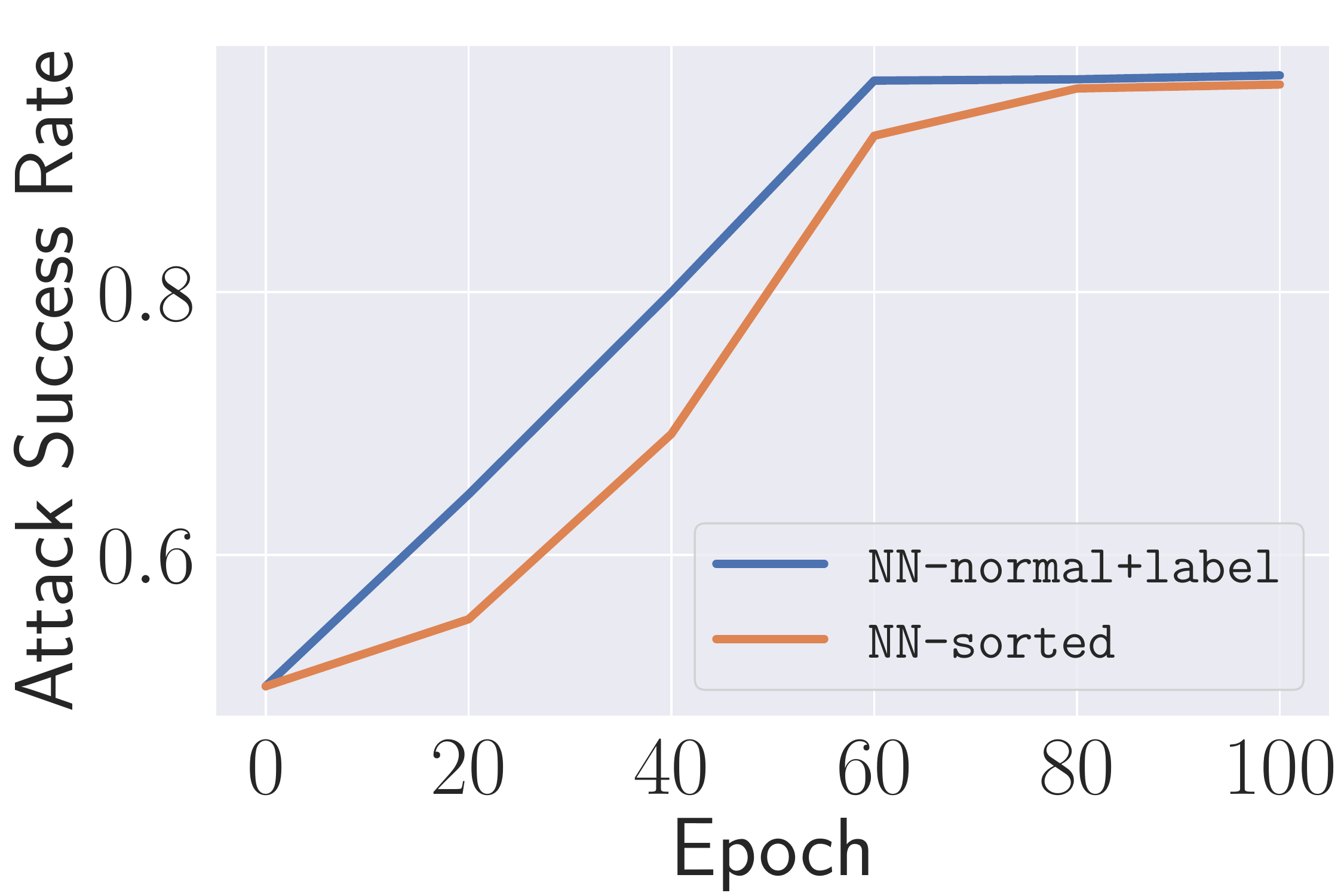}
\caption{Place40}
\label{figure:RQ3_attack_mobilenetv2_Place40_no_aug}
\end{subfigure}
\begin{subfigure}{0.45\columnwidth}
\includegraphics[width=\columnwidth]{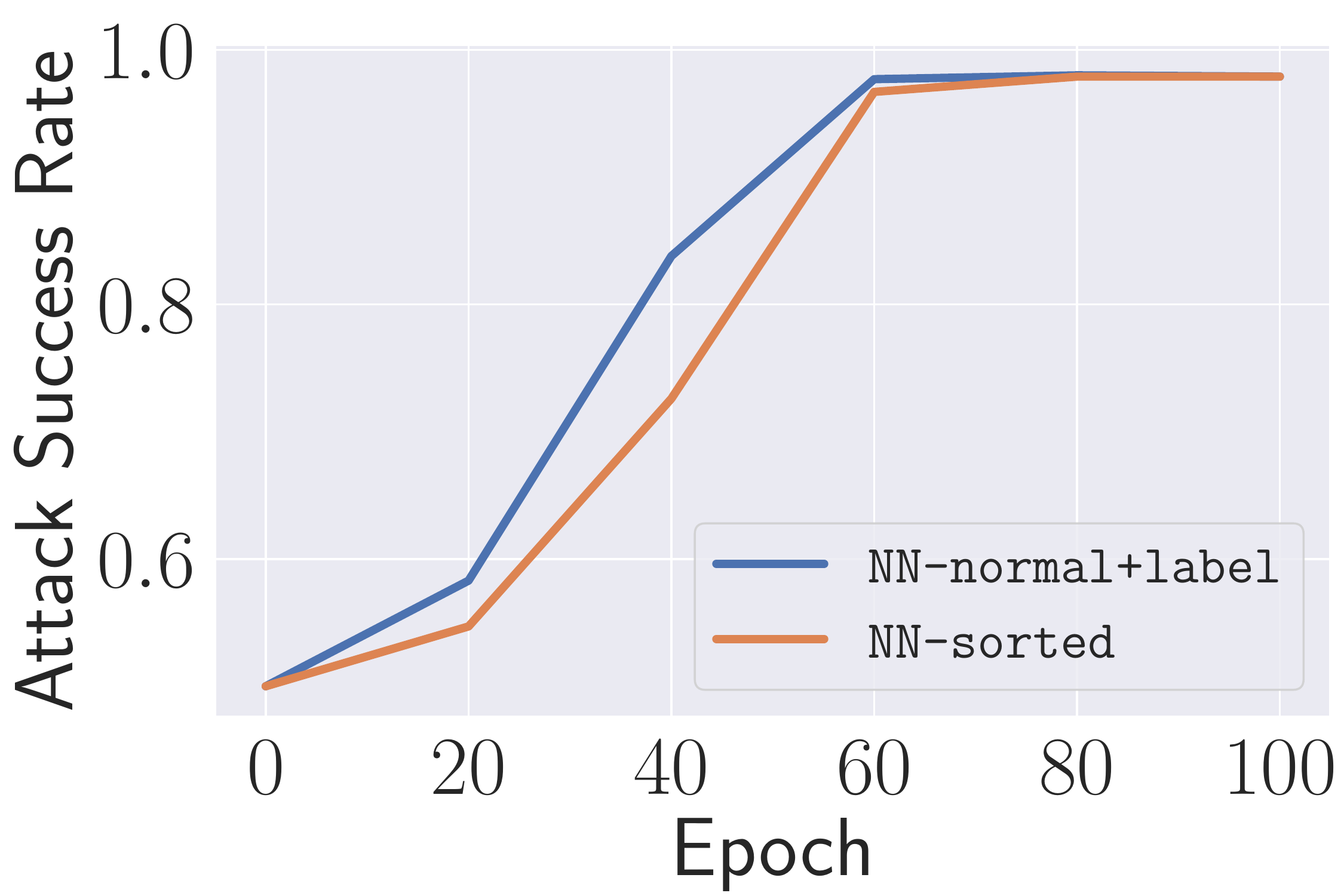}
\caption{Place60}
\label{figure:RQ3_attack_mobilenetv2_Place60_no_aug}
\end{subfigure}
\begin{subfigure}{0.45\columnwidth}
\includegraphics[width=\columnwidth]{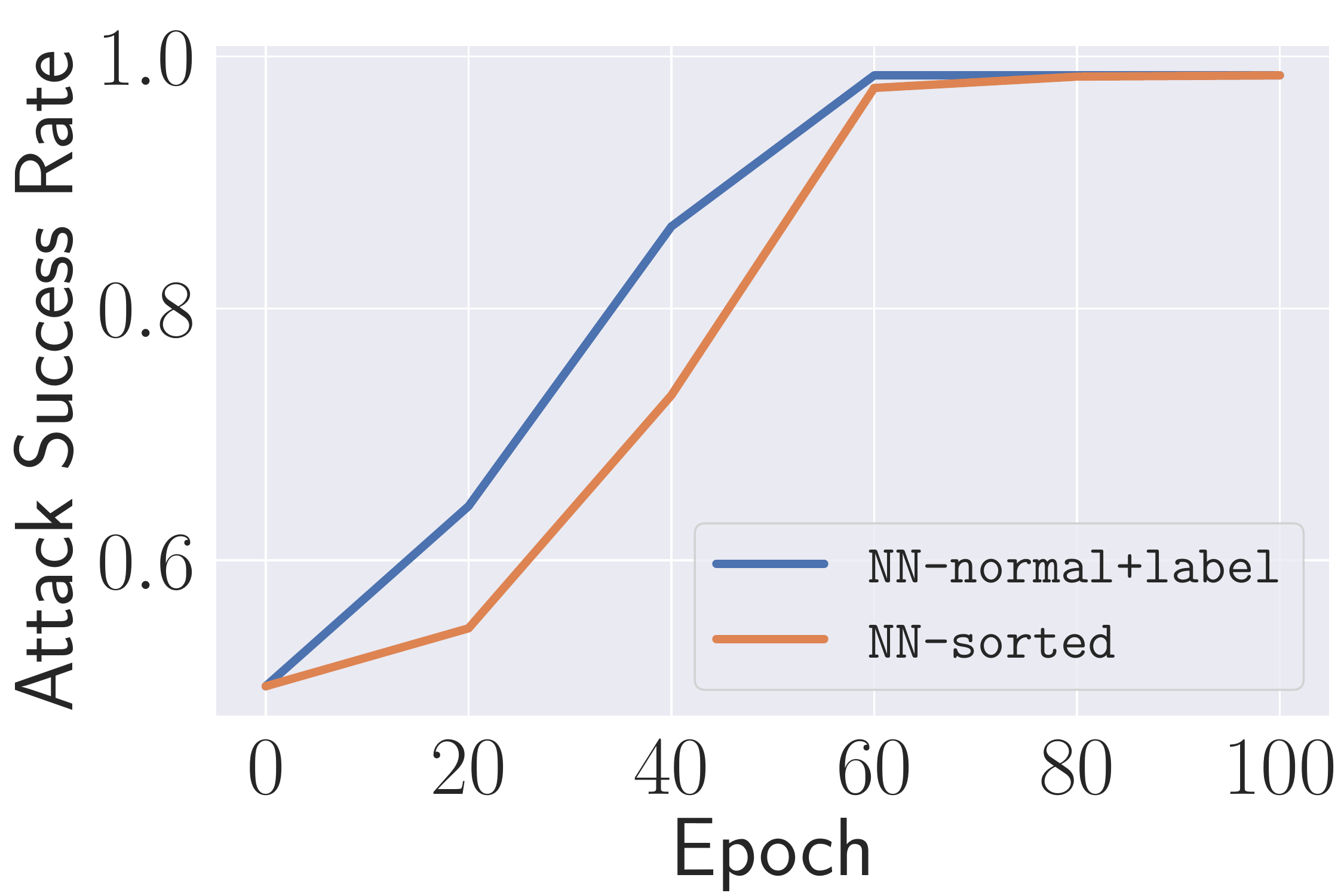}
\caption{Place80}
\label{figure:RQ3_attack_mobilenetv2_Place80_no_aug}
\end{subfigure}
\caption{The distance and attack performance against MobileNetV2 on Place20, Place40, Place60, and Place80 under different numbers of epochs for model training. The first (second) row denotes the distance (attack performance).}
\label{figure:RQ3_mobilenetv2_Place_series_no_aug} 
\end{figure*}

\begin{figure*}[!ht]
\centering
\begin{subfigure}{0.45\columnwidth}
\includegraphics[width=\columnwidth]{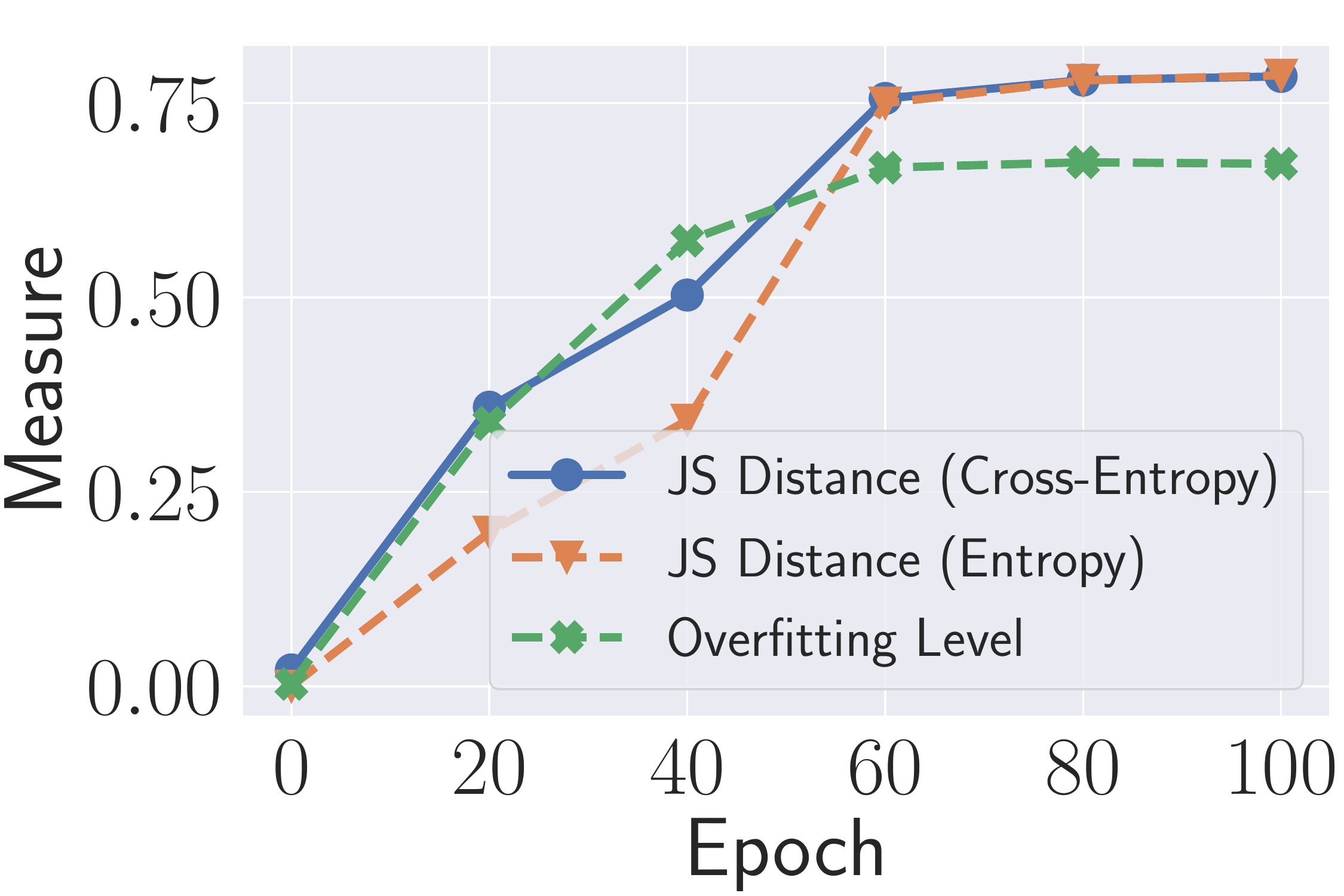}
\caption{Distance}
\label{figure:RQ3_target_mobilenetv2_CIFAR100_no_aug}
\end{subfigure}
\begin{subfigure}{0.45\columnwidth}
\includegraphics[width=\columnwidth]{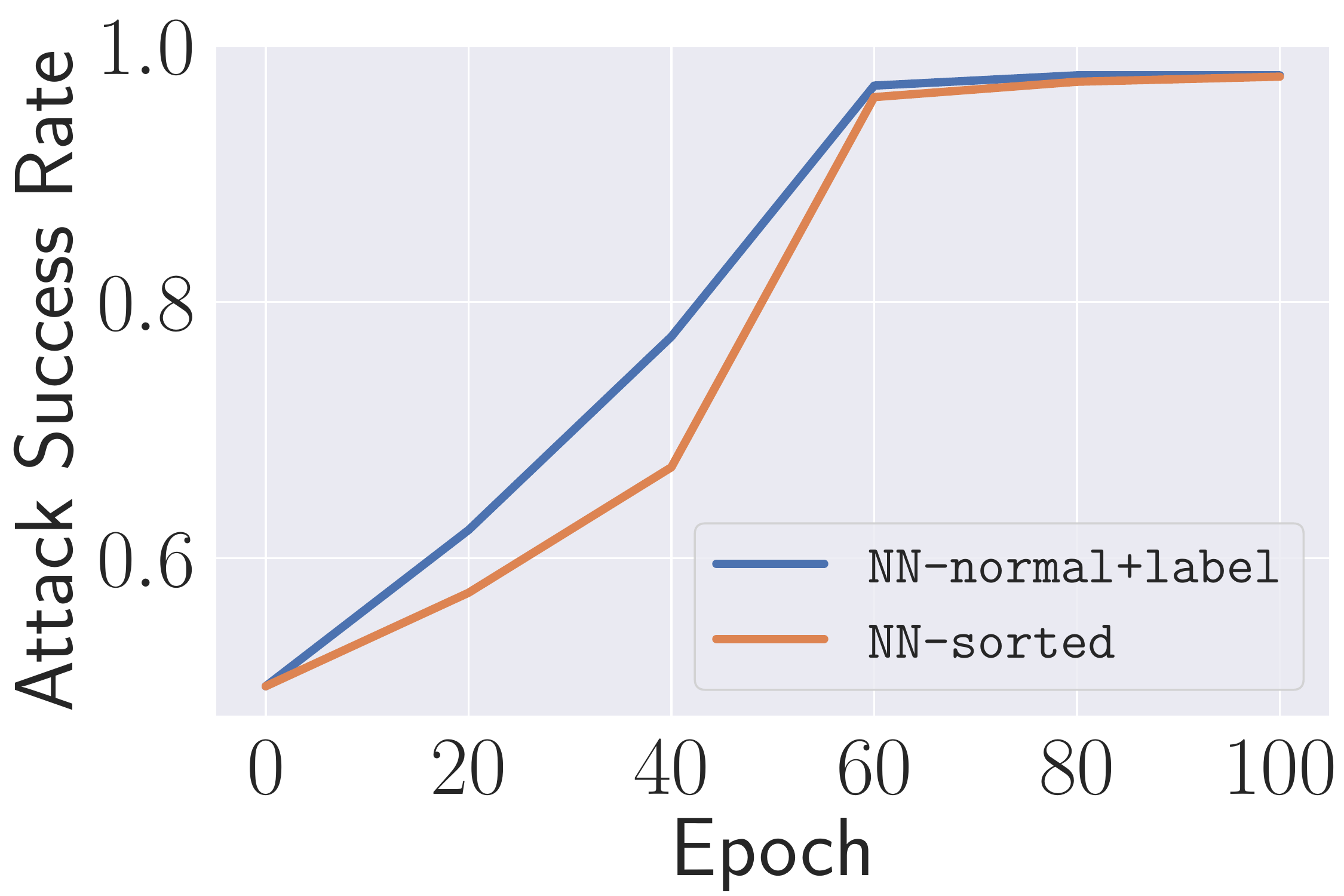}
\caption{Attack Performance}
\label{figure:RQ3_attack_mobilenetv2_CIFAR100_no_aug}
\end{subfigure}
\caption{The distance and attack performance against MobileNetV2 on CIFAR100 under different numbers of epochs for model training.}
\label{figure:RQ3_mobilenetv2_CIFAR100_no_aug} 
\end{figure*}

The distance and attack performance against MobileNetV2 on Place60, Place80, and CIFAR100 are shown in~\Cref{figure:RQ3_mobilenetv2_Place_series_no_aug} and ~\Cref{figure:RQ3_mobilenetv2_CIFAR100_no_aug}, respectively.

\subsection{Defense Effectiveness and Utility under Models Trained with Data Augmentation}
\label{appendix:subsection_defense}

\begin{figure*}[!ht]
\centering
\begin{subfigure}{0.48\columnwidth}
\includegraphics[width=\columnwidth]{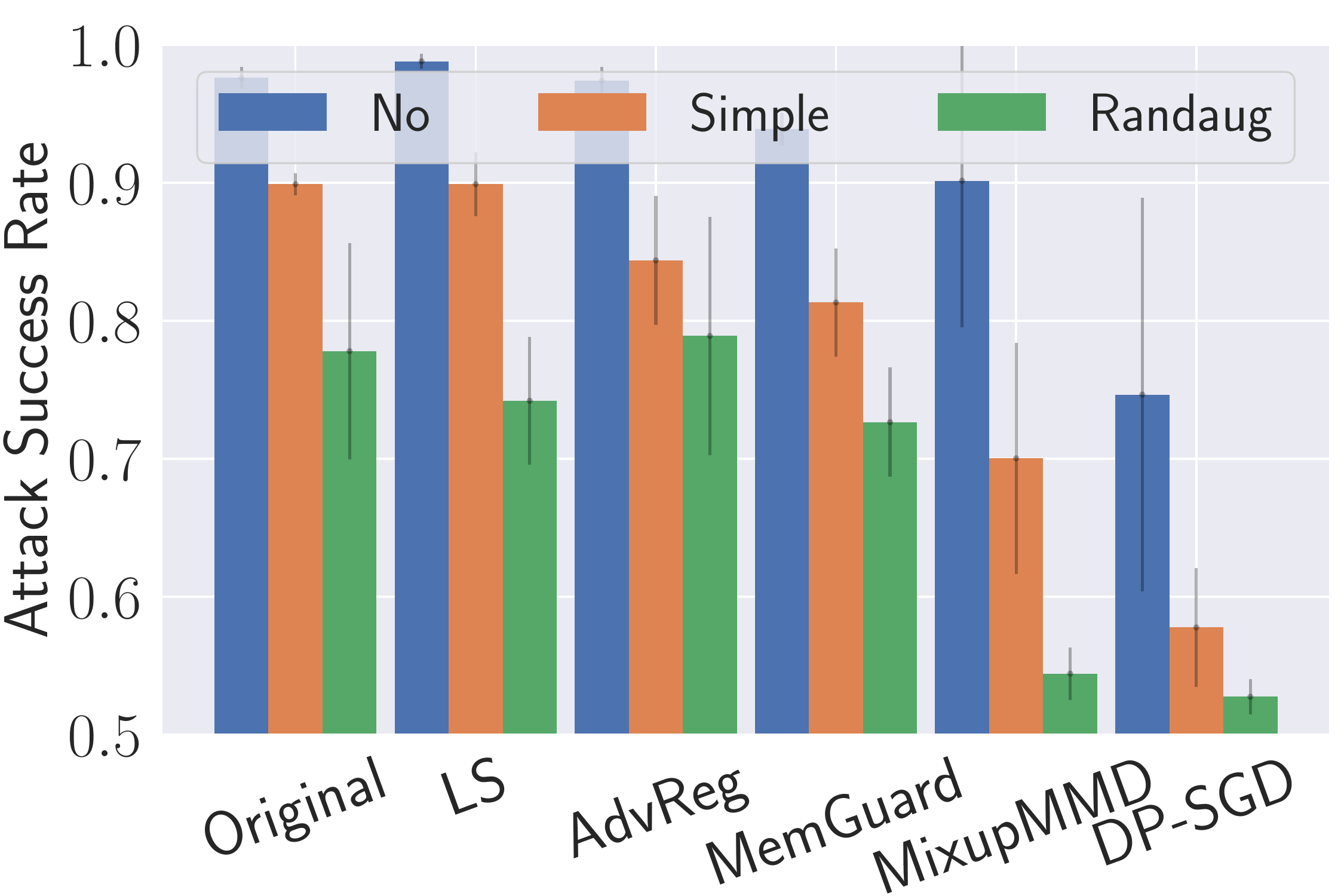}
\caption{Defense Effectiveness}
\label{figure:RQ4_CIFAR100_attack}
\end{subfigure}
\begin{subfigure}{0.48\columnwidth}
\includegraphics[width=\columnwidth]{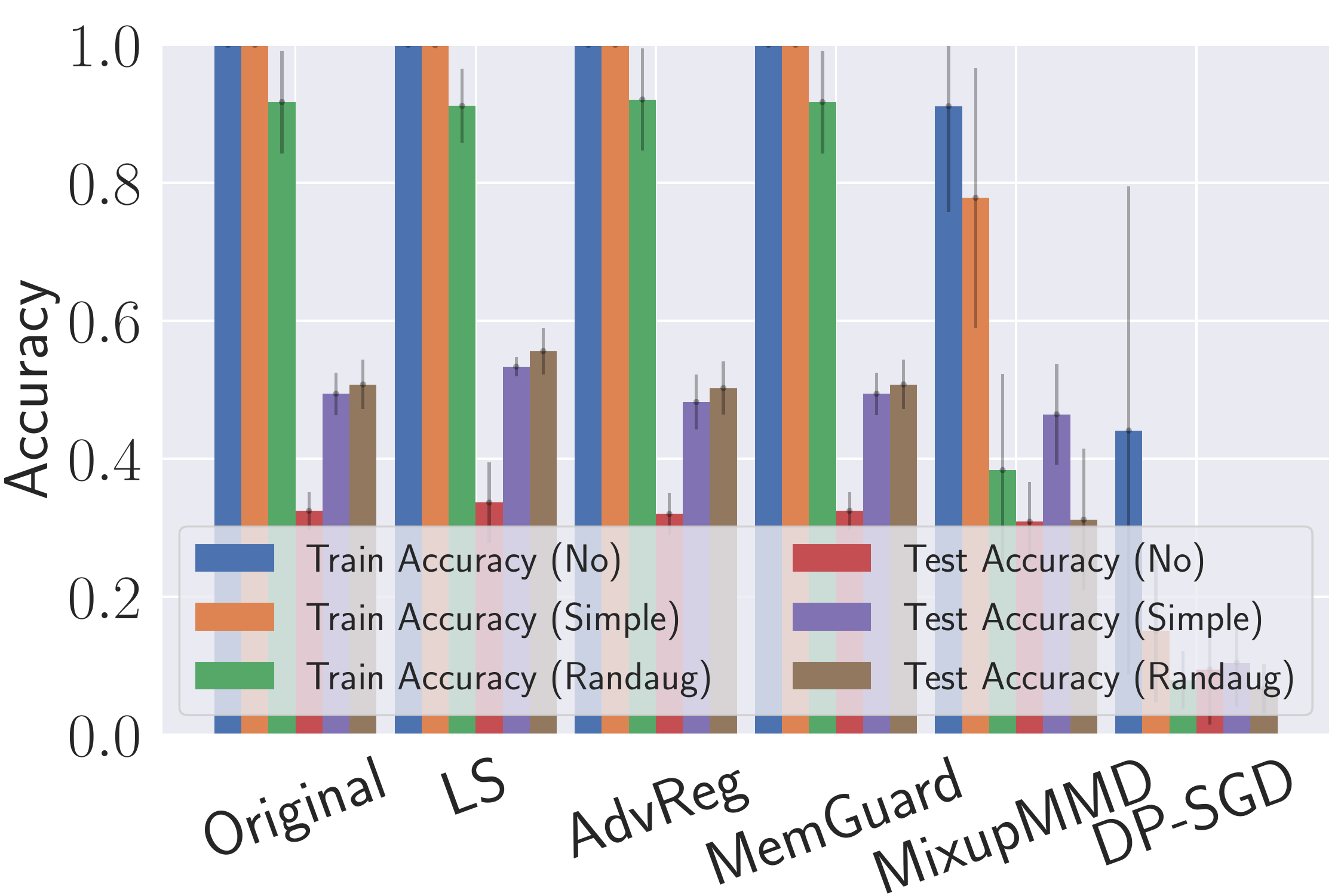}
\caption{Utility}
\label{figure:RQ4_CIFAR100_target}
\end{subfigure}
\caption{Defense effectiveness against the best attacks and utility in the original classification tasks on CIFAR100. Note that we average the performance for different model architectures and report the standard deviations as well.}
\label{figure:RQ4_defense_effectiveness_and_utility_CIFAR100} 
\end{figure*}

\begin{figure*}[!ht]
\centering
\begin{subfigure}{0.45\columnwidth}
\includegraphics[width=\columnwidth]{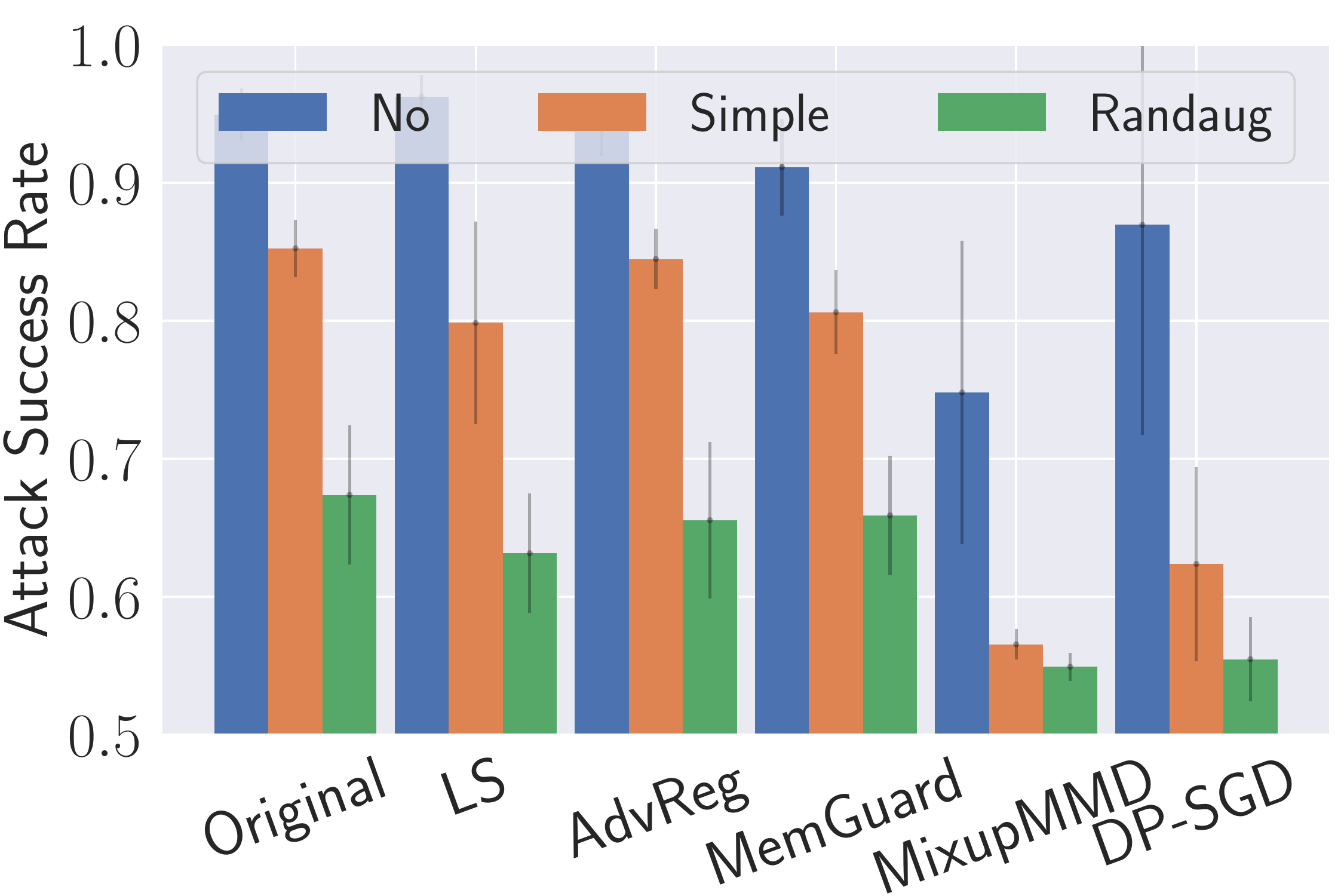}
\caption{Place20}
\label{figure:RQ4_Place20_attack}
\end{subfigure}
\begin{subfigure}{0.45\columnwidth}
\includegraphics[width=\columnwidth]{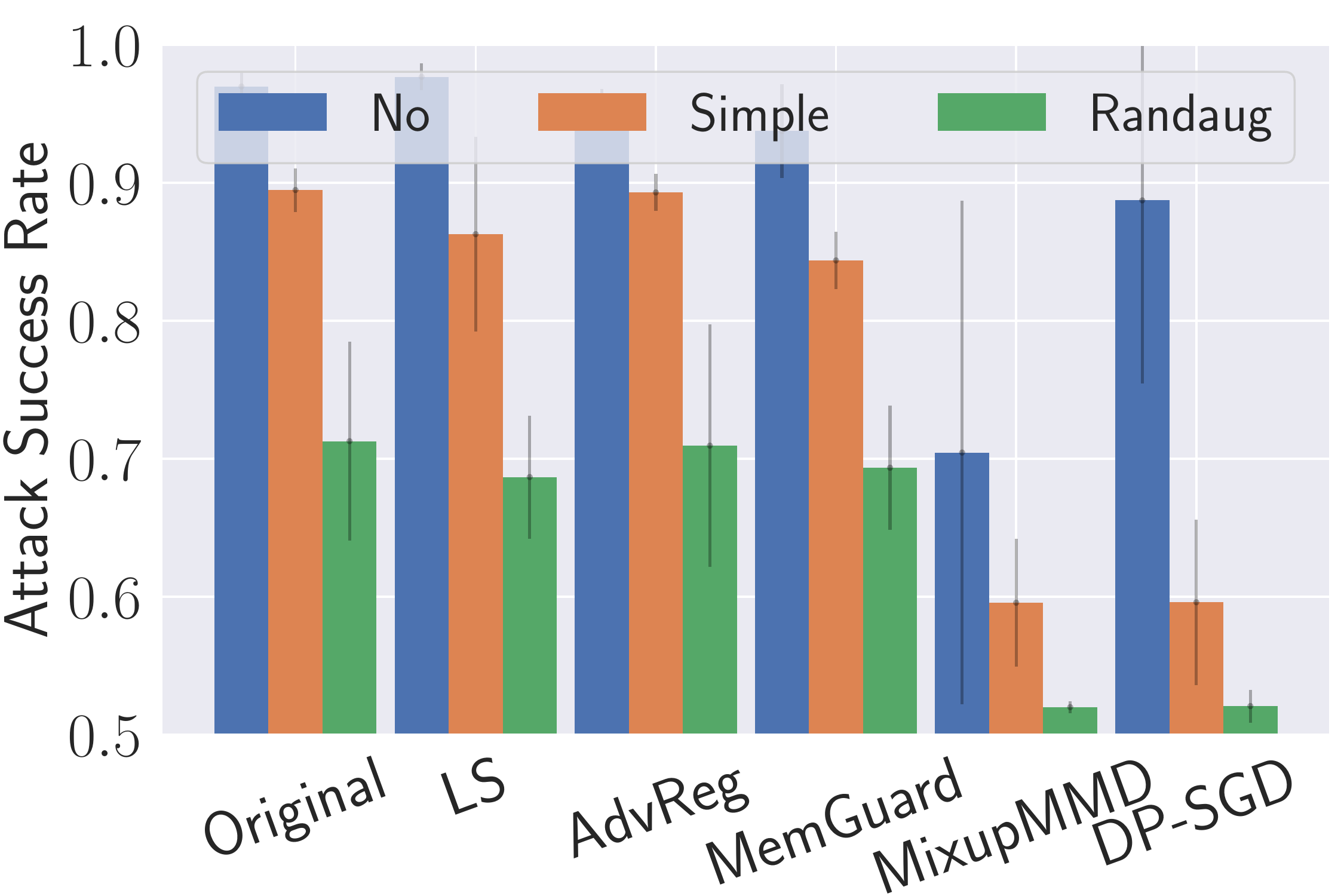}
\caption{Place40}
\label{figure:RQ4_Place40_attack}
\end{subfigure}
\begin{subfigure}{0.45\columnwidth}
\includegraphics[width=\columnwidth]{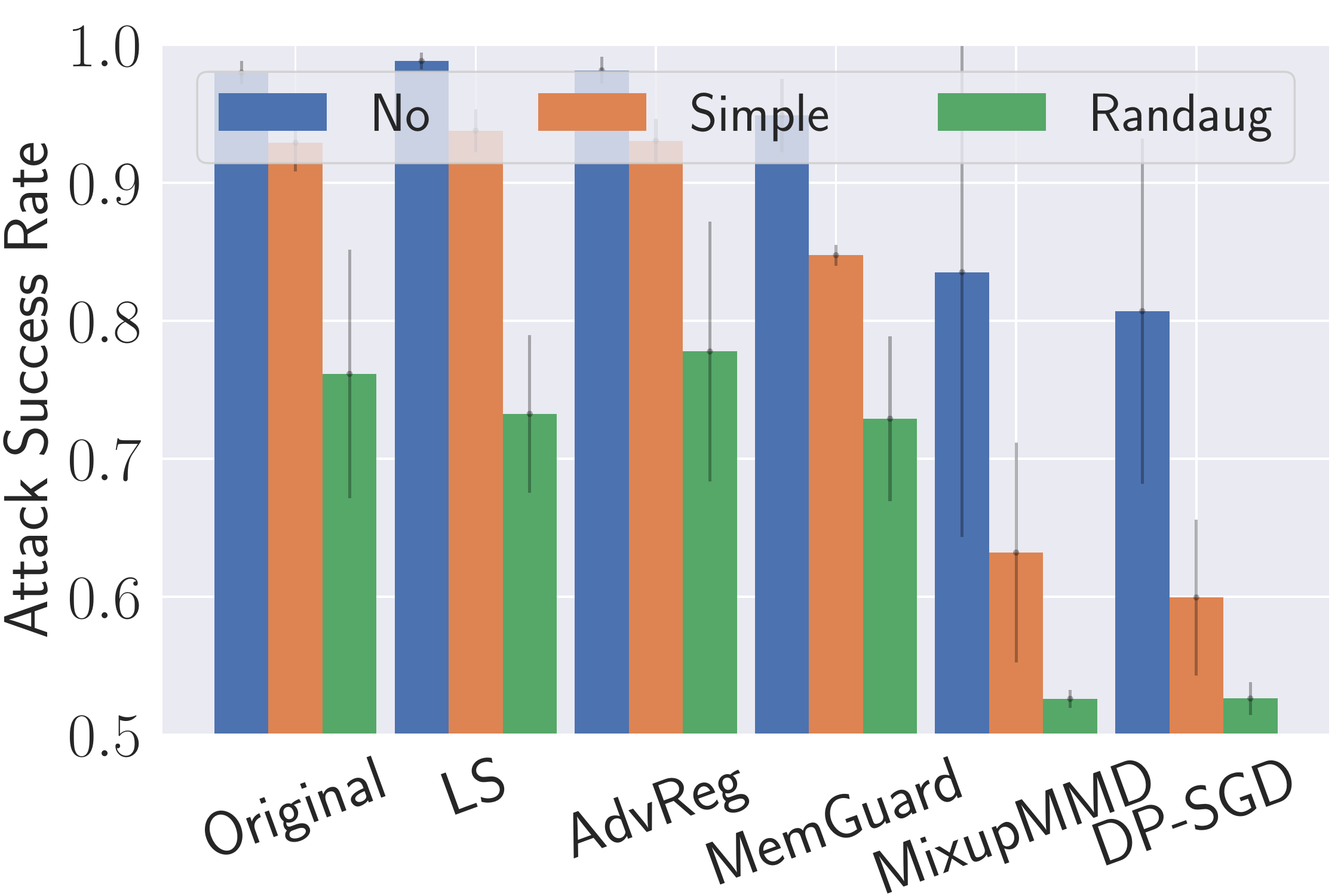}
\caption{Place60}
\label{figure:RQ4_Place60_attack}
\end{subfigure}
\begin{subfigure}{0.45\columnwidth}
\includegraphics[width=\columnwidth]{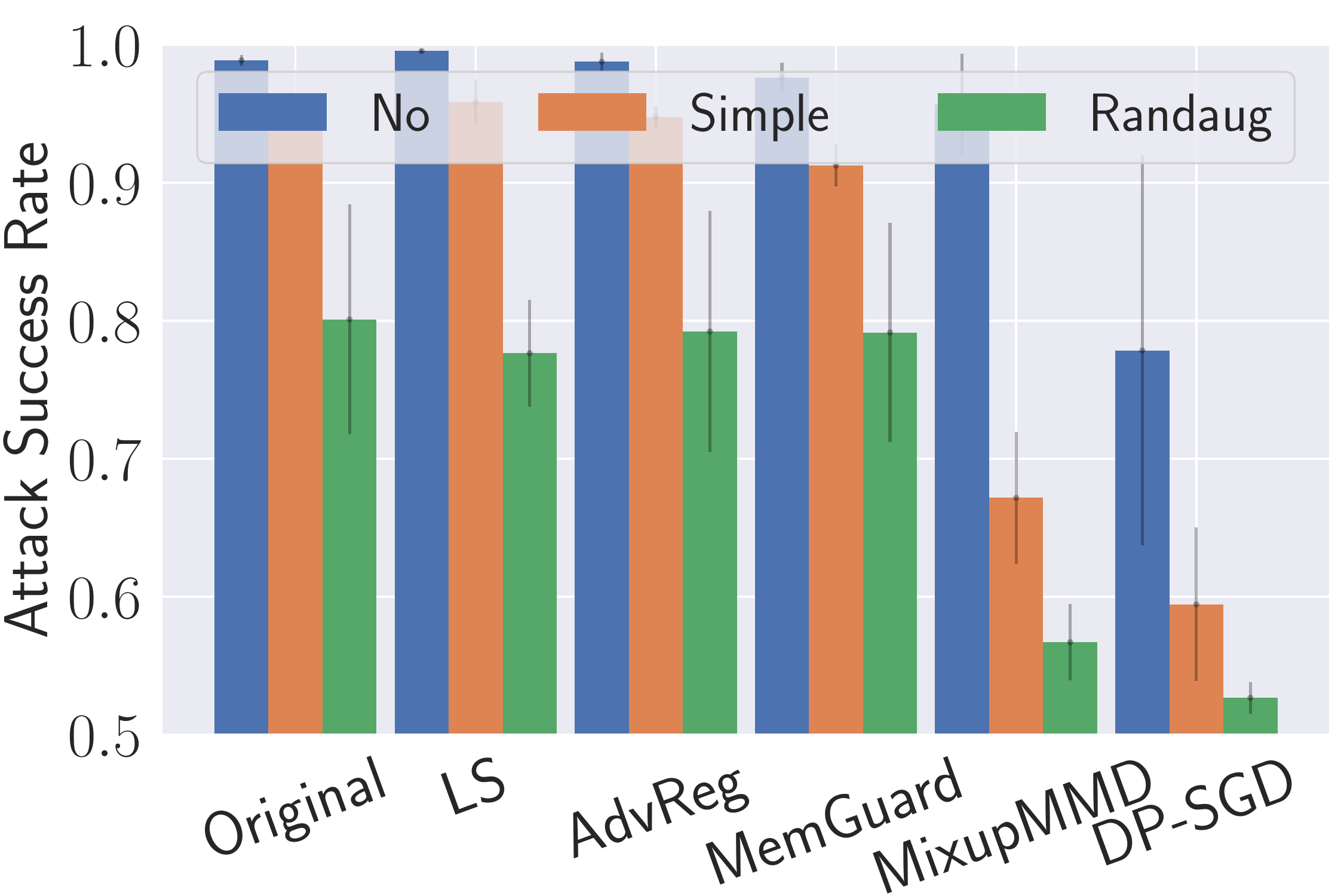}
\caption{Place80}
\label{figure:RQ4_Place80_attack}
\end{subfigure}
\begin{subfigure}{0.45\columnwidth}
\includegraphics[width=\columnwidth]{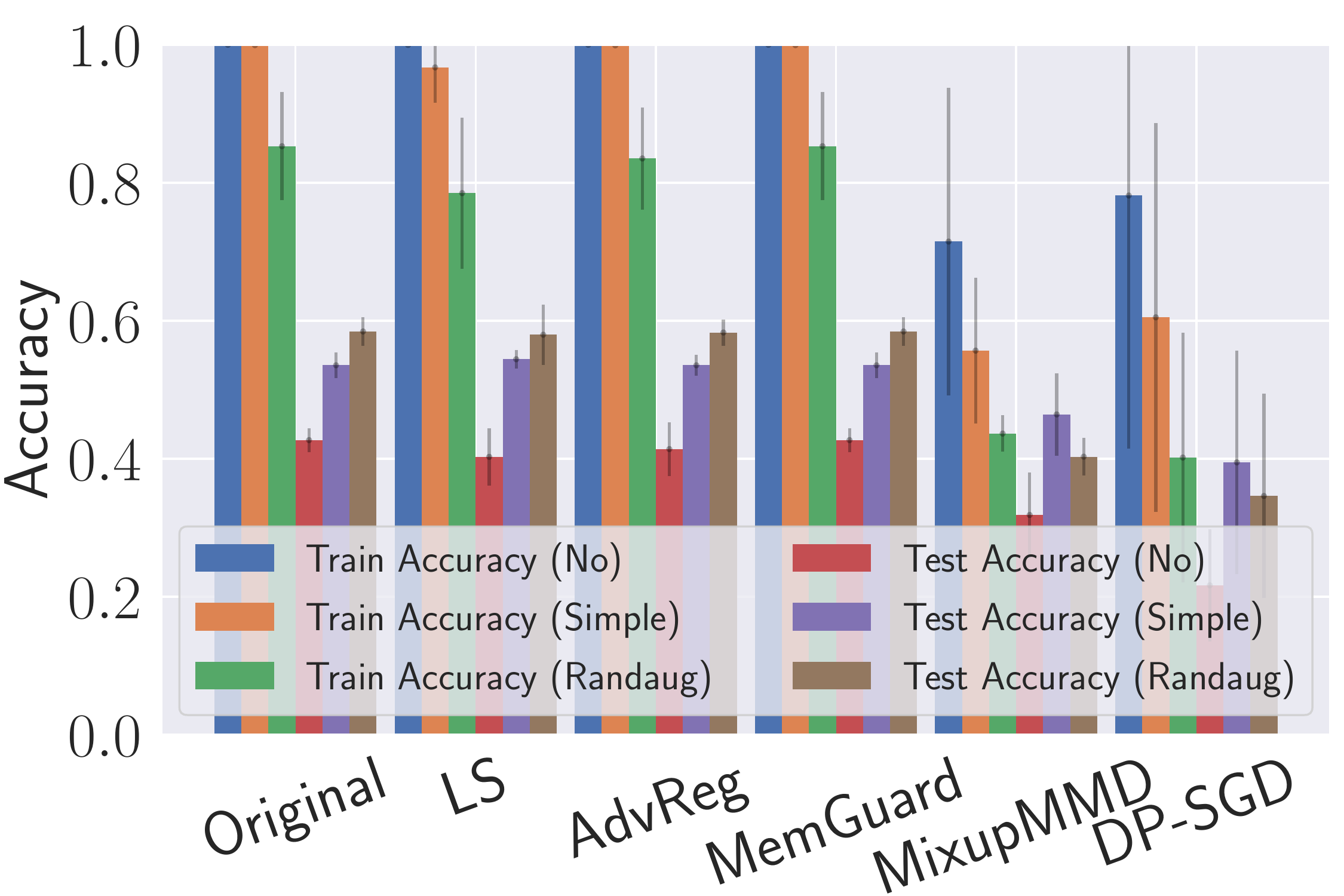}
\caption{Place20}
\label{figure:RQ4_Place20_target}
\end{subfigure}
\begin{subfigure}{0.45\columnwidth}
\includegraphics[width=\columnwidth]{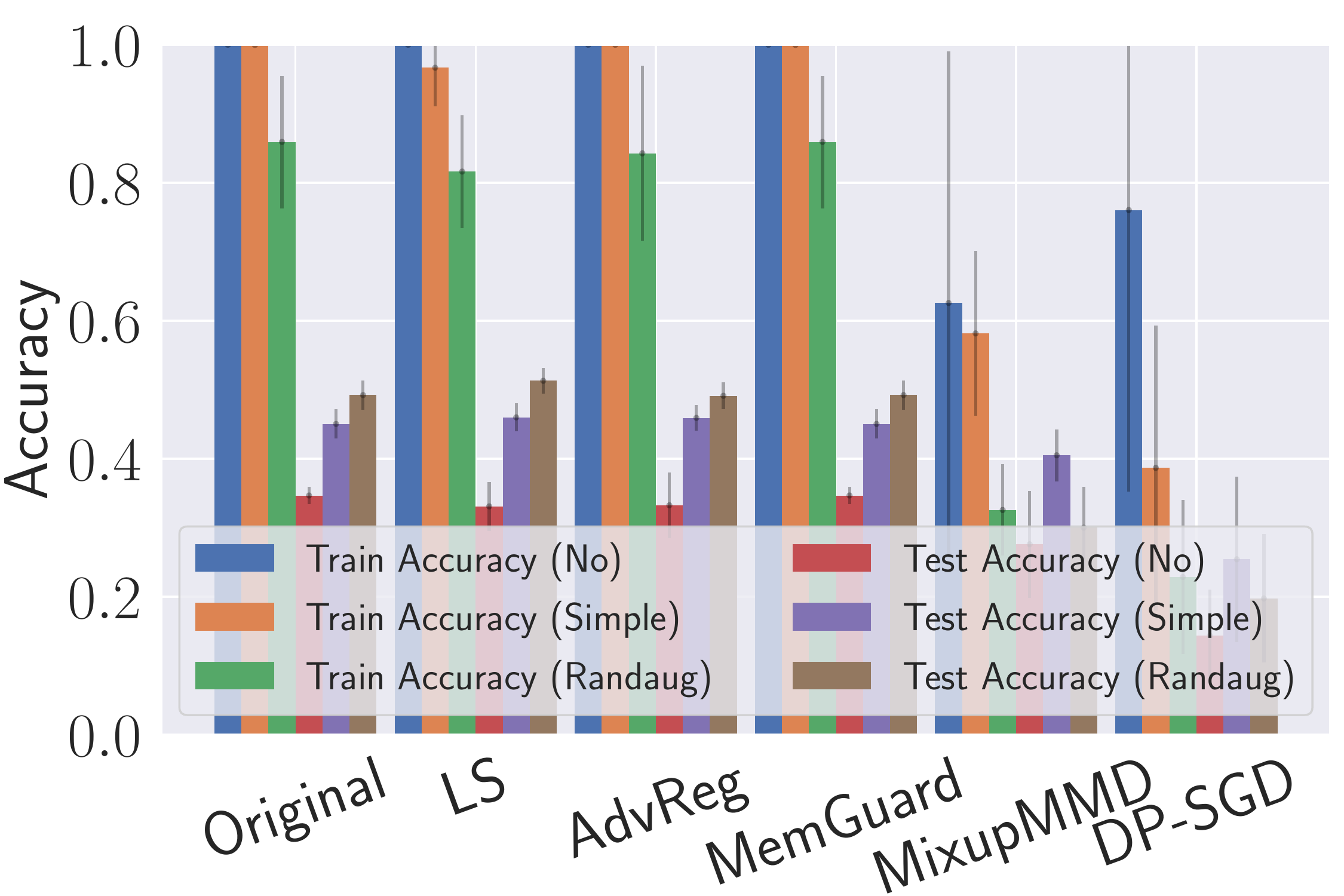}
\caption{Place40}
\label{figure:RQ4_Place40_target}
\end{subfigure}
\begin{subfigure}{0.45\columnwidth}
\includegraphics[width=\columnwidth]{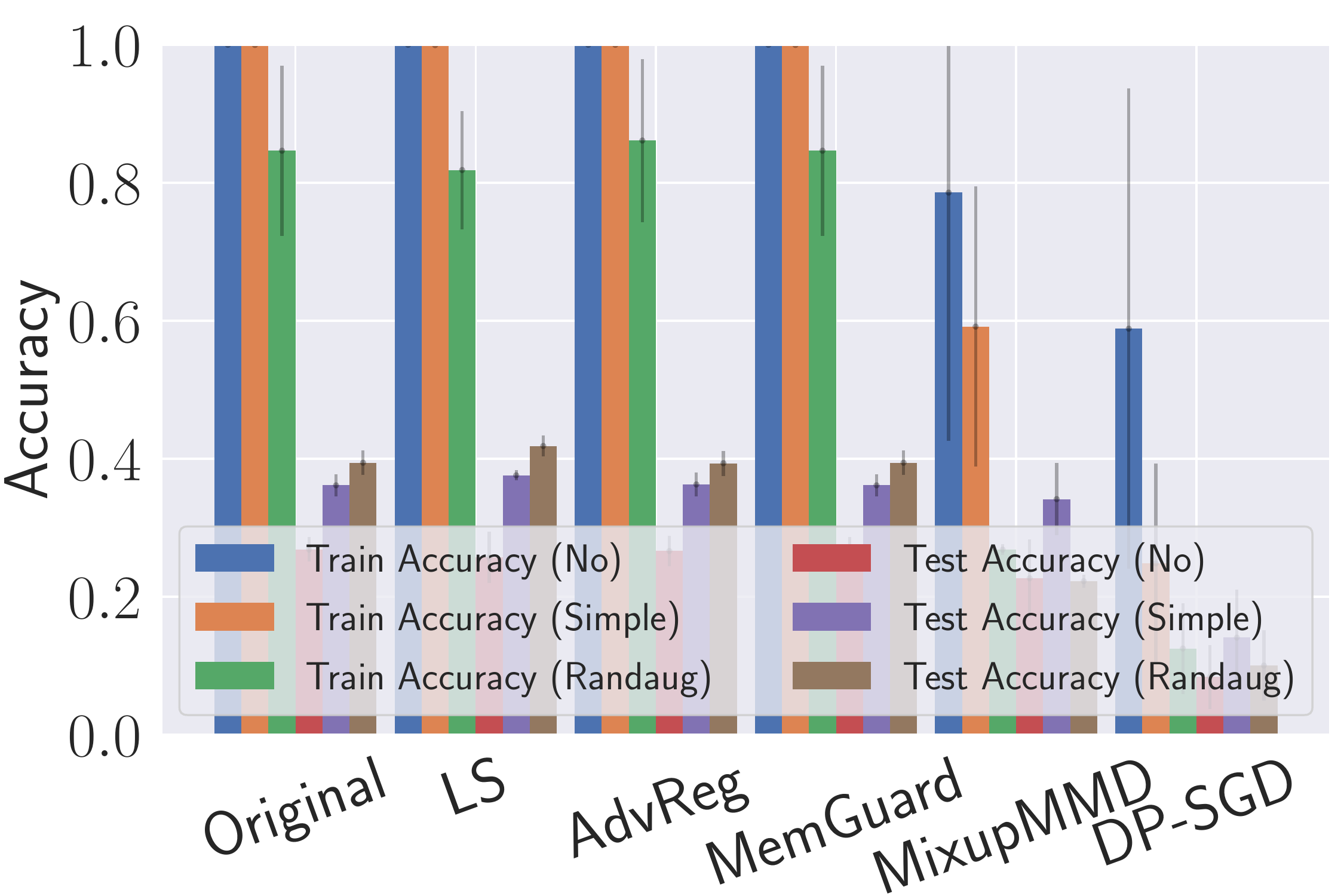}
\caption{Place60}
\label{figure:RQ4_Place60_target}
\end{subfigure}
\begin{subfigure}{0.45\columnwidth}
\includegraphics[width=\columnwidth]{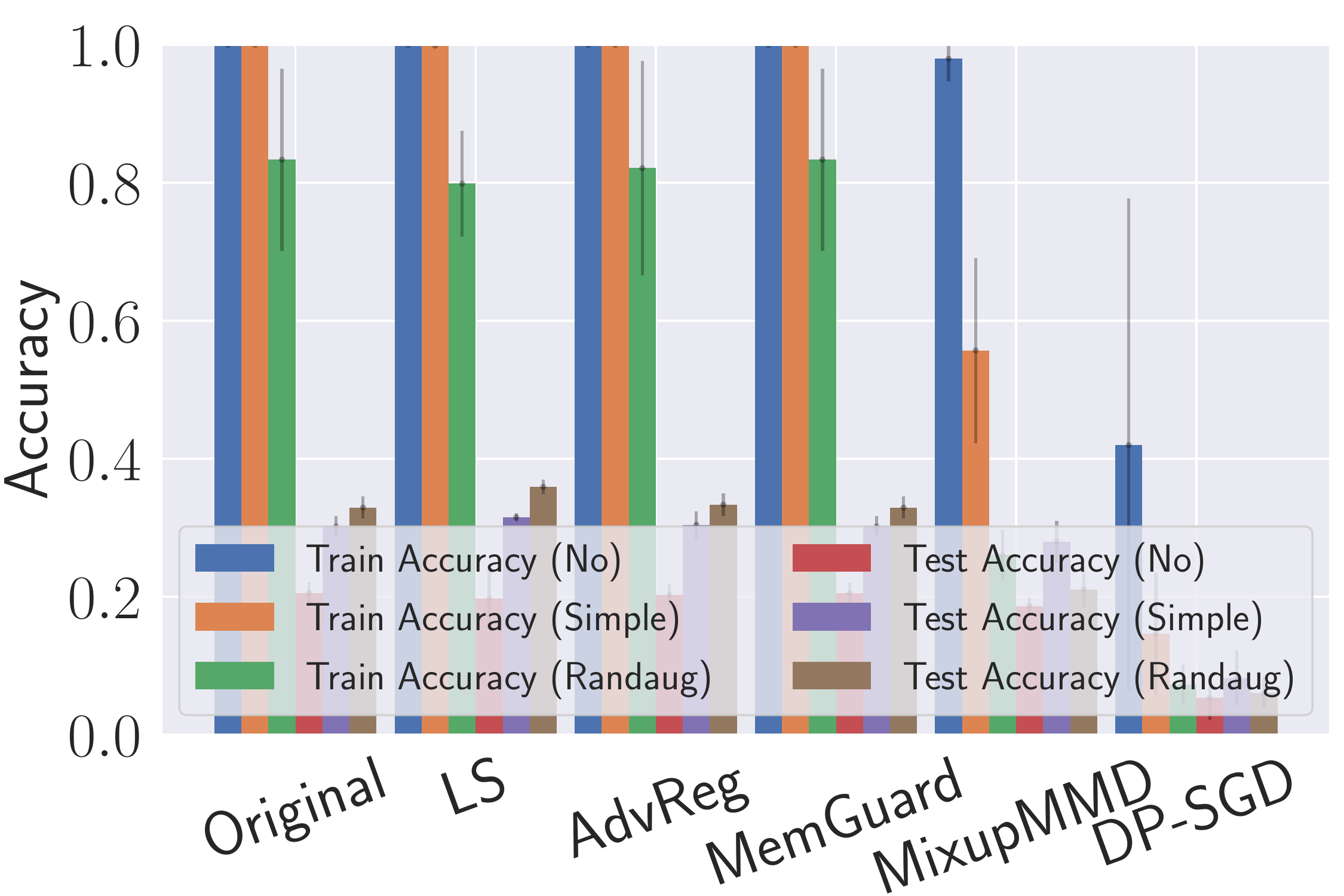}
\caption{Place80}
\label{figure:RQ4_Place80_target}
\end{subfigure}
\caption{Defense effectiveness against the best attacks and utility in the original classification tasks on Place20, Place40, Place60, and Place80. Note that we average the performance for different model architectures and report the standard deviations as well. The first (second) row denotes the defense effectiveness (utility).}
\label{figure:RQ4_defense_effectiveness_and_utility_Place_series} 
\end{figure*}

The defense effectiveness against the best attacks and utility in the original classification tasks on Place20, Place40, Place60, Place80, and CIFAR100 are shown in~\Cref{figure:RQ4_defense_effectiveness_and_utility_Place_series} and~\Cref{figure:RQ4_defense_effectiveness_and_utility_CIFAR100}, respectively.

\begin{table}[!ht]
\centering
\caption{The class-wised overfitting level and the attack success rate of membership inference attack (\texttt{NN-normal+label}) when the target model is VGG11 trained on CIFAR10. Note that the table is sorted by the attack accuracy in descending order.}
\begin{tabular}{c|c c}
\toprule
Class Index & Overfitting Level & Attack Success Rate \\
\midrule
3 & 0.409 & 0.845 \\
4 & 0.352 & 0.832 \\
5 & 0.432 & 0.812 \\
2 & 0.350 & 0.806 \\
0 & 0.238 & 0.794 \\
7 & 0.256 & 0.772 \\
6 & 0.280 & 0.766 \\
1 & 0.164 & 0.757 \\
9 & 0.153 & 0.746 \\
8 & 0.169 & 0.734 \\
\bottomrule
\end{tabular}
\label{table:class_wised_performance}
\end{table}

\begin{figure*}[!ht]
\centering
\begin{subfigure}{0.45\columnwidth}
\includegraphics[width=\columnwidth]{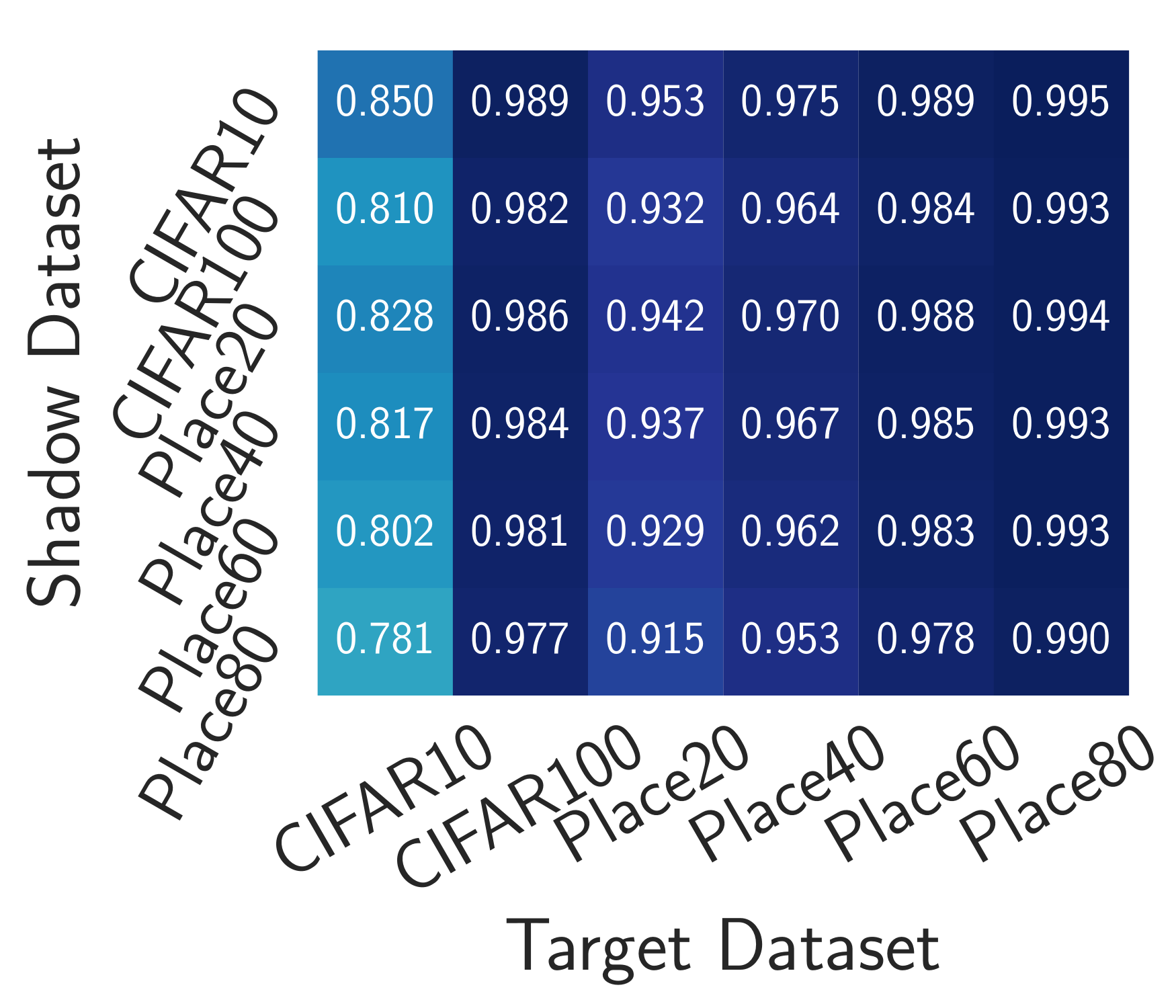}
\caption{ResNet18}
\label{figure:RQ2_attack_resnet18_no_aug}
\end{subfigure}
\begin{subfigure}{0.45\columnwidth}
\includegraphics[width=\columnwidth]{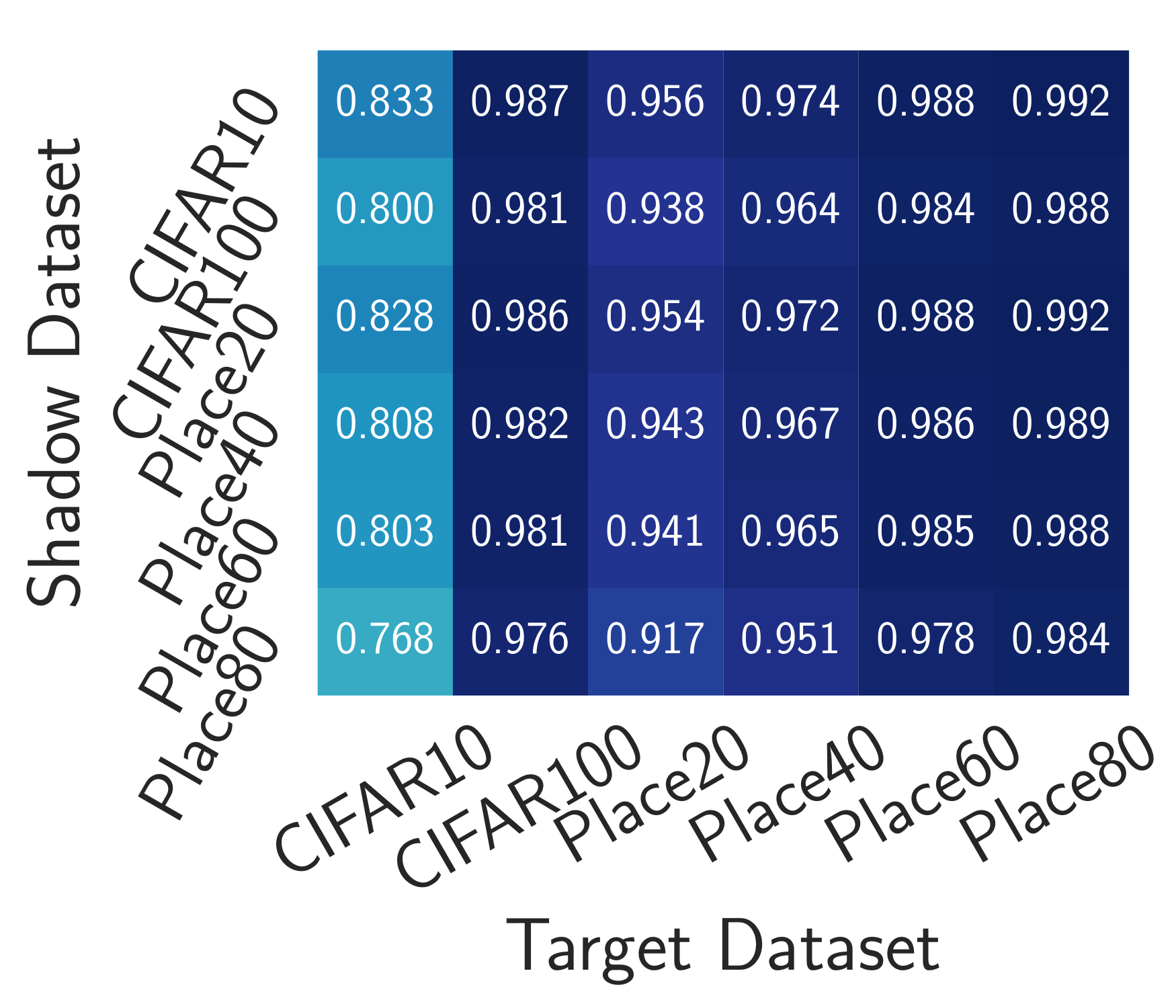}
\caption{ResNet34}
\label{figure:RQ2_attack_resnet34_no_aug}
\end{subfigure}
\caption{The performance of membership inference attack (\texttt{NN-top3}) when the shadow dataset comes from different distributions of the target dataset.}
\label{figure:RQ2_relax_assumption_shadow_dataset_no_aug_appendix} 
\end{figure*}

\begin{figure*}[!ht]
\centering
\begin{subfigure}{0.5\columnwidth}
\includegraphics[width=\columnwidth]{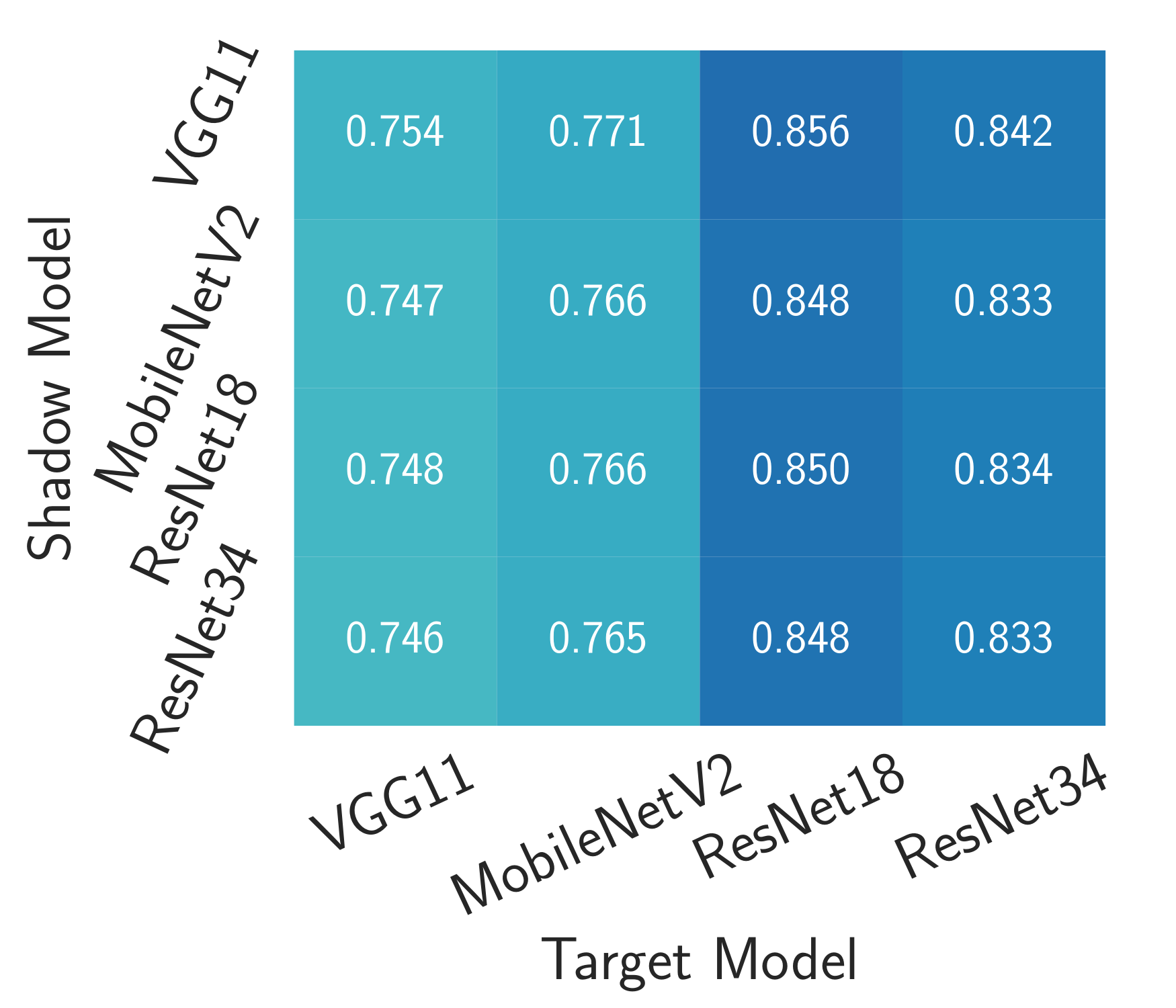}
\caption{CIFAR10}
\label{figure:RQ2_attack_CIFAR10_no_aug}
\end{subfigure}
\begin{subfigure}{0.5\columnwidth}
\includegraphics[width=\columnwidth]{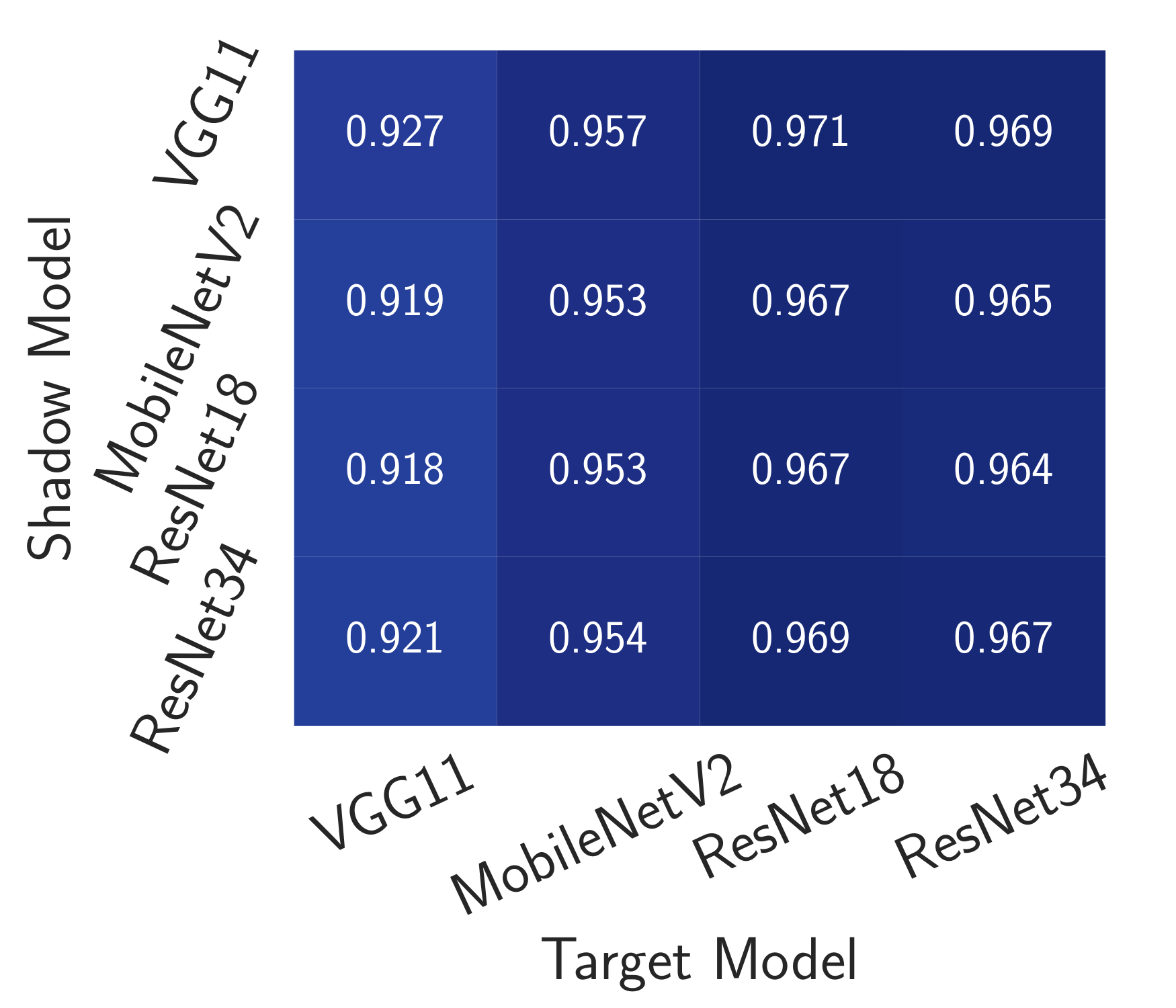}
\caption{Place40}
\label{figure:RQ2_attack_Place40_no_aug}
\end{subfigure}
\begin{subfigure}{0.5\columnwidth}
\includegraphics[width=\columnwidth]{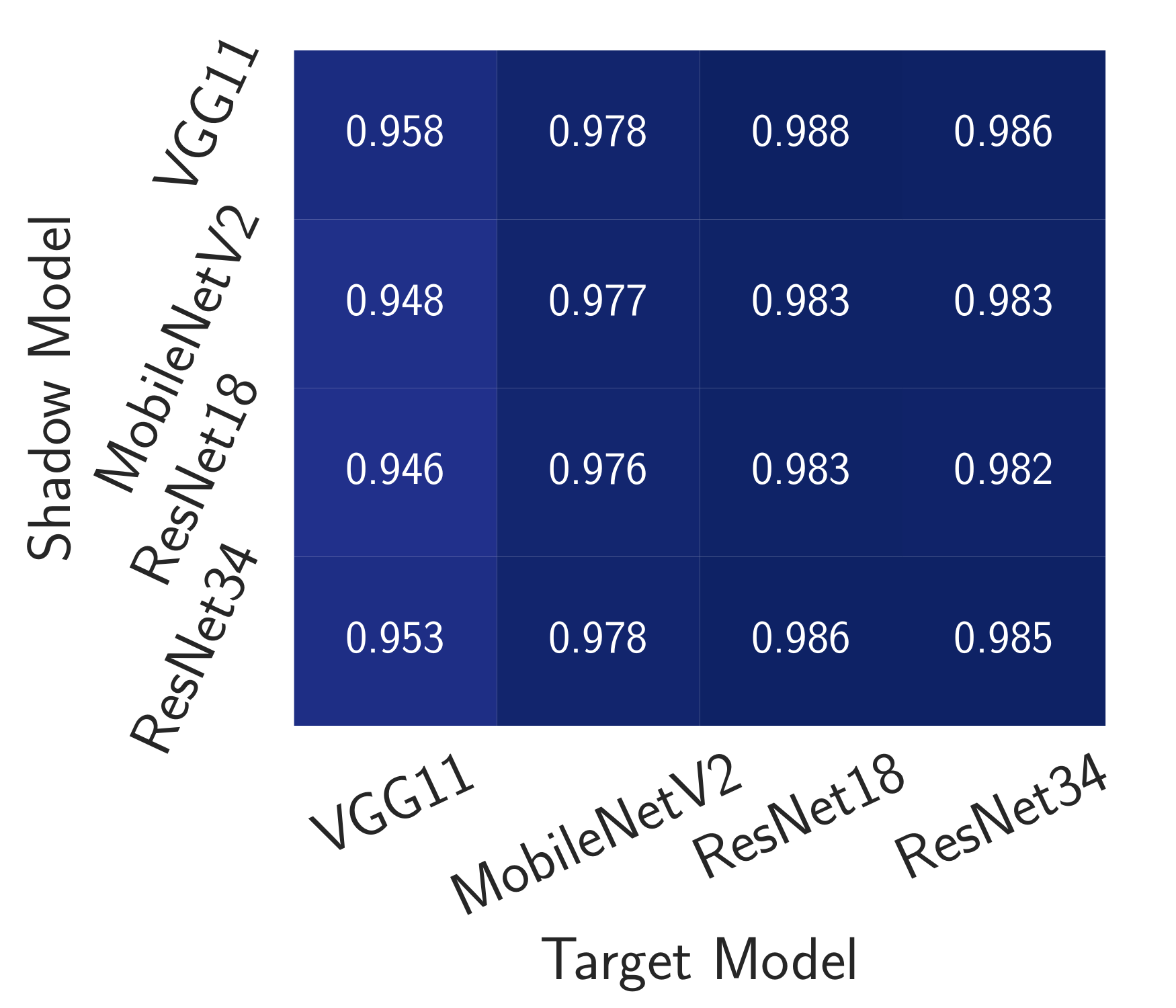}
\caption{Place60}
\label{figure:RQ2_attack_Place60_no_aug}
\end{subfigure}
\begin{subfigure}{0.5\columnwidth}
\includegraphics[width=\columnwidth]{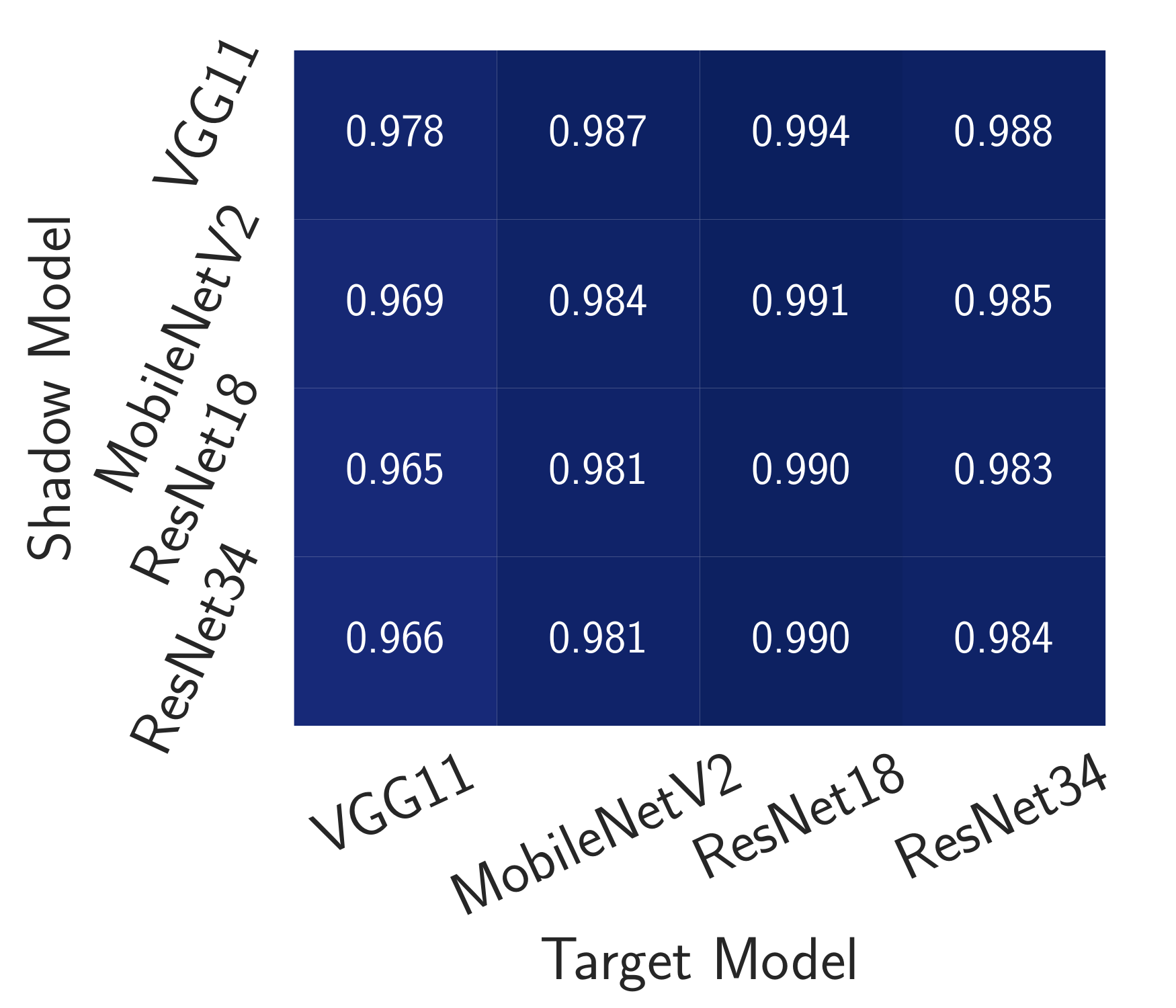}
\caption{Place80}
\label{figure:RQ2_attack_Place80_no_aug}
\end{subfigure}
\caption{The performance of membership inference attack (\texttt{NN-top3}) when the shadow model has different architecture compared to the target model.}
\label{figure:RQ2_attack_relax_assumption_shadow_model_no_aug_appendix} 
\end{figure*}

\end{document}